\documentclass[reprint, amsmath, amssymb, aip, jcp, onecolumn]{revtex4-2}

\usepackage{graphicx}
\usepackage{newtxtext}
\usepackage{newtxmath}
\usepackage{natbib}
\usepackage{hyperref}
\hypersetup{
    colorlinks = true,
    urlcolor   = blue,
    citecolor  = black,
}

\newcommand{\RomanNumeralCaps}[1]
\linenumbers

\usepackage[T1]{fontenc}
\usepackage[utf8]{inputenc}

\usepackage[version=3]{mhchem} 

\usepackage{multirow}
\usepackage{dcolumn}
\usepackage{bm}
\usepackage[caption=false]{subfig}
\usepackage[normalem]{ulem} 
\usepackage{makecell} 
\usepackage{booktabs} 
\usepackage{xspace}

\usepackage[per-mode=reciprocal, range-units=single, range-phrase=--]{siunitx} 
\DeclareSIUnit[number-unit-product = {}]\sinumber{{\scriptstyle\#}}

\usepackage{cleveref}
\crefname{figure}{figure}{figures}
\Crefname{figure}{Figure}{Figures}
\crefname{table}{table}{tables}
\Crefname{table}{Table}{Tables}
\crefname{equation}{equation}{equations}
\Crefname{equation}{Equation}{Equations}

\usepackage[textsize=tiny,prependcaption,colorinlistoftodos,textwidth=1.5cm]{todonotes}   

\def \rb {\mathbf{r}}
\def \dr {\, \mathrm{d}\mathbf{r}}

\DeclareSIUnit\bar{bar}
\DeclareSIUnit\angstrom{\text{\AA}}

\def \vb {\mathbf{v}}
\def \d {\, \mathrm{d}}

\def \kB {k_\mathrm{B}}

\def \ext {\mathrm{ext}}

\def \mw {M}

\def \divd {\nabla \!\boldsymbol{\cdot}\!}

\makeatletter
\newcommand{\thickhline}{%
    \noalign {\ifnum 0=`}\fi \hrule height 1pt
    \futurelet \reserved@a \@xhline
}
\newcolumntype{"}{@{\hskip\tabcolsep\vrule width 1pt\hskip\tabcolsep}}
\makeatother

\newcommand{\Dumux}{{Du\-Mu$^\text{x}$}\xspace}
\newcommand{\feos}{{FeO$_\mathrm{s}$}\xspace}
\newcommand{\epssf}{{\varepsilon_\mathrm{sf}^*}\xspace}
\def\CC{{C\nolinebreak[4]\hspace{-.05em}\raisebox{.4ex}{\tiny\bf ++}}\xspace}

\begin{document}

\title{Modelling Interfacial Dynamics Using Hydrodynamic Density Functional Theory: Dynamic Contact Angles and the Role of Local Viscosity
}

 \author{Benjamin Bursik}
 \affiliation{Institute of Thermodynamics and Thermal Process Engineering, University of Stuttgart, Pfaffenwaldring~9, 70569~Stuttgart, Germany}
 \author{Rolf Stierle} 
 \affiliation{Institute of Thermodynamics and Thermal Process Engineering, University of Stuttgart, Pfaffenwaldring~9, 70569~Stuttgart, Germany}
 \author{Hamza Oukili} 
 \affiliation{Institute for Modelling Hydraulic and Environmental Systems, University of Stuttgart, Pfaffenwaldring~61, 70569~Stuttgart, Germany}
 \author{Martin Schneider} 
 \affiliation{Institute for Modelling Hydraulic and Environmental Systems, University of Stuttgart, Pfaffenwaldring~61, 70569~Stuttgart, Germany}
 \author{Gernot Bauer} 
 \affiliation{Institute of Thermodynamics and Thermal Process Engineering, University of Stuttgart, Pfaffenwaldring~9, 70569~Stuttgart, Germany}
 \author{Joachim Gross} 
 \affiliation{Institute of Thermodynamics and Thermal Process Engineering, University of Stuttgart, Pfaffenwaldring~9, 70569~Stuttgart, Germany}
 
\begin{abstract}
  
  Hydrodynamic density functional theory (DFT) is applied to analyse dynamic contact angles of droplets in order to assess its predictive capability regarding wetting phenomena at the microscopic scale and to evaluate its feasibility for multiscale modelling. 
  Hydrodynamic DFT  incorporates the influence of fluid-fluid and solid-fluid interfaces into a hydrodynamic theory by including a thermodynamic model based on classical DFT for the chemical potential of inhomogeneous fluids. It simplifies to the isothermal Navier-Stokes equations far away from interfaces, thus connecting microscopic molecular modelling and continuum fluid dynamics. 
  In this work we use a Helmholtz energy functional based on the perturbed-chain  statistical associating fluid theory (PC-SAFT) and the viscosity is obtained from generalised entropy scaling, a one-parameter model which takes microscopic information of the fluid and solid phase into account. 
  Deterministic (noise-free) density and velocity profiles reveal wetting phenomena including 
  different advancing and receding contact angles,
   the transition from equilibrium to steady state and the rolling motion of droplets. Compared to a viscosity model based on bulk values, generalised entropy scaling provides more accurate results, which stresses the importance of including microscopic information in the local viscosity model. Hydrodynamic DFT is transferable as it captures the influence of different external forces, wetting strengths and 
   (molecular) 
   solid roughness. For all results good quantitative agreement with non-equilibrium molecular dynamics simulations is found, which  
   emphasises that hydrodynamic DFT is able to predict wetting phenomena at the microscopic scale. 
  \end{abstract}
  \maketitle

\section{Introduction}\label{sec:intro}
Wetting is defined as the collective set of phenomena that occur when a solid is exposed to a fluid and it comprises a number of interfacial effects. These include the interplay of capillary forces, which are net forces on the fluid, especially in the three-phase contact region, resulting from solid-fluid and fluid-fluid interactions as well as external (e.g.\ gravitational) forces. In the dynamic case viscous forces additionally influence the wetting behaviour. 
Wetting is important in several industrial areas \citep{deGennes2004capillarity,bonn2009wetting}, such as in the chemical industry for the spreading of paints \citep{schoff1992wettability,prajapati2024role}, in soil science for the study of the penetration of liquids into porous rocks \citep{melrose1965wettability,garfi2022determination}, or in the construction industry for waterproofing of concrete \citep{mccarter1998surface,muhammad2015waterproof}. It is also ubiquitous in life sciences, e.g.\ the rise of sap in plants, the adhesion of parasites on wet surfaces or even the wetting of the eye aided by special proteins \citep{deGennes2004capillarity,holly1971wettability}. 
The static (equilibrium) contact angle $\theta$ of sessile droplets is an important measure of the static wetting behaviour and has been the subject of several theoretical \citep{becker2014contact,graham2000contact,jasper2019generalized,vafei2005theoretical, pereira2012equilibrium, yatsyshin2021surface} and experimental \citep{vafei2006effect,sharp2011contact,schuster2015influence,ghasemi2010sessile} studies. 
In the case of total wetting $\theta=\SI{0}{\degree}$ and for total dewetting $\theta= \SI{180}{\degree}$, with partial wetting between these limits. 

\begin{figure}
  \centering
  \includegraphics[width=0.55\textwidth]{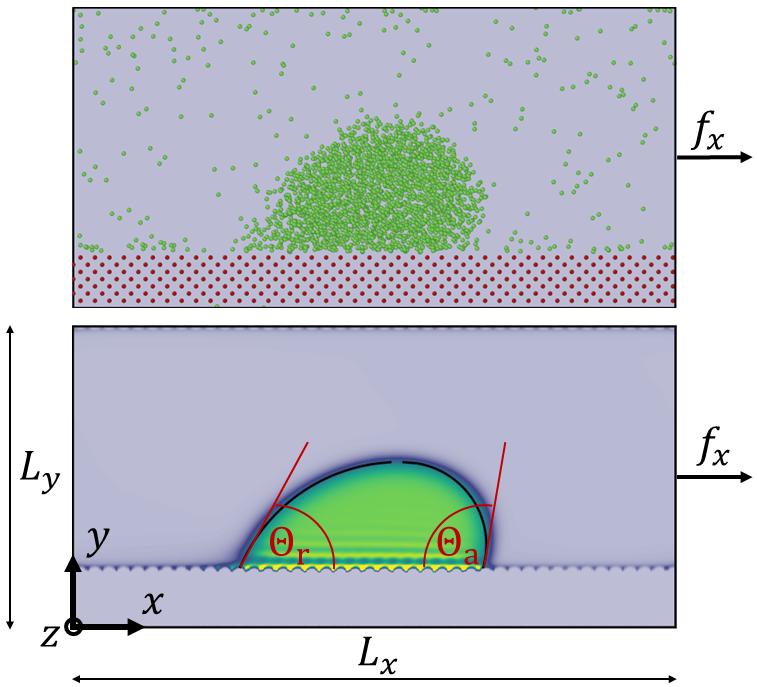}
  \caption{Snapshot of a droplet moving parallel to the solid-fluid interface due to an external force $f_x$ in a system with dimensions $L_x$ and $L_y$. Top: Atomistic model used for equilibrium and non-equilibrium molecular dynamic (NEMD) simulations with individual solid (red) and fluid (green) particles. Bottom: Density profile from hydrodynamic DFT with molecular layering, surface roughness as well as advancing and receding dynamic contact angles,  $\Theta_\mathrm{a}$ and $\Theta_\mathrm{r}$, respectively.}
  \label{fig:system}
\end{figure}

Analogously to the equilibrium case, the dynamic wetting behaviour, i.e.\ when a droplet placed on a solid spreads or is driven by an external force (e.g.\ gravity), can be characterised by dynamic contact angles \citep{bonn2009wetting}. 
Importantly, depending on the direction of the movement of the three-phase contact region, an advancing or receding dynamic contact angle,  $\Theta_\mathrm{a}$ or $\Theta_\mathrm{r}$, which generally differ, are observed as visualised in the lower part of \cref{fig:system}. Dynamic contact angles depend on velocity or, equivalently, on the driving force causing the movement of the contact region ($f_x$ in \cref{fig:system}) \citep{voinov1976hydrodynamics,cox1986dynamics}. 
The difference between advancing and receding contact angles is commonly attributed to chemical  or spatial inhomogeneities (roughness) of the solid surface and 
sometimes termed  dynamic contact angle hysteresis in literature \citep{butt2022contact}.

The dynamics close to the three-phase contact region (in macroscopic studies often called contact line) are at the centre of research in many works \citep{jacqmin2000contact,narhe2004contact,decker1997contact,koplik1988molecular,lee2022contact}. This can be attributed to the inherent challenges of modelling the time evolution of contact lines, where continuum fluid dynamics fails due to a stress singularity. Particularly, the widely used no-slip boundary condition requires the contact line to be at rest, which is obviously not correct for spreading droplets \citep{huh1971hydrodynamic}. 
Since the dynamic behaviour of the contact line is strongly affected by molecular interactions, it needs to be studied at the microscopic scale of individual molecules (i.e.\ a few \SI{}{\angstrom} to \SI{}{\nano\meter}). 

Several routes have been taken to address this challenge. Microscopic information can be utilised as an input for macroscopic models. This procedure is followed, for example, by sharp interface models, where interfaces are taken to be infinitely thin. These models describe a slip length based on microscopic considerations, which is then included in continuum fluid dynamics models to avoid the stress singularity \citep{hocking1983spreading,hocking1992rival,eggers2004hydrodynamic}.
A second route is through (atomistic) molecular simulations, where equilibrium molecular dynamics and non-equilibrium  molecular dynamics (NEMD) simulations provide insights into a variety of microscopic phenomena related to wetting \citep{hong2009static,koplik1988molecular,lee2022contact,li2018dynamic,li2019dynamic,yuan2013multiscale}.
MD provides detailed results, but is limited to small length- and time-scales. A further challenge for these models is that MD provides quantities with statistical uncertainties whereas macroscopic (continuum) models are deterministic. This effectively limits models to using averaged or fitted parameters, which are determined from MD, in macroscopic models instead of density and velocity fields \citep{nakamura2013dynamic,diewald2020molecular,qian2006variational}. 
A third approach are diffuse interface models, which consider the interfaces to be of finite thickness. This physical thickness is typically in the order of few molecular diameters \citep{bonn2009wetting}, which renders numerical resolution of macroscopic problems computationally expensive \citep{shokrpour2019binary}. In many studies, however, these diffuse interface models make no attempt to resemble the physical interface; rather, the diffuse interface is a numerical mean to distinguish and track the interface between fluid phases. 
These models avoid the stress singularity at the contact line and provide evolution equations for density or composition and sometimes also for velocity fields. Successful examples are models based on the coupled Cahn-Hilliard/Navier-Stokes \citep{cahn1958free,novickcohen2008cahn, demont2023numerical},  Navier-Stokes-Korteweg equations \citep{dunn1985thermomechanics,heida2010compressible,diehl2016numerical} or dynamic density functional theory (DDFT) \citep{marconi1999dynamic,marconi2000dynamic}.     

The Cahn-Hilliard equation was initially derived for the study of phase separation in a two-component system, where a phase-field variable or order parameter is employed to differentiate between the two immiscible phases. The coupled Cahn-Hilliard/Navier-Stokes equations include the gradient of the order parameter in the momentum balance to describe the time-evolution of the velocity and the order parameter for incompressible fluids. 
While the Cahn-Hilliard/Navier-Stokes equations have been applied to the study of contact region dynamics in incompressible systems \citep{yue2011wall,yue2010sharp,jacqmin2000contact,qian2006variational}, the approach is not applicable to compressible fluids. 

For compressible two-phase systems, the Navier-Stokes-Korteweg equations can be employed \citep{korteweg1901sur,dunn1985thermomechanics,heida2010compressible,diehl2016numerical}. The so-called Korteweg tensor, a constitutive relation which includes density gradients and accounts for capillarity effects, is incorporated into the momentum balance of the Navier-Stokes equations in addition to viscous stresses. 
Microscopic investigations of dynamic contact angles using the Navier-Stokes-Korteweg model yield accurate results when compared to MD simulations \citep{diewald2019navier,diewald2020molecular}. The contact angle in the Navier-Stokes-Korteweg model, however, is prescribed by an input parameter chosen such that the contact angle agrees with the MD simulations. Additionally, in order to obtain the correct vapour-liquid surface tension, a second parameter, $\kappa$, needs to be determined, e.g.\ from MD simulations. 
The Helmholtz energy functional, which is used in Cahn-Hilliard/Navier-Stokes and Navier-Stokes-Korteweg to model phase behaviour and capillary forces, only contains information of the immediate surroundings (by means of density gradients), and  molecular layering at the  solid-fluid interface can therefore not be resolved \citep{diewald2020molecular,sauer2017classical}.

Dynamic density functional theory (DDFT) is an alternative diffuse interface model. It can be viewed as a dynamic extension of classical density functional theory (DFT), which in general employs non-local Helmholtz energy functionals, to non-equilibrium situations \citep{teVrugt2020classical}. 
It provides time evolution equations for the  density and in some extensions also for the momentum density. 
It was initially derived by \citet{marconi1999dynamic,marconi2000dynamic} based on the Langevin equation.  While stochastic and deterministic versions of DDFT can be obtained \citep{archer2004dynamical_stochastic}, this paper focuses on the deterministic version.  
 Alternative routes for the derivation of DDFT are based on integrating the Smoluchowski equation
(see e.g.\ \citet{archer2004dynamical}) or using the projection operator formalism \citep{teVrugt2019mori}, 
where
the phase-space probability density is projected onto the density (see e.g.\ \citet{yoshimori2005microscopic,espanol2009derivation}).
A theoretical existence proof based on the Liouville equation was provided by \citet{chan2005timedependent}. \Citet{archer2009dynamical} derived the first DDFT for atomic/molecular fluids including a momentum balance, whereas \citet{goddard2013multispecies} extended DDFT towards mixtures of colloids. 
DDFT was applied to spinodal decomposition \citep{archer2004dynamical}, to phase separation of colloids confined in a cavity \citep{archer2005dynamical}, to determine the van Hove correlation function \citep{hopkins2010vanHove,stopper2015modeling,stopper2015dynamical,stopper2016structural} and to study the phase transition of colloidal systems under shear influence \citep{stopper2018nonequilibrium}.
In addition, DDFTs including inertia \citep{archer2006dynamical}, hydrodynamic interactions between colloids \citep{goddard2012general,goddard2013unification,duran2016dynamical,goddard2016dynamical} or fluctuations \citep{donev2014dynamic,archer2004dynamical,duran2017general,russo2021finite,russo2020memory} were developed. 

In contrast to  equilibrium DFT, DDFT is not an exact theory, since the so-called adiabatic approximation is employed:  The non-equilibrium two-body spatial correlation function is assumed to be the same as in equilibrium (for the same density profile) \citep{marconi1999dynamic,marconi2000dynamic,archer2009dynamical,teVrugt2020classical}. Consequently, the equilibrium Helmholtz energy functional  can be applied to the non-equilibrium system \citep{stierle2021hydrodynamic}. A drawback of this approximation is that history-dependent superadiabatic forces (e.g.\ memory effects) can not be described by DDFT. 
An exact version of DDFT which includes superadiabatic forces is given by power functional theory \citep{schmidt2013power,brader2015power,deLasHeras2018velocity,schmidt2018power}.

\Citet{stierle2021hydrodynamic} proposed hydrodynamic DFT based on a variational principle \citep{herivel1955derivation,serrin1959mathematical}. 
It describes the time-evolution of density and momentum including inertial effects and viscous dissipation for pure substances and mixtures. 
The Cauchy pressure tensor is modelled assuming a Newtonian fluid with a spatially varying viscosity coefficient. 
A DFT term with the functional derivative of the Helmholtz energy functional is included in the momentum balance and non-local Helmholtz energy functionals can be employed. 
In a recent work, \citet{nold2024hydrodynamic} employ a similar approach to study the fluid close to the contact line. The no-slip boundary condition is assumed between the solid and the fluid. The evolution from a non-equilibrium contact angle to equilibrium is investigated thereby analysing the influence of compression and shear on the slip length and contact line friction. 
 The DFT term, which appears in hydrodynamic DFT captures molecular forces, such as capillary forces, in a predictive manner. The predictive power is not limited to vapour-liquid interfaces, but also captures adsorption and wetting effects. While not restricted to a certain Helmholtz energy functional, it was combined with a Helmholtz energy functional based on the PC-SAFT \citep{gross2000application,gross2001perturbed,gross2002application,gross2003modeling} equation of state, which shows good agreement with experimental data in a wide range of equilibrium applications \citep{gross2009density,sauer2019prediction,rehner2018surface,rehner2021surfactant,nitzke2023,bursik2024predicting,stierle2024classical}. Specifically, density profiles in confined systems \citep{sauer2017classical,bursik2024viscosities}, adsorption isotherms \citep{sauer2019prediction,stierle2024classical} and contact angles of sessile droplets including molecular layering effects \citep{sauer2018prediction} are accurately predicted. 
Other Helmholtz energy functionals, such as the one describing density gradient theory and leading to the Navier-Stokes-Korteweg equation (with a less accurate description of solid-fluid interfaces), can also be employed.
An important feature of hydrodynamic DFT is that in a bulk phase, i.e.\ far enough away from interfaces, the model is equal to the isothermal (continuum) Navier-Stokes equation \citep{stierle2021hydrodynamic}.

For these reasons we propose hydrodynamic DFT as a suitable candidate for a unified model applicable to the description of the dynamic behaviour of fluids -- and especially wetting phenomena -- from the microscopic to the macroscopic scale.
However, to date hydrodynamic DFT has only been applied in a qualitative study of coalescence phenomena in one dimension \citep{stierle2021hydrodynamic}. The contribution of this paper is, thus, to conduct a (i) higher-dimensional, (ii) quantitative, (iii) wetting-related investigation of the predictive capabilities of hydrodynamic DFT in order to assess its potential for modelling dynamic wetting behaviour.
In particular, we apply hydrodynamic DFT to predict dynamic contact angles of two-dimensional microscopic sessile droplets driven by an external force. 
We use the already mentioned Helmholtz energy functional based on the PC-SAFT equation of state to model methane as an exemplary fluid.  We obtain local values for the shear viscosity from a recently developed generalised entropy scaling model \citep{bursik2024viscosities}. We analyse the importance of this local transport coefficient model by contrasting the results obtained from a model for shear viscosities in bulk phases, as it is typically employed in continuum approaches. All results are assessed by comparison to NEMD simulations for a Lennard--Jones fluid using PC-SAFT parameters for methane.
Because this study focuses on the dynamic behaviour of droplets, with emphasis on the viscosity model, we choose molecular models for which reasonable agreement for (static) equilibrium properties can be expected. The PC-SAFT model has been shown to provide rather satisfactory predictions of the equilibrium properties of the Lennard--Jones fluid  \citep{sauer2017classical}. We consider a Lennard-Jones fluid using methane parameters to generate illustrative results that are more readily interpretable.    

This paper is structured as follows: We present the balance equations of hydrodynamic DFT, the  Helmholtz energy functional based on PC-SAFT  and the shear viscosity model in \cref{sec:theory}. Numerical details on the discretization, a description of NEMD simulations and the procedure for determining contact angles from density profiles
are given in \cref{sec:methods}. In \cref{sec:results} we present and discuss results for density and velocity profiles including advancing and receding contact angles from hydrodynamic DFT, using the generalised entropy scaling model and a viscosity model for bulk phases, and compare them to results from NEMD simulations.

\section{\label{sec:theory} Hydrodynamic Density Functional Theory}

First, we introduce classical (equilibrium) DFT, followed by a short presentation of the hydrodynamic DFT model. The local shear viscosity is modelled using generalised entropy scaling \citep{bursik2024viscosities}, which is summarised subsequently.   

\subsection{Classical (Equilibrium) Density Functional Theory \label{sec:theoryDFT} }
  Classical DFT is an exact theory which describes inhomogeneous fluids in equilibrium using a grand canonical density functional $\Omega$ at constant chemical potential $\mu$, volume  $V$ and temperature $T$ according to 
  \begin{equation}\label{eq:Omega}
  \Omega\left[\rho(\rb)\right]=F\left[\rho(\rb)\right]-\int \rho(\rb) \left(\mu-V^\mathrm{ext}(\rb)\right)\dr
\end{equation}
where $F$ is the Helmholtz energy functional and $\rho$ is the density profile which depends on the spatial coordinate $\rb$. The square brackets denote that $F$ and $\Omega$ are functionals of the (number) density $\rho$.  
In this work, we split the external potential $V^\mathrm{ext}$ into two contributions according to 
\begin{equation} \label{eq:vext}
  V^\mathrm{ext}= V^\mathrm{ext,g} + V^\mathrm{ext,sf}
\end{equation}
where $V^\mathrm{ext,g}$ captures the effect of external potentials, such as gravity, on the fluid. The quantity $V^\mathrm{ext,sf}$ captures the interactions of solid interaction sites with the fluid and therewith the effect of confinement. From the perspective of a fluid, the solid acts as a (solid-fluid) external potential $V^\mathrm{ext,sf}$.

At equilibrium the grand canonical functional from \cref{eq:Omega} has a minimum and becomes  the grand potential. Mathematically, this leads to the Euler-Lagrange equation  
\begin{equation}  \label{eq:ELE}
  \frac{\delta F[\rho]}{\delta \rho(\rb)}-\mu+V^\mathrm{ext}(\rb)=0 \qquad\qquad
\end{equation}
which includes the functional derivative $\frac{\delta F[\rho]}{\delta \rho(\rb)}$ and allows to calculate the equilibrium density profile. 
In the context of \cref{eq:ELE}, we use $\rho(\rb)$ to denote the equilibrium  density, while in latter chapters $\rho(\rb)$ is used for non-equilibrium density fields.

In this work, the Helmholtz energy functional is based on the PC-SAFT model, i.e.\ for homogeneous phases the PC-SAFT equation of state is obtained. In PC-SAFT fluid molecules are assumed to be chains of $m_\mathrm{f}$ tangentially bonded spheres, which are called segments. $m_\mathrm{f}=1$ is used for the molecules in this work, but an extension to non-spherical molecules is readily available \citep{sauer2017classical}.
For this choice of parameters the PC-SAFT model reduces to a model for spherical molecules, which provides accurate results for the Lennard--Jones fluid. 
Each segment has a diameter parameter $\sigma_\mathrm{ff}$ and an energy interaction parameter $\varepsilon_\mathrm{ff}$. The PC-SAFT parameters, which were previously fitted to vapour-liquid equilibrium data \citep{gross2001perturbed}, are taken from literature and given in \cref{sec:methods}. The Helmholtz energy functional, similarly to the PC-SAFT equation of state, comprises of additive contributions. For the molecules studied here (i.e.\ with $m_\mathrm{f}=1$),  the ideal gas, hard-sphere and dispersion contributions are employed as
\begin{subequations}
	\begin{align}\label{eq:F}
  F\left[\rho(\rb)\right]&=F_\mathrm{ig}\left[\rho(\rb)\right]+F_\mathrm{hs}\left[\rho(\rb)\right]
  +F_\mathrm{disp}\left[\rho(\rb)\right]\\
  &\equiv F_\mathrm{ig}\left[\rho(\rb)\right] + F_\mathrm{res}\left[\rho(\rb)\right]
\end{align}
\end{subequations}
defining the residual Helmholtz energy $F_\mathrm{res}\left[\rho(\rb)\right]$. 

The ideal gas contribution can be derived exactly from statistical mechanics. For the hard-sphere contribution, which models each segment as an impenetrable sphere, fundamental measure theory \citep{yu2002structures,roth2002fundamental} is employed. The dispersion contribution describes the attractive, non-polar interactions of chain molecules modelled here by a functional developed by \citet{sauer2017classical}. This model can be extended to chain molecules \citep{tripathi2005microstructure,tripathi2005microstructureII},  associating molecules \citep{yu2002fundamental,sauer2017classical} and polar molecules \citep{sauer2017classical,gross2005equation,gross2006equation,vrabec2008vapor}.   

The interactions between solid and fluid, are accounted for in the (solid-fluid) external potential $V^\mathrm{ext,sf}$. If the solid consists of atomistic Lennard--Jones interaction sites, the external potential due to dispersive interactions between the solid and the fluid can be calculated according to  
\begin{equation} \label{eq:vext_sf}
  V^\mathrm{ext,sf}(\rb)=\sum_{\zeta=1}^{N_\mathrm{s}} 4\varepsilon_{\zeta \mathrm{f}} \left(\left(\frac{\sigma_{\zeta \mathrm{f}}}{|\rb_\zeta-\rb|}\right)^{12}-\left(\frac{\sigma_{\zeta \mathrm{f}}}{|\rb_\zeta-\rb|}\right)^6\right)
\end{equation}
with the index of the Lennard--Jones interaction sites $\zeta$ and their position $\rb_\zeta$ as well as the total number of solid interaction sites $N_\mathrm{s}$. In this work, the solid and the fluid consist of only one species respectively, such that the solid-fluid interaction parameters are equal for each solid-fluid pair ($\sigma_{\zeta \mathrm{f}}=\sigma_{\mathrm{sf}}$ and $\varepsilon_{\zeta \mathrm{f}}=\varepsilon_{ \mathrm{sf}}$). Rather than defining $\sigma_{\mathrm{sf}}$ and  $\varepsilon_{ \mathrm{sf}}$, we prefer to specify solid interaction site parameters $\sigma_{\mathrm{ss}}$ and $\varepsilon_{ \mathrm{ss}}$ assuming the Berthelot--Lorentz combining rules as 
\begin{equation} \label{eq:combining}
  \begin{aligned}
      \sigma_{\mathrm{sf}}      & = \frac{\left(\sigma_\mathrm{ss} + \sigma_{\mathrm{ff}}\right)}{2} \\
      \varepsilon_{\mathrm{sf}} & = \sqrt{\varepsilon_\mathrm{ss} \varepsilon_{\mathrm{ff}}}
  \end{aligned}
\end{equation}
Knowing the positions and parameters of the solid interaction sites, the solid-fluid external potential $V^\mathrm{ext,sf}$ as given in \cref{eq:vext_sf}, can be evaluated and used in equilibrium and hydrodynamic DFT. To obtain a two-dimensional external potential $V^\mathrm{ext,sf}(x,y)$, the three-dimensional external potential is first calculated for the depth of a single unit cell of the solid in $z$-direction (cf. \cref{fig:system}) and then free-energy averaged in the same direction \citep{eller2021free}. 
The resulting external potential varies strongly perpendicular to the solid-fluid interface due to decaying solid-fluid interactions. Minor variations occur also in the direction parallel to the solid-fluid interface, which resemble the molecular roughness of the solid. More pronounced roughnesses are investigated by removing solid interaction sites at the interface to create stronger variations of the external potential parallel to the solid.  
Other external fields, such as gravity, are included by means of the gravitational external potential $V^\mathrm{ext,g}$. In this work, an external force is applied in hydrodynamic DFT simulations to induce the movement of liquid droplets. An external force is equivalent to a negative gradient in the (gravitational) external potential. 

DFT calculations are carried out using \feos \citep{rehner2023feos,rehner2021application}, a framework for equations of state and DFT. The system is discretized in a two-dimensional Cartesian grid with 256 grid points and periodic boundary conditions  in each direction. Initial density profiles for sessile droplets can be constructed from equilibrated droplets in a surrounding vapour phase or taken from MD results; both approaches provide the same final density profile within numerical accuracy. The same number of molecules as in the MD simulations projected to the two-dimensional system (i.e.\ the same average density), is chosen. 

 Since in the grand-canonical ($\mu,V,T$) ensemble, different (correct) solutions can be obtained from the Euler-Lagrange equation \labelcref{eq:ELE}, the number of molecules in the system is kept constant using a mathematical reformulation proposed by \citet{rehner2018surface}. Essentially, a minimisation with constraints is performed using Lagrange multipliers, during which the chemical potential is adjusted to achieve the desired ensemble-averaged number of molecules in the system (see \cref{sec:appendix_Nconst}). Picard iterations and an Anderson mixing scheme \citep{anderson1965iterative,anderson2019comments} are used to obtain the equilibrium density profile from \cref{eq:ELE}. The convolutions which appear in the Helmholtz energy functionals and in \cref{eq:rho_bar_ES_tildepsi}, are implemented using fast Fourier transforms \citep{stierle2020guide,sermoud2024classical}. 

\subsection{Model Equations}

We present the hydrodynamic DFT model following \citet{stierle2021hydrodynamic}, originally developed for pure fluids by \citet{archer2009dynamical}. Even though it is applicable to mixtures, we solely provide equations for pure substances to improve clarity. The equations comprise a mass balance (continuity equation) and momentum balance according to \citep{stierle2021hydrodynamic} 
\begin{subequations}
	\begin{align}
		\frac{\partial \left(\mw\rho\right)}{\partial t} + \divd (\mw\rho \vb) &=                                 0  \label{eq:ContinuityDDFT}                                                                                                                                            \\
		\frac{\partial (\mw\rho \vb)}{\partial t} + \divd \left(\mw\rho \vb \vb^\intercal \right) &=  - \rho \nabla  \left( \frac{\delta  F}{\delta \rho} +  V^\ext \right) - \divd \boldsymbol{\tau}  \label{eq:MomentumBalanceDDFT}
	\end{align}
	\label{eq:DDFTModel}
\end{subequations}

where $\mw$ is the molecular mass and $\vb$ is the velocity vector. The Helmholtz energy functional $F$ captures intermolecular fluid-fluid interactions and is described in \cref{sec:theoryDFT}. 
The DFT term $\rho \nabla \left( \frac{\delta F}{\delta \rho} \right)$ includes the functional derivative $\frac{\delta F}{\delta\rho}$ and captures the influence of fluid-fluid interfaces on the momentum, whereas solid-fluid interactions, e.g.\ in a confined system, and gravitational external  potentials are included using the external potential $ V^\mathrm{ext}$ according to \cref{eq:vext}. 
In bulk phases, i.e.\ sufficiently far from solid-fluid or fluid-fluid interfaces, the solid-fluid external potential $\rho \nabla V^\mathrm{ext,sf}$ is zero, the gravitational external potential term $\rho \nabla V^\mathrm{ext,g}$ becomes an external force $\rho f^\mathrm{ext,g}$ and the DFT term   $\rho \nabla \left( \frac{\delta  F}{\delta \rho} \right)$ simplifies to the pressure gradient $\nabla p$ using the Gibbs-Duhem equation. This results in the momentum balance for homogeneous fluids known from the isothermal Navier-Stokes equations.

The viscous stresses in \cref{eq:MomentumBalanceDDFT} are modelled assuming a Newtonian fluid according to

	\begin{equation}
		\boldsymbol{\tau}  = -\chi \left( \divd \vb \right) \mathbb{I} - \eta \left( \nabla\vb + (\nabla\vb)^\intercal - \frac{2}{3} \left( \divd \vb \right) \mathbb{I} \right)  \label{eq:CauchyPressureTensorSecondVisc}
	\end{equation}
where $\chi$ is the volume viscosity and $\eta$ is the shear viscosity. The first term in \cref{eq:CauchyPressureTensorSecondVisc} describes the dilatation, while the second, traceless term describes the viscous shear contribution \citep{stierle2021hydrodynamic}. In this work, the volume viscosity  $\chi$ is neglected as often done for liquid systems and more detail on the shear viscosity $\eta$ is given in \cref{sec:theoryES}.

\subsection{Modelling the Shear Viscosity \label{sec:theoryES}}

For hydrodynamic DFT, i.e.\ in inhomogeneous systems, local values for the shear viscosity can be determined by applying entropy scaling locally following our previous work \citep{bursik2024viscosities}, which was based on the homogeneous approach of \citet{loetgeringlin2018pure}. Instead of a residual entropy as in the homogeneous case, local values for the residual entropy density  $\tilde{s}_\mathrm{res}(\rb)$, an entropy per unit volume, can be calculated. From the residual Helmholtz energy density $f_\mathrm{res}(\rb)$, which is related to the Helmholtz energy by $F_\mathrm{res} = \int f_\mathrm{res}(\rb) \d \rb$, the residual entropy density is derived according to  
\begin{equation} \label{eq:sres_density}
  \tilde{s}_\mathrm{res}(\rb) = -\left(\frac{\partial f_\mathrm{res}(\rb)}{\partial T}\right)_{\rho,V}
\end{equation}

We define an dimensionless residual entropy profile per molecule $s_\mathrm{res}^\#(\rb)$ as
\begin{equation} \label{eq:sres}
    s_\mathrm{res}^{\#}(\rb) = \frac{\tilde{s}_\mathrm{res}(\rb)}{\bar{\rho}^\mathrm{ES}(\rb)k_\mathrm{B}}
\end{equation}
where a weighted density, an average density that takes non-local effects into account, is used as
\begin{equation} \label{eq:rho_bar_ES_tildepsi}
    \bar{\rho}^\mathrm{ES}(\rb) = \frac{3}{4\pi\left(\psi\varepsilon_{\mathrm{sf}}d\right)^3} \int \rho(\rb) \Theta(\psi\varepsilon_{\mathrm{sf}} d - |\rb-\rb'|) \d \rb
\end{equation}
with the temperature dependent, effective hard-sphere diameter \citep{gross2001perturbed} $d(T) = \sigma_\mathrm{ff} \bigl(1-0.12\exp(-3\frac{\varepsilon_\mathrm{ff}}{k_\mathrm{B}T}) \bigr)$ and the Heaviside step function $\Theta$. 
 An adjustable parameter $\psi$ is required to quantitatively capture the influence of solid-fluid interactions on the viscosity in the close vicinity of the interface  by varying the convolution radius $\psi\varepsilon_{\mathrm{sf}} d$. It was shown that once adjusted for a given system (by a single fluid-phase MD simulation), the parameter $\psi$ is transferable to different temperatures, densities, shear rates and solid-fluid interactions described by $\varepsilon_\mathrm{sf}$, which influences the wetting behaviour \citep{bursik2024viscosities}. In this work, $\psi=1$ is adjusted to the velocity profile of liquid-phase flow in the same geometry from a single steady state NEMD simulation and no wetting information enters the parameter adjustment (see \cref{sec:appendix_psi}).

A third-order polynomial ansatz function \citep{loetgeringlin2018pure,bursik2024viscosities} is then evaluated locally 
\begin{equation} \label{eq:poly3}
  \ln \left(\frac{\eta^\mathrm{Entr. Scal.}(\rb)}{\eta_\mathrm{ref}(\rb)}\right)=A + B s_\mathrm{res}^\#(\rb) + C  \left(s_\mathrm{res}^\#(\rb)\right)^2  + D \left(s_\mathrm{res}^\#(\rb)\right)^3
\end{equation}
where $A$, $B$, $C$ and $D$ are substance specific parameters, that were previously fitted to experimental data of pure substances in homogeneous phases by \citet{loetgeringlin2018pure}. 
The reference viscosity is defined by 
\begin{equation}
  \eta_\mathrm{ref}(\rb) = \eta_\mathrm{CE} + \bar{\rho}_\mathrm{s}(\rb)\sigma_\mathrm{ss}^3\eta_{\mathrm{s},\infty}
  \label{eq:inhomogeneousRefViscosity}
\end{equation}
The Chapman-Enskog viscosity $\eta_\mathrm{CE}$ \citep{hirschefelder1954molecular} is used as the reference for calculating the dimensionless viscosity in homogeneous systems and inhomogeneous systems without solid-fluid interfaces. It is calculated as  
\begin{equation}
  \eta_{\mathrm{CE}} = \frac{5}{16}\frac{\sqrt{M k_\mathrm{B}T/(N_\mathrm{A}\pi)}}{\sigma^2 \Omega^{(2,2)\#}}
  \label{eq:CEViscosity}
\end{equation}
where an empirical correlation \citep{neufeld1972empirical} is used for the dimensionless collision integral $\Omega^{(2,2)\#}$.
In the vicinity and within the solid, the reference viscosity needs to be adjusted, since the Chapman--Enskog viscosity is valid only for dilute vapour phases. The second contribution in \cref{eq:inhomogeneousRefViscosity} accounts for the low mobility of fluid molecules within the solid and the influence of the solid-fluid interface on viscosity \citep{bursik2024viscosities}. $\bar{\rho}_\mathrm{s} $ is the weighted density of solid interaction sites determined from 
\begin{equation}
  \bar{\rho}_\mathrm{s} =\int \rho_\mathrm{s}\left(\mathbf{r}'\right) \Theta\left(R-|\mathbf{r}-\mathbf{r}'|\right) \mathrm{d} \mathbf{r}'
\end{equation}
where $\rho_\mathrm{s}$ is the density of solid interaction sites.
 $\eta_{\mathrm{s},\infty}$ is the hypothetical viscosity of the fluid inside the solid far away from the solid-fluid interface. Since the latter is very difficult to obtain and in general, will be much larger than the fluid viscosity, it is here set to a large positive value.   

\section{Case Study and Numerical Methods \label{sec:methods}}

\subsection{Setup of the Contact Angle Study}

An exemplary snapshot of a moving droplet is provided in \cref{fig:system}. The individual Lennard--Jones particles from NEMD are shown in the top part, where the solid consists of Lennard--Jones particles in a lattice. The density profile of the fluid from hydrodynamic DFT is given in the lower part. An external force in $x$-direction $f_x$ is applied to induce the movement of the droplet and the surrounding fluid in both approaches. This is analogous to the gravitational force causing the movement of droplets on a vertical wall in macroscopic systems. Advancing and receding contact angles $\Theta_\mathrm{a}$ and $\Theta_\mathrm{r}$ were reported for NEMD simulations of droplets in similar systems \citep{hong2009static} and can be determined from the density profiles of hydrodynamic DFT\@. The equilibrium MD and NEMD simulations are carried out in three dimensions where the average density does not change in the $z$-direction (a side view is provided in \cref{fig:system}). Hydrodynamic DFT is simulated in a two-dimensional system with lengths $L_x$ and $L_y$. 

Since at the microscopic scale capillary forces have a much stronger influence on the droplet movement compared to the macroscopic scale, the external (body) force must be large compared to macroscopic driving forces such as gravity (see \cref{sec:appendix_externalForce}).  
Here, external forces ranging from $f_x=\SI{0.056}{\pico\newton}$ to $f_x=\SI{0.224}{\pico\newton}$ per particle are chosen, which for methane corresponds to an acceleration of about \SI{2.1e12}{\meter\per\second\squared} and \SI{8.4e12}{\meter\per\second\squared}, respectively. 
The dynamic contact angles take on values that can be observed for larger (macroscopic) droplets in earth's gravity field. 
All parameters used for the generation of results are listed in \cref{table:params}. 
The temperature is set to $T=\SI{120.02}{\kelvin}$ for all simulations in this work, which is well below the critical temperature
and a pressure of $p=\SI{0.191}{\mega\pascal}$ is obtained from the PC-SAFT model. 
 
\begin{table}
  \centering
  \renewcommand{\arraystretch}{1.2}
  \begin{tabular}{c  @{\hspace{5\tabcolsep}} c  @{\hspace{5\tabcolsep}} c }
      parameter & value & source \\
      \hline
      $\sigma_\mathrm{ff}$ & \SI{3.7039 }{\angstrom} & \citet{gross2001perturbed} \\
      $\varepsilon_\mathrm{ff}/\kB$ & \SI{150.03 }{\kelvin} & \citet{gross2001perturbed} \\
      $m_\mathrm{f}$ & \SI{1.0 }{} & \citet{gross2001perturbed} \\
      $\mw$ & \SI{16.043 }{\gram\per\mol} & \citet{gross2001perturbed} \\
      $A$ & \SI{-0.0595}{} & \citet{loetgeringlin2018pure} \\
      $B$ & \SI{-0.8908}{} & \citet{loetgeringlin2018pure} \\
      $C$ & \SI{-0.0348}{} & \citet{loetgeringlin2018pure} \\
      $D$ & \SI{-0.0177}{} & \citet{loetgeringlin2018pure} \\
      $\rho_\mathrm{cut}$ & \SI{5e-6}{\kilo\mol\per\cubic\meter} &  comparable to Entr. Scal.\\
      $\eta_\mathrm{l}$ & \SI{93.406}{\micro\pascal\second} & VLE methane \\
      $\eta_\mathrm{v}$ & \SI{4.633}{\micro\pascal\second} & VLE methane \\
      $\rho_\mathrm{l}$ & \SI{25.59}{\kilo\mol\per\cubic\meter}  & VLE methane \\
      $\rho_\mathrm{v}$ & \SI{0.2023}{\kilo\mol\per\cubic\meter}  & VLE methane \\     
      $\psi$ & \SI{1}{} & \makecell{adjusted to single NEMD \\simulation of liquid-phase\\  Poiseuille flow}  \\
  \end{tabular}
  \caption{PC-SAFT, generalised entropy scaling and bulk viscosity model parameters for methane. }
  \label{table:params}
\end{table}

\subsection{ Implementation of Hydrodynamic DFT in \Dumux }

In the following, the numerical model for solving \cref{eq:DDFTModel} is briefly described. For the discretization of temporal and spatial differential operators 
and for solving nonlinear and linear systems of equations, we rely on the open-source software package \citep{Koch2021Dumux} \Dumux.
\Dumux is based on the numerics framework DUNE \citep{Dune2021} and is a simulator for flow and transport processes, adaptable to various multiphysics problems. 
For the implementation of hydrodynamic DFT, its  modular architecture  allows \Dumux  to be coupled  with advanced thermodynamic models, such as DFT based on PC-SAFT, 
to predict molecular interactions that influence the fluid behaviour.

DFT calculations are conducted using \feos \citep{rehner2021application,rehner2023feos}, i.e.\ the evaluation of the Helmholtz energy functional $F$ and the functional derivative $\frac{\delta F}{\delta \rho}$ as well as  generalised entropy scaling for the viscosity calculation. To couple \Dumux and \feos, which is written in the Rust programming language, we build a \CC interface for the required \feos functions. This interface can then be dynamically linked and  accessed directly from \Dumux.  

For the discretization of coupled partial differential equations as in \cref{eq:DDFTModel}, it is well-known that collocated finite-volume methods or standard finite-element schemes lead to numerical instabilities. 
In this work, a staggered-grid finite-volume scheme \citep{harlow1965numerical} is applied, where densities are defined at element centres, while velocity components are placed on element faces, around which
dual control volumes are constructed. Such staggering of degrees of freedom naturally leads to a stable scheme. 
Further details and a compact notation of the discrete equations can be found in \citet{schneider2020}.
The DFT term $\rho \nabla \left( \frac{\delta F}{\delta \rho} + V^\ext\right)$ 
is discretized with a central difference approximation related to the dual control volumes, leading to its evaluation on element centres. 

When using an implicit time discretization scheme, such as an implicit Euler method, the convolutions appearing in the Helmholtz energy functional $F$ 
lead to a non-local stencil for this DFT term and consequently to dense matrices when considering the related functional derivatives within nonlinear solvers. 
As a result, this term is treated semi-implicitly, i.e.\ the derivatives are not directly accounted for within the nonlinear solver (in this case, Newton's method). 
This approach maintains the sparsity pattern of the matrices.

\subsection{ Non-equilibrium Molecular Dynamics Simulations } \label{sec:methods_MD}

Hydrodynamic DFT is assessed by comparing results for dynamic contact angles to NEMD simulations. Simulations are conducted using the Large-scale Atomic/Molecular Massively Parallel Simulator (LAMMPS, stable release 2 Aug 2023) \citep{thompson2022lammps} in a configuration analogous to that employed in our previous  development of generalised entropy scaling \citep{bursik2024viscosities}. 
The equations of motion in the canonical ensemble (constant $N,V,T$) are integrated by means of a velocity-Verlet method. The time step is set to $\Delta t^*=0.005t^*$ with $t^*= \frac{t}{\sigma_\mathrm{ff} \sqrt{\mw / \varepsilon_\mathrm{ff}}}$, where the asterisk~$*$ denotes dimensionless quantities. For the conversion from dimensionless numbers to real units, 
the PC-SAFT parameters of the fluid $\sigma_\mathrm{ff}$ and $\varepsilon_\mathrm{ff}$ and the molecular mass $\mw$ is employed. Effectively, we compare results for methane from hydrodynamic DFT using the PC-SAFT model to results from NEMD for the Lennard--Jones fluid, where the PC-SAFT parameters are used to convert dimensionless results to real units.

Three Nose-Hoover chains \citep{tuckermann2006liouville} with a damping time constant $t_\mathrm{D}^*=100\Delta t^*$ are employed as a thermostat for the temperature in the system. 
The usage of a global thermostat in an inhomogeneous non-equilibrium system can lead to temperature variations due to large local shear stresses. For the system studied here, the temperature is found to be constant throughout the majority of the droplet, with temperature deviations of up to 10\% in the contact regions as discussed in \cref{sec:appendix_temperature}.
The solid is modelled as a frozen rigid wall. 

As described above and visualised in \cref{fig:system}, NEMD simulations are performed in three dimensions, whereas a two-dimensional approach is employed for hydrodynamic DFT\@. A sessile cylinder is simulated with both methods. The same length $L_x$ and the same height $L_y$ are chosen for NEMD and hydrodynamic DFT, respectively. The depth of the third-dimension in NEMD is chosen as $L_z=L_x/3$ in order to balance computational time and statistics of the results. Periodic boundary conditions are applied in all directions for NEMD in accordance with hydrodynamic DFT. 
A solid block is introduced at the bottom of the system, which consists of Lennard--Jones sites at their equilibrium distance in a bcc-lattice with dimensions $L_x \times    5\sigma_\mathrm{ff}\times L_z$. Due to the periodic boundary conditions, fluid at the upper end of the system  interacts with the lower end of the solid. 
The Lennard--Jones parameters for solid-solid and fluid-fluid interactions are set to $\sigma_\mathrm{ff}=\sigma_\mathrm{ss}$ and $\varepsilon_\mathrm{ff}=\varepsilon_\mathrm{ss}$, which for the  solid-fluid interactions leads to  $\sigma_\mathrm{sf}=\sigma_\mathrm{ff}=\sigma_\mathrm{ss}$ using the Lorentz--Berthelot combining rules. However, the dimensionless solid-fluid energy interaction parameter $\epssf = \frac{\varepsilon_\mathrm{sf}}{\varepsilon_\mathrm{ff}}$ is varied as noted in the respective paragraphs or figures, which allows for the study of different wetting behaviours (see e.g.\ \citet{becker2014contact}).
The fluid-fluid and solid-fluid interactions are cutoff and shifted at $r_\mathrm{c,ff} = 5 \sigma_\mathrm{ff}$ and at $r_{\mathrm{c,sf}} = 4 \sigma_\mathrm{ff}$, respectively.

For the simulation of static contact angles, equilibrium MD simulations are performed. Following \citet{becker2014contact}, fluid particles were initially arranged in a cuboid on top of the solid. The initial density in the cuboid and in the gas need to be chosen such that a stable droplet is formed. These densities were taken from \citet{vrabec2006comprehensive} who studied droplets in a gas phase (without solid) and Lennard--Jones interactions with a cutoff at $r_\mathrm{c,ff} = 2.5 \sigma_\mathrm{ff}$. While not representing exactly the same system as in this work, these densities proved to be sufficient as an initial estimate for obtaining stable sessile droplets.   
A conjugate gradient method and an equilibration of \num{2.5e5} steps were used to obtain an equilibrated sessile droplet. Production runs of \num{4.75e6} steps were carried out where atom positions were written out every 1000 steps  to determine equilibrium/static density profiles and contact angles. Accumulation of net momentum due to numerical inaccuracies is avoided by removing the momentum in $x$- and $z$-direction of the centre of mass every time step. 

For dynamic contact angles the equilibrium system is utilised as an initial condition in the NEMD simulations and an external driving force is applied to induce the movement of the droplet. The external driving force parallel to the solid-fluid interface $f_x$ is added to the $x$-component of the force vector for each individual fluid atom in each time step. Therefore, the  $x$-component of the particle velocities is not included in the calculation of the fluid temperature needed for thermostatting. Particle positions and velocities are written to a file every 100 time steps. All errorbars provided in this work  denote the $95\%$ confidence interval. 

\subsection{Contact Angles from Density Profiles} \label{sec:methods_contactAngleMethodology}

\begin{figure}
  \centering
  \includegraphics[width=0.55\textwidth]{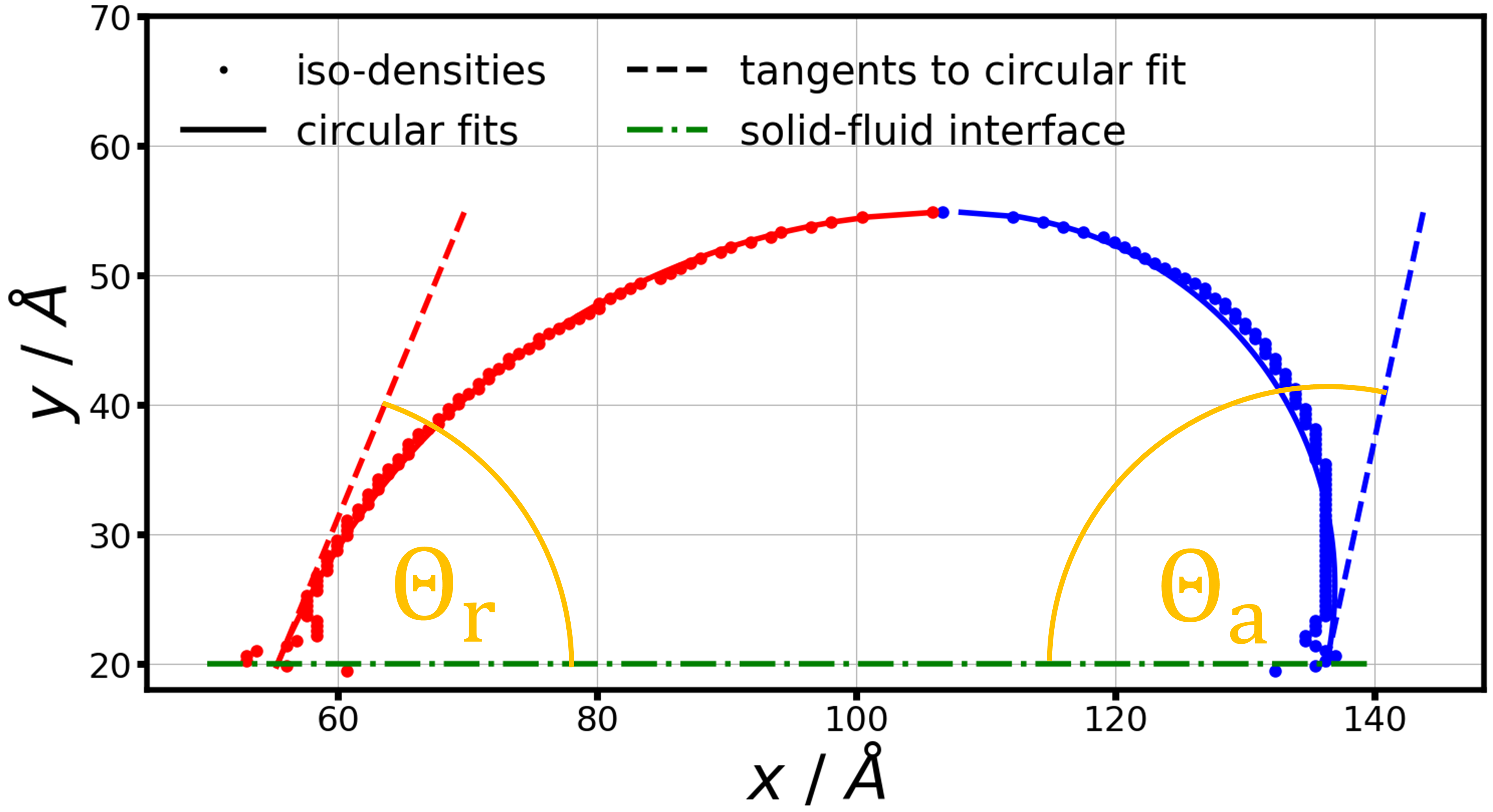}
  \caption{Visualisation of the methodology for determining contact angles from density profiles. The red and blue iso-density points represent the vapour-liquid interface of the droplet, the red and blue solid lines shows the circular fit, the red and blue dashed lines are the tangents to the circle at the solid-fluid interface. The latter is shown as a green dashed line.}
  \label{fig:contactAngleMethodology}
\end{figure}

The static and dynamic contact angles are determined from density profiles for both, hydrodynamic DFT and NEMD. 
For diffuse interface models the exact location of the vapour-liquid interface of the droplet is not unambiguously defined. One possible approach is the Gaussian convolution method, which can be employed to determine the vapour-liquid interface from atomic coordinates \citep{willard2010instantaneous}. Here we follow a procedure similar to \citet{sauer2018prediction} and \citet{heier2021molecular}, which is summarised in \cref{fig:contactAngleMethodology}. 
We also note that \citet{sauer2018prediction} performed a finite size study for static contact angles and confirmed convergence towards the contact angle determined from Young's equation. First, a 
dividing density $\rho_\mathrm{iso}=\SI{11.44}{\kilo\mol\per\cubic\meter}$ is chosen. From this, iso-density points are obtained, which represent the vapour-liquid interface of the droplet (red and blue points in \cref{fig:contactAngleMethodology}). Half circles are fitted to it for each half of the droplet (red and blue lines).  
This is necessary as the density profiles of moving droplets are, in general, not axis-symmetric. The contact angle can be calculated from the tangent to each half of the circle (red and blue dashed lines) at the solid-fluid interface (green dash-dotted line). 
The  molecular layering at the solid-liquid interface leads to oscillations in the density. As noted in the literature, an improvement of the fit can be obtained by disregarding the region very close to the solid-fluid interface \citep{sauer2018prediction,heier2021molecular}.  In addition, in the present geometry the specific location of the solid-fluid interface is not uniquely defined due to the molecular roughness of the solid. Minor differences in the contact angle value were observed by varying either the location of the solid-fluid interface, disregarding a certain region close the  interface or changing the dividing density $\rho_\mathrm{iso}$. Because this study is concerned with the comparison of hydrodynamic DFT and NEMD, we apply the same procedure for hydrodynamic DFT and NEMD results and thereby ensure comparability between both modelling frameworks.

\section{\label{sec:results}Results and Discussion}
In this section we present results from equilibrium and hydrodynamic DFT for sessile droplets on a solid and validate the results by comparison with equilibrium and non-equilibrium MD simulations, respectively.  

\subsection{ Equilibrium Droplets }\label{sec:results_static}
\begin{figure}
  \centering
  \subfloat{
      \centering
      \includegraphics[width=0.475\textwidth]{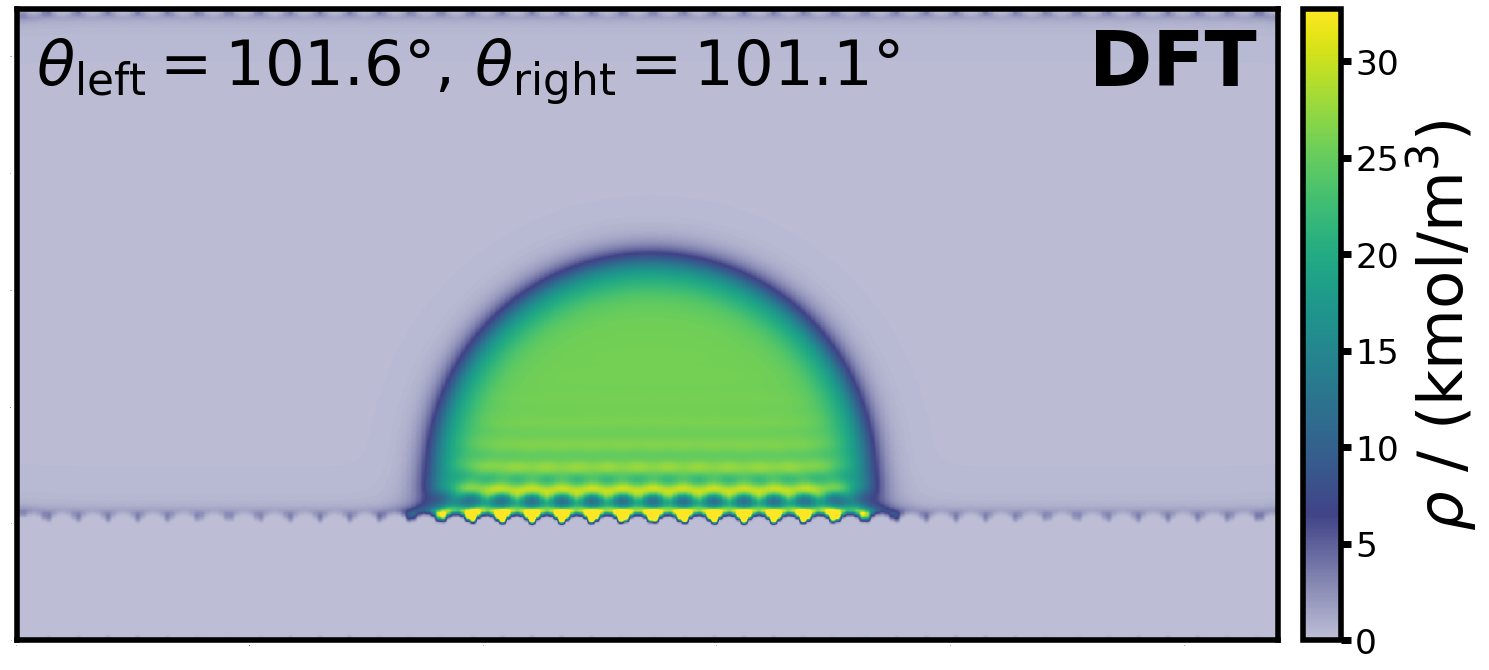}\label{fig:rho_static_dft}
 }
 \hfill
  \subfloat{
      \centering
      \includegraphics[width=0.475\textwidth]{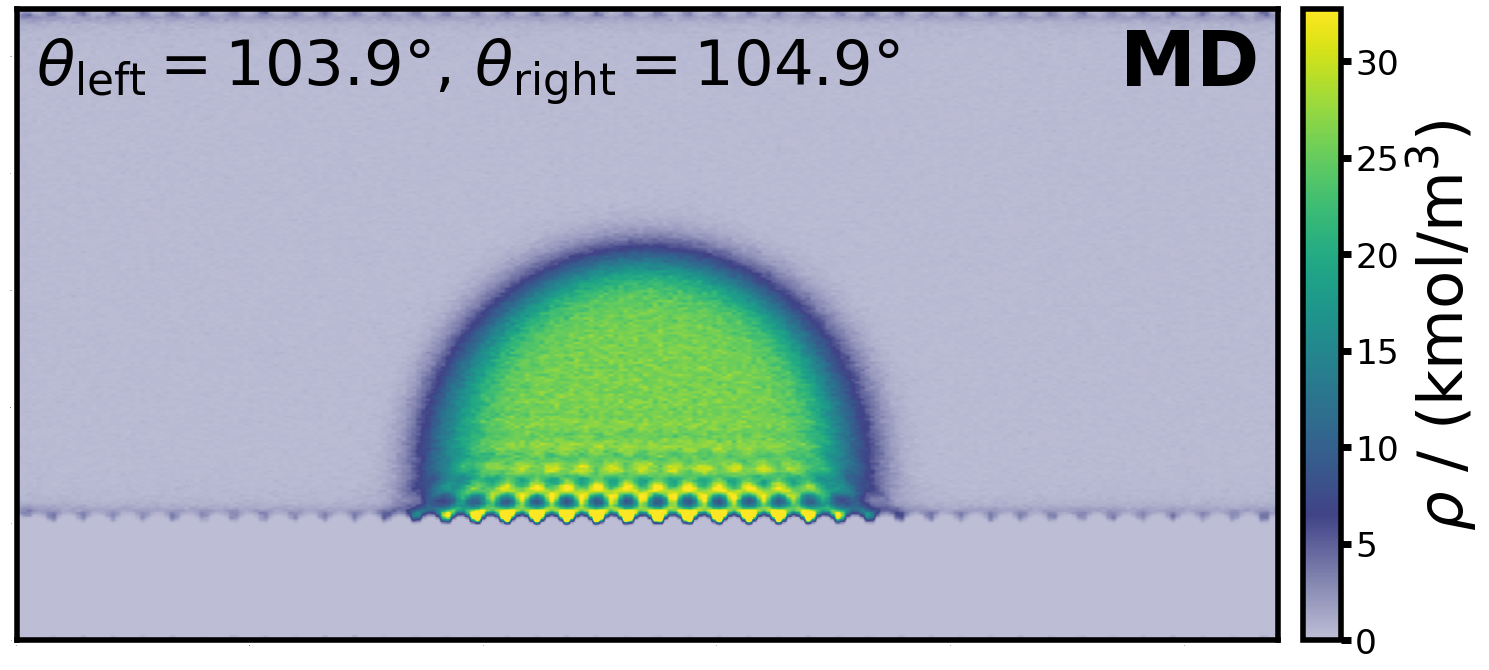}\label{fig:rho_static_emd}
 } 
 \caption{Density profile of equilibrium droplet ($f_x=0$) from DFT (left) and equilibrium MD (right) at $T=\SI{120.02}{\kelvin}$ with $\varepsilon_\mathrm{sf}^*=0.5$.}
  \label{fig:rho_static}
\end{figure}

Equilibrium droplets from DFT and equilibrium MD are used as initial profiles for the dynamic simulations in hydrodynamic DFT and NEMD, respectively. In order to achieve good agreement of the dynamic simulations, it is essential that the equilibrium densities agree. 
\Cref{fig:rho_static} visualises the two-dimensional density profiles of the droplets from equilibrium DFT (left) and equilibrium MD (right) for a solid-fluid energy interaction parameter $\varepsilon_\mathrm{sf}^*=0.5$. 
We note that the DFT calculations are performed in two-dimensions, whereas for MD a three-dimensional simulation was performed. In the MD simulations, the simulation box dimensions and the number of methane molecules were chosen to ensure that a stable or moving droplet adopts the shape of a cylindrical cap (entirely comparable to the geometry considered in the hydrodynamic DFT).
Results from DFT are predictions without adjustable parameters. 
Sessile droplets with a finite contact angle are obtained in both cases 
and two important phenomena are observed: 
First, inside the droplet alternating layers of varying density are observed along the normal direction of the solid-fluid interface. This molecular layering (as reported e.g.\ in \citet{sauer2018prediction,becker2014contact,lee2022contact}) extends for several layers into the lower part of the droplet. 
Second, the density profile exhibits a wave-like inhomogeneity in direction parallel to the solid-fluid interface, which is caused by the molecular roughness of the solid. This is true even outside of the droplets, which can be explained by the adsorption of particles from the vapour phase.

The droplets are nearly symmetrical and the small asymmetry is caused by the solid roughness. The mean contact angles from DFT are in good agreement to equilibrium MD with  \SI{101.4}{\degree}  and \SI{104.4}{\degree}, respectively. A similar minor underestimation of contact angles from DFT compared to molecular simulations was previously reported \citep{sauer2018prediction}.
Several reasons may contribute to it: The PC-SAFT model (here with methane parameters) used in DFT does not exactly reproduce the properties of the Lennard--Jones fluid, which is employed in MD \citep{sauer2017classical}. In addition, the different treatment of solid-fluid interactions (explicit pair interactions vs.\ external potential) and the different dimensionalities of the simulations (two- vs.\ three-dimensional) may add to the deviation. 
Overall, we find very satisfactory agreement of DFT compared to MD calculations, both, regarding density profiles and contact angles, thus providing a suitable basis for the investigation of dynamic phenomena.

\subsection{ Importance of Local Viscosity Model } \label{sec:results_eta}

\begin{figure}
  \centering
  \subfloat{
      \centering
      \includegraphics[width=0.475\textwidth]{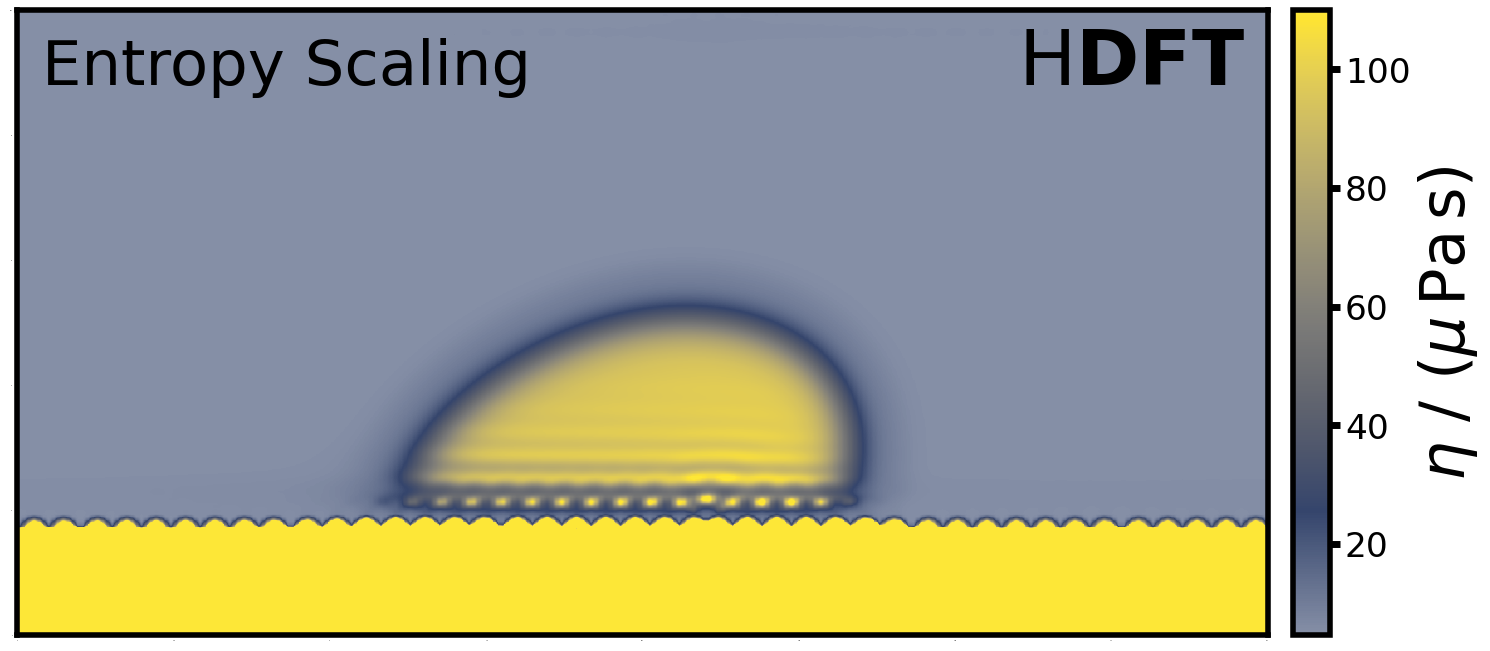}\label{fig:eta_es}
 }
 \hfill
  \subfloat{
      \centering
      \includegraphics[width=0.475\textwidth]{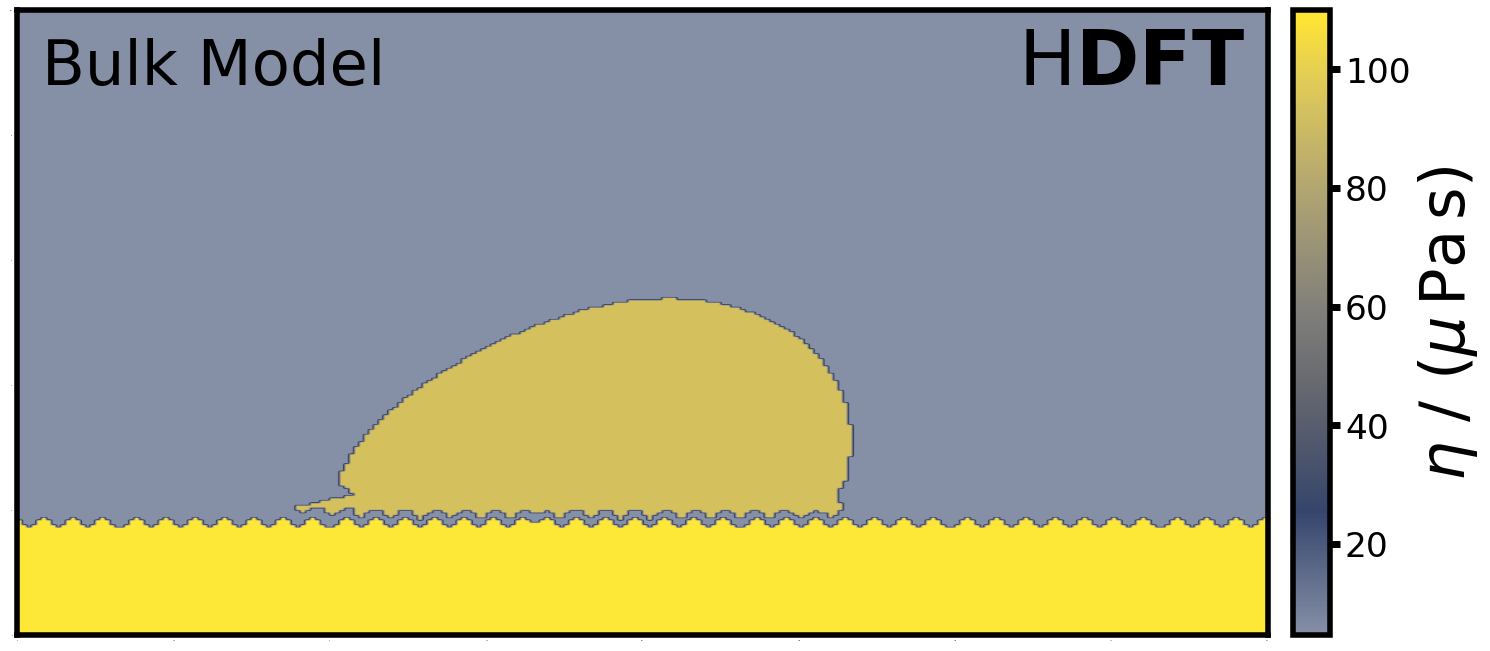}\label{fig:eta_bulk}
 }
  \caption{Viscosity profiles from local entropy scaling model (left) and  bulk viscosity model (right) at  $f_x=\SI{0.112}{\pico\newton}/\mathrm{particle}$ and $T=\SI{120.02}{\kelvin}$ after $t=\SI{1000}{\pico\second}$.}
  \label{fig:eta}
\end{figure}

This section assesses the role of a local viscosity model. The goal is to ascertain whether incorporating molecular details in the viscosity model by using generalised entropy scaling is essential for accurately modelling wetting phenomena or if continuum models are sufficient.
Therefore, we compare generalised entropy scaling to a different viscosity model, which represents a typical continuum fluid dynamics approach and is based on viscosities from bulk phases (thus called ``bulk'' model). 

The bulk viscosity model uses viscosities from bulk phases depending on the local density $\rho(\rb)$ according to 
\begin{equation} \label{eq:eta}
  \eta^\mathrm{bulk}(\rb) = 
    \begin{cases}
      \eta_\mathrm{l},  & \rho(\rb)> 0.5(\rho_\mathrm{l}+\rho_\mathrm{v})+\rho_\mathrm{v}\\
      \eta_\mathrm{v},  & \rho(\rb)<= 0.5(\rho_\mathrm{l}+\rho_\mathrm{v})+\rho_\mathrm{v}\\
      \eta_\mathrm{s},  & \rho(\rb)<\rho_\mathrm{cut}\\
  \end{cases}
\end{equation}
where $\rho_\mathrm{l}$ and $\rho_\mathrm{v}$ are densities from a vapour-liquid equilibrium at $T=\SI{120.02}{\kelvin}$ and the corresponding bulk viscosities $\eta_\mathrm{l}$ and $\eta_\mathrm{v}$ are determined using entropy scaling for homogeneous phases. 
The procedure is as follows: first, the density of the liquid and the vapour phase are determined from the phase equilibrium of pure methane at $T=\SI{120.02}{\kelvin}$. The residual entropy is then determined from the PC-SAFT model as the partial derivative of the Helmholtz energy with respect to the temperature. Using the appropriate dimensionless form of the entropy, the value for $\ln \left(\eta^\mathrm{Entr. Scal.}/\eta_\mathrm{ref}\right)$ is determined from the ansatz function in \cref{eq:poly3}, where the parameters $A$, $B$, $C$ and $D$ were previously adjusted to experimental data \citep{loetgeringlin2018pure}. The Chapman-Enskog reference viscosity $\eta_\mathrm{CE}$ is calculated according to \cref{eq:CEViscosity}, which finally allows to calculate the viscosity $\eta_\mathrm{l}$ and $\eta_\mathrm{v}$, respectively. 
A very large viscosity~$\eta_\mathrm{s}$ is chosen for the liquid within the solid domain, which is defined by a cutoff density~$\rho_\mathrm{cut}$, to ensure comparability with the generalised entropy scaling approach, but the impact of the viscosity in this region on velocity profiles is negligible. All parameters are provided in \cref{table:params}.

The resulting viscosity profile is compared to the one from generalised entropy scaling in \cref{fig:eta}.  
The viscosity from entropy scaling (left) exhibits a more elaborate behaviour with strong oscillations perpendicular but also parallel to the solid-fluid interface. This is expected since the viscosity from entropy scaling is a non-linear and non-local function of the density. In contrast, the viscosity from the bulk model (right) assumes only two values in the vapour-liquid domain and a different constant value in the solid domain. While the entropy scaling model provides a  continuous transition between liquid-like and vapour-like viscosity values, discontinuities are found at the transition between the different regions for the bulk viscosity model. 

\begin{figure}
\centering
{
      \centering
      \includegraphics[width=0.475\textwidth]{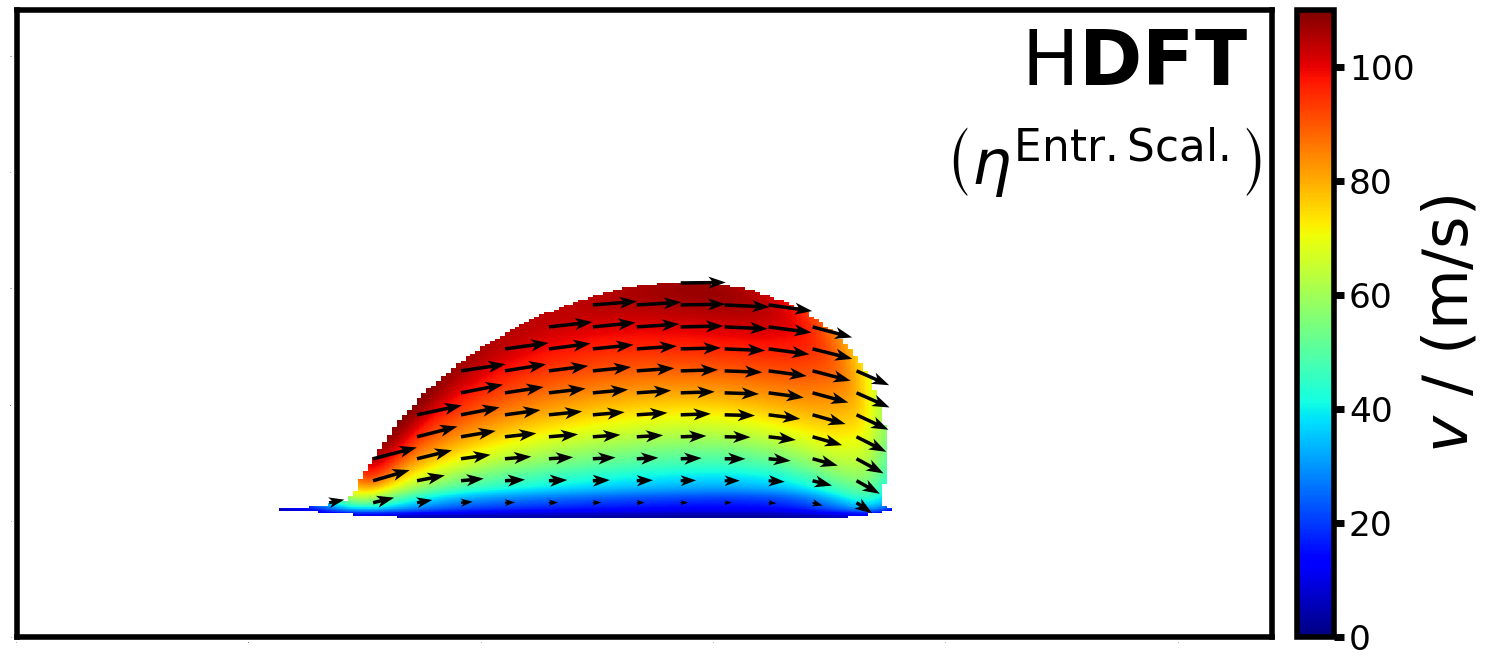} \label{fig:v_02_hdft}
 }
 \hfill
 {
     \centering
     \includegraphics[width=0.475\textwidth]{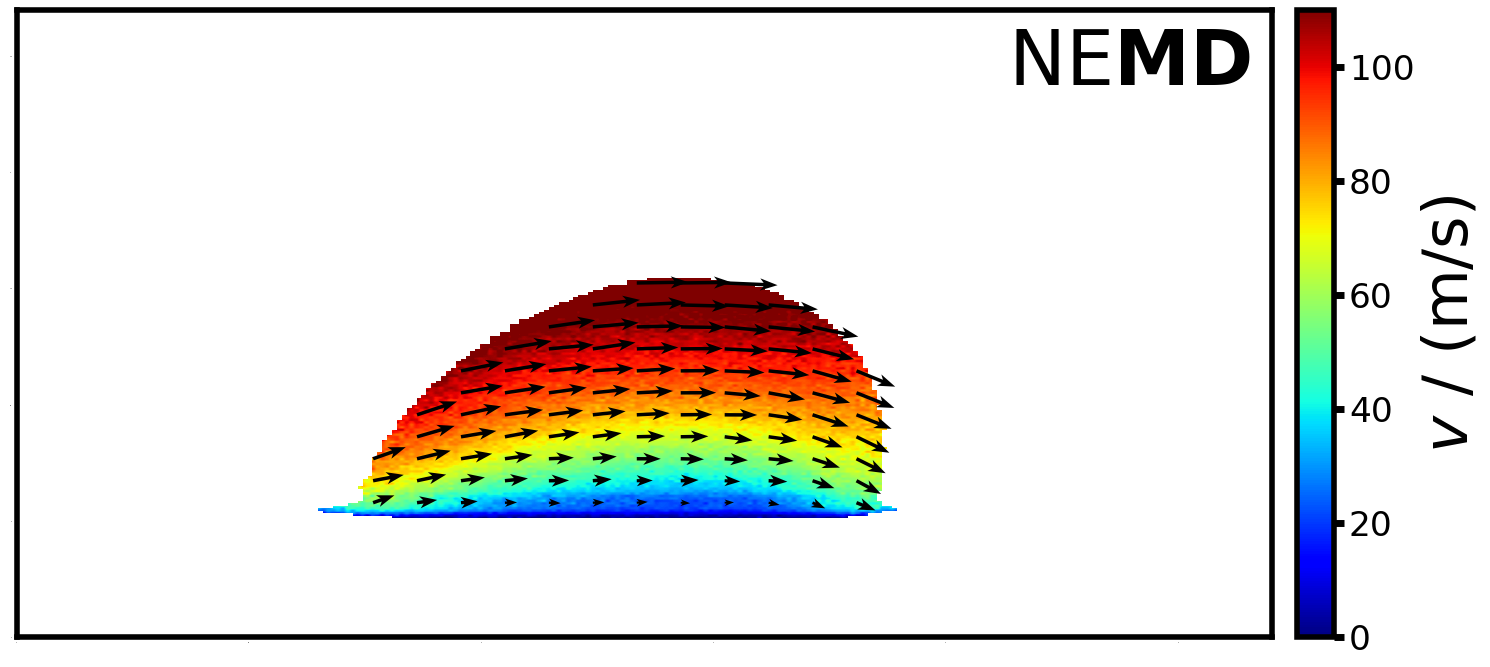} \label{fig:v_02_md_eta}
}
 \hfill
  {
      \centering
      \includegraphics[width=0.475\textwidth]{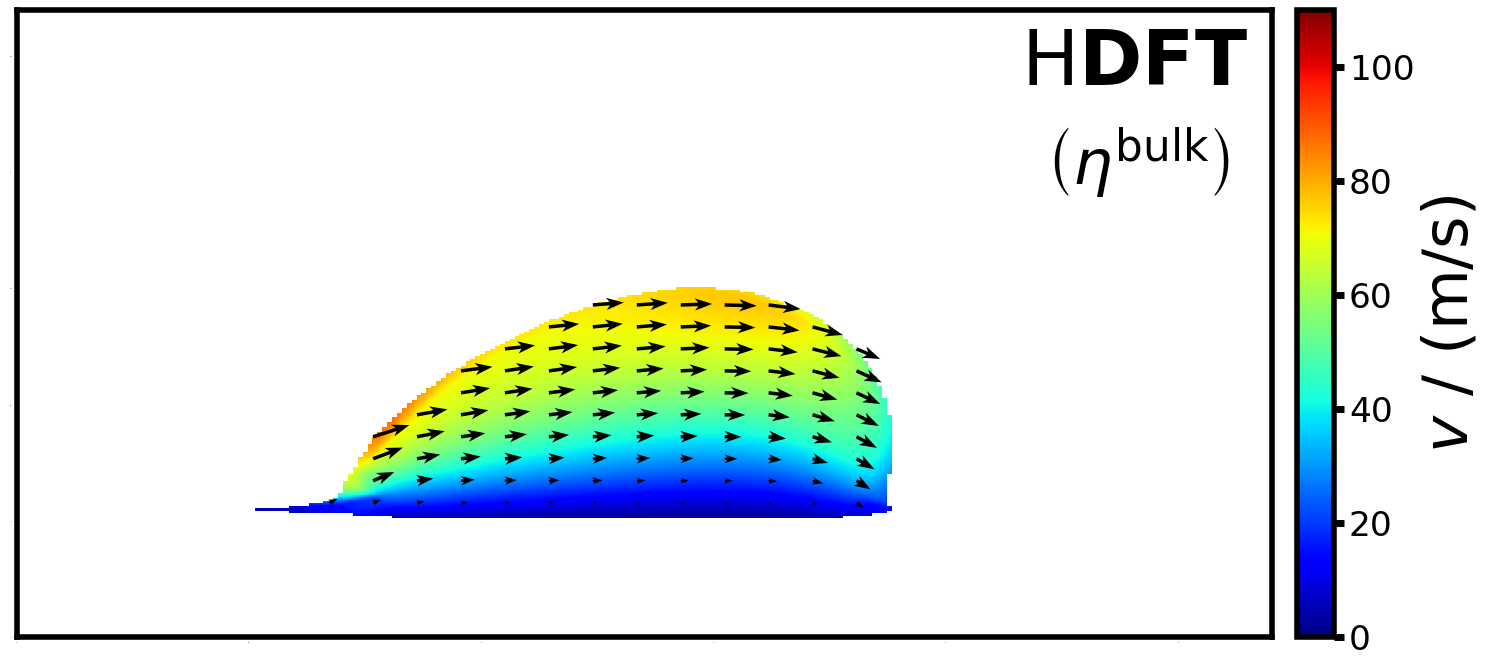} \label{fig:v_02_hdft_eta}
 } 
 \caption{Velocities inside the droplet from hydrodynamic DFT using the generalised entropy scaling viscosity model (top) and the bulk viscosity model (bottom) as well as from NEMD (middle) with the solid as the frame of reference for $f_x=\SI{0.112}{\pico\newton}/\mathrm{particle}$ and $T=\SI{120.02}{\kelvin}$  averaged over at least $\SI{700}{\pico\second}$ after a steady state is reached. Arrows denote the direction and magnitude of the velocity, whereas the colours correspond to the magnitude of the velocity.}
  \label{fig:v_eta}
\end{figure}
In order to assess the accuracy of the viscosity models, we compare velocity profiles determined from hydrodynamic DFT using the two viscosity models and validate the results by NEMD simulations as provided in \cref{fig:v_eta} for the medium force~$f_x=\SI{0.112}{\pico\newton}$, where the droplet moves into the positive $x$-direction (to the right). The velocities are averaged over at least \SI{700}{\pico\second} in hydrodynamic DFT and \SI{10000}{\pico\second} in NEMD after a steady state is reached.
For the NEMD results, in addition to averaging velocities in the third dimension, the centre of mass of the droplet is determined for each configuration entering the calculation of the velocity profile. Using  coordinates relative to this centre of mass, substantial time-averaging  of the velocity is performed to obtain a reasonable signal-to-noise ratio. For the initial acceleration of the droplet the velocity results have considerable uncertainty and we do not report velocities for the acceleration period of the droplet. Noise-free velocity profiles are a significant advantage of hydrodynamic DFT over NEMD.
For all three cases, the largest velocity relative to the solid is found at the top of the droplet and the smallest velocity at the solid-fluid interface. Good quantitative agreement is observed between hydrodynamic DFT with generalised entropy scaling (top) and NEMD (middle). For the bulk viscosity model (bottom), significantly lower velocities are found throughout the droplet, where especially the velocity close to the solid-fluid interface in the centre of the droplet is much lower than in NEMD and hydrodynamic DFT with generalised entropy scaling. The substantially different behaviour of the viscosity in the first molecular layers at the solid-fluid interface for the bulk and entropy scaling viscosity models (cf. \cref{fig:eta}) explains these results. This emphasises that molecular details, which are captured in the entropy scaling viscosity model, have a severe effect on the microscopic wetting behaviour. 

\begin{figure}
  \centering
  \includegraphics[width=0.55\textwidth]{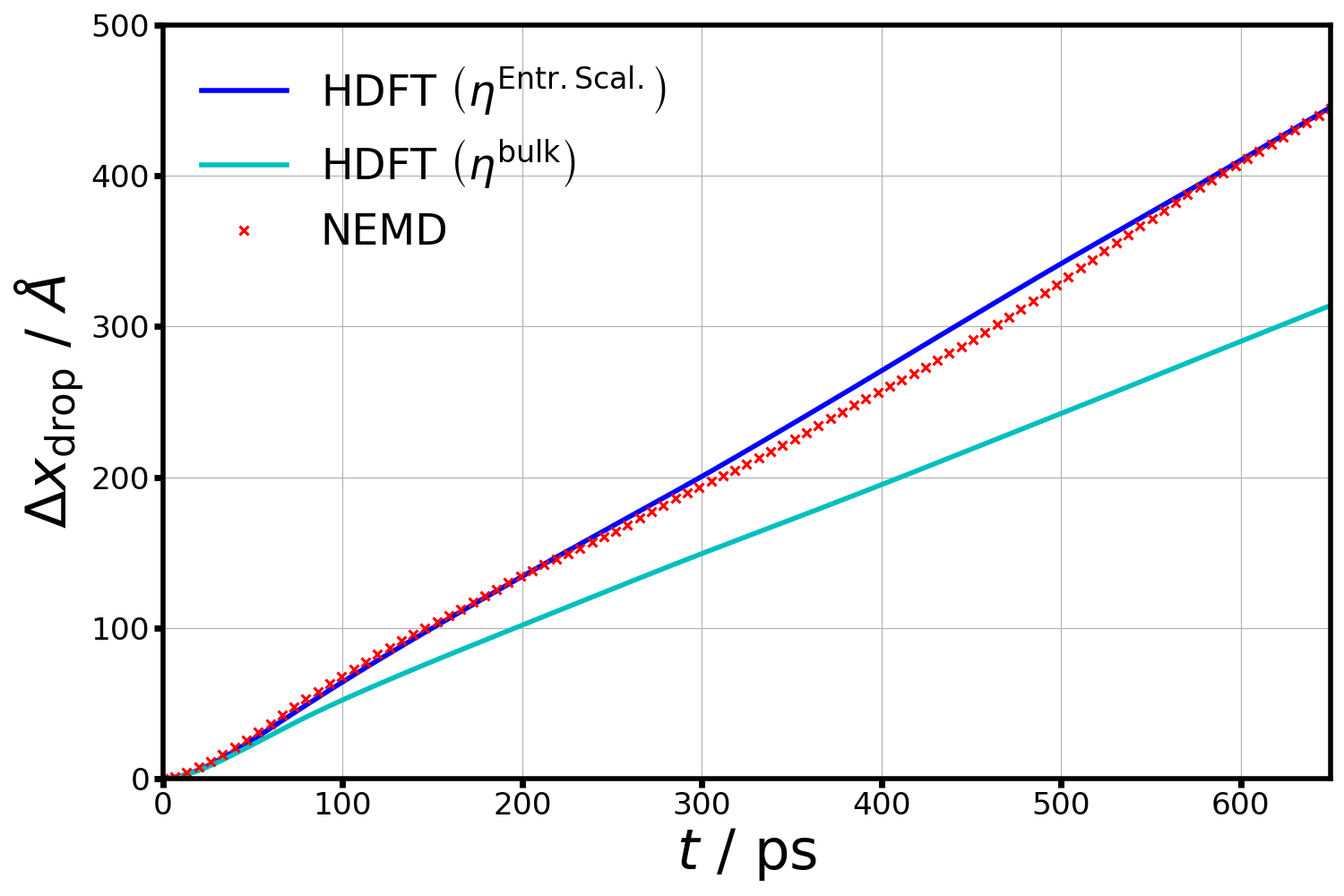}
  \caption{Distance travelled by the centre of mass of the droplet from hydrodynamic DFT (HDFT) with entropy scaling viscosity model (blue line) and bulk viscosity model (light blue line) as well as from NEMD (red crosses) for $f_x=\SI{0.112}{\pico\newton}/\mathrm{particle}$ and $T=\SI{120.02}{\kelvin}$.}
  \label{fig:xcom_eta}
\end{figure}
Furthermore, the influence of the viscosity models on the velocity of the entire droplet is analysed by plotting the distance travelled by the centre of mass of the droplet versus the simulation time (\cref{fig:xcom_eta}). 
In the initial phase of the simulation until about \SI{100}{\pico\second}, the distance covered by the droplet increases with increasing slope, i.e.\ the droplet moves with increasing velocity for all three models. After this initial phase, the distance travelled by the  droplet increases almost linearly, which corresponds to a constant velocity~$v^\mathrm{avg}_\mathrm{drop}$ of the droplet and shows that a steady state is reached. 
At the very beginning of the simulation (until about \SI{50}{\pico\second}), the velocity of the droplet is small and results from all models agree. After about \SI{50}{\pico\second}, the results from the bulk viscosity model increasingly deviate from NEMD results, whereas good agreement between hydrodynamic DFT with generalised entropy scaling and NEMD is obtained. 
   
\begin{figure}
  \centering
  \includegraphics[width=0.55\textwidth]{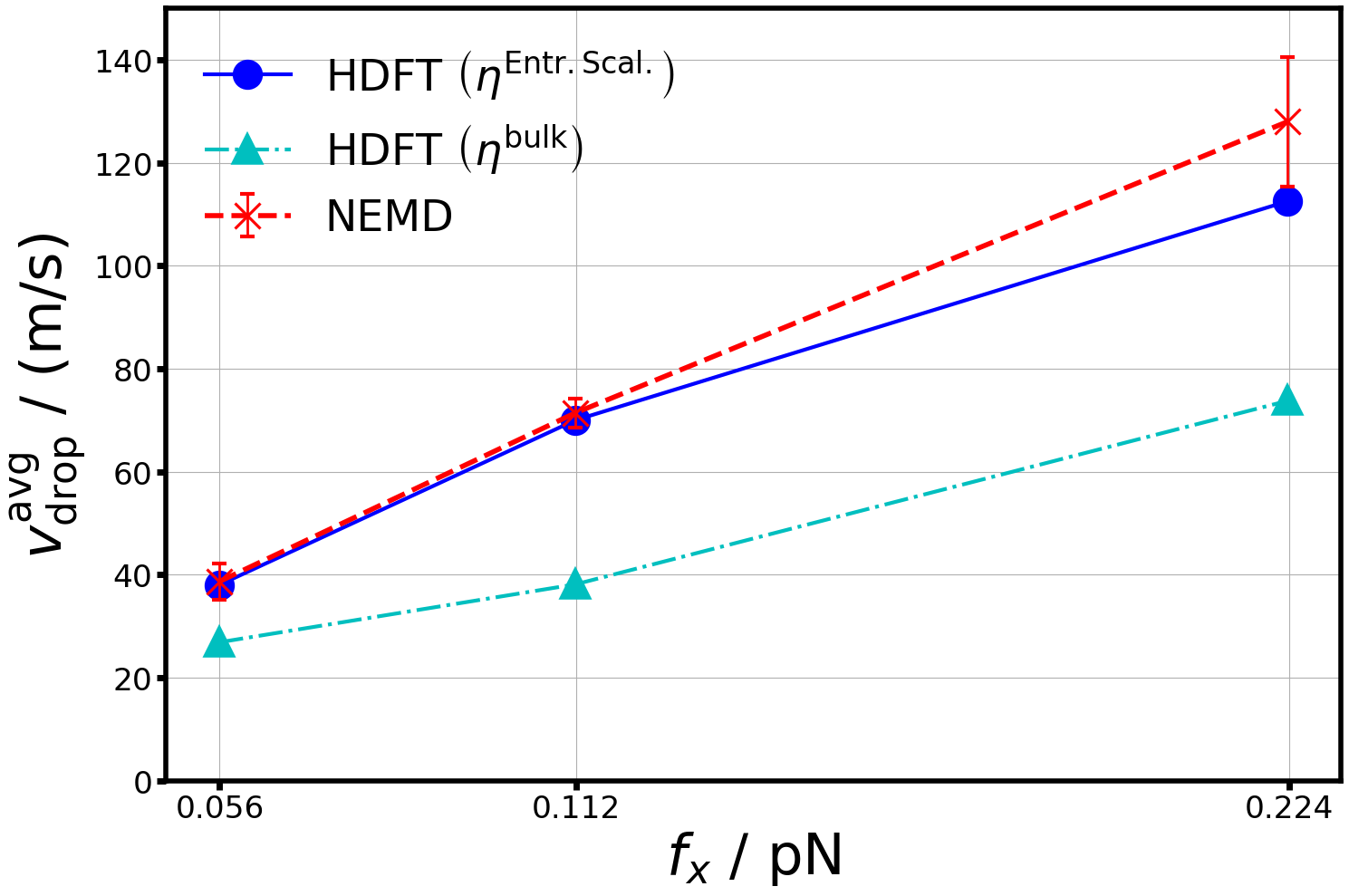}
  \caption{Steady state velocity of the centre of mass of the moving droplet for  different external forces (per particle) from hydrodynamic DFT (HDFT) with entropy scaling viscosity model (blue circles) and bulk viscosity model (light blue triangles) as well as from NEMD (red crosses) at $T=\SI{120.02}{\kelvin}$. }
  \label{fig:vavg_eta}
\end{figure}
The steady state velocities of the droplets~$v^\mathrm{avg}_\mathrm{drop}$ are summarised in \cref{fig:vavg_eta} for different forces. For all models, the steady-state velocities increase with increasing force. 
For the low and medium forces ($f_x=\SI{0.056}{\pico\newton}$ and $f_x=\SI{0.112}{\pico\newton}$) the velocity of the droplet predicted from hydrodynamic DFT with generalised entropy scaling accurately represents the results from NEMD. A small underestimation is observed for the largest force~$f_x=\SI{0.224}{\pico\newton}$, noting however that the statistical uncertainty of the velocity from NEMD is also large compared to the lower forces.
In contrast, the bulk viscosity model substantially underestimates the steady state velocity for all forces, where the underestimation is largest for the largest force~$f_x=\SI{0.224}{\pico\newton}$. This increasing influence of the viscosity at larger forces can be explained by the presence of larger velocities, which in turn lead to larger viscous dissipation. 

Compared to macroscopic systems, our study investigates high velocity-gradients that are caused by large external forces needed to overcome the surface tension on the atomistic scale. The transition to macroscopic systems could best be investigated using a size-scaling study, where an extrapolation to macroscopically large droplets is performed. We note however, that our study does not reveal any indication about non-linear shear-strain relations. Our results are in that sense also applicable to low deformation tensors and low stress tensors.
We conclude that using generalised entropy scaling provides more accurate results than the simpler model based on bulk viscosities. This is likely due to the importance of molecular details close to the solid-fluid interface, which are captured by entropy scaling. Thus, in the remainder of this work, we present results only for hydrodynamic DFT using the generalised entropy scaling model and we will exclude the bulk viscosity model.

\subsection{ Droplet Movement: Slip and  Rolling Motion} \label{sec:results_v}
In this section, the velocity profiles of droplets (some of which were already presented above in \cref{fig:v_eta}) are discussed in more detail to evaluate whether hydrodynamic DFT with generalised entropy scaling captures mechanisms of wetting.  
Studies of microscopic models, such as diffuse interface models or MD simulations, usually report slip, i.e.\ a non-vanishing velocity at the solid-fluid interface in the contact region  \citep{fernandez2019molecular,blake2015forced, nakamura2013dynamic,yue2011wall,yue2010sharp,lee2022contact}. 
Furthermore, it was reported from MD simulations \citep{fernandez2019molecular,li2018dynamic,li2019dynamic} that microscopic droplets move along the solid-fluid interface by a `rolling' mechanism and
 that this mechanism requires the presence of slip within the contact region. 

\begin{figure}
  \centering
  {
      \centering
      \includegraphics[width=0.475\textwidth]{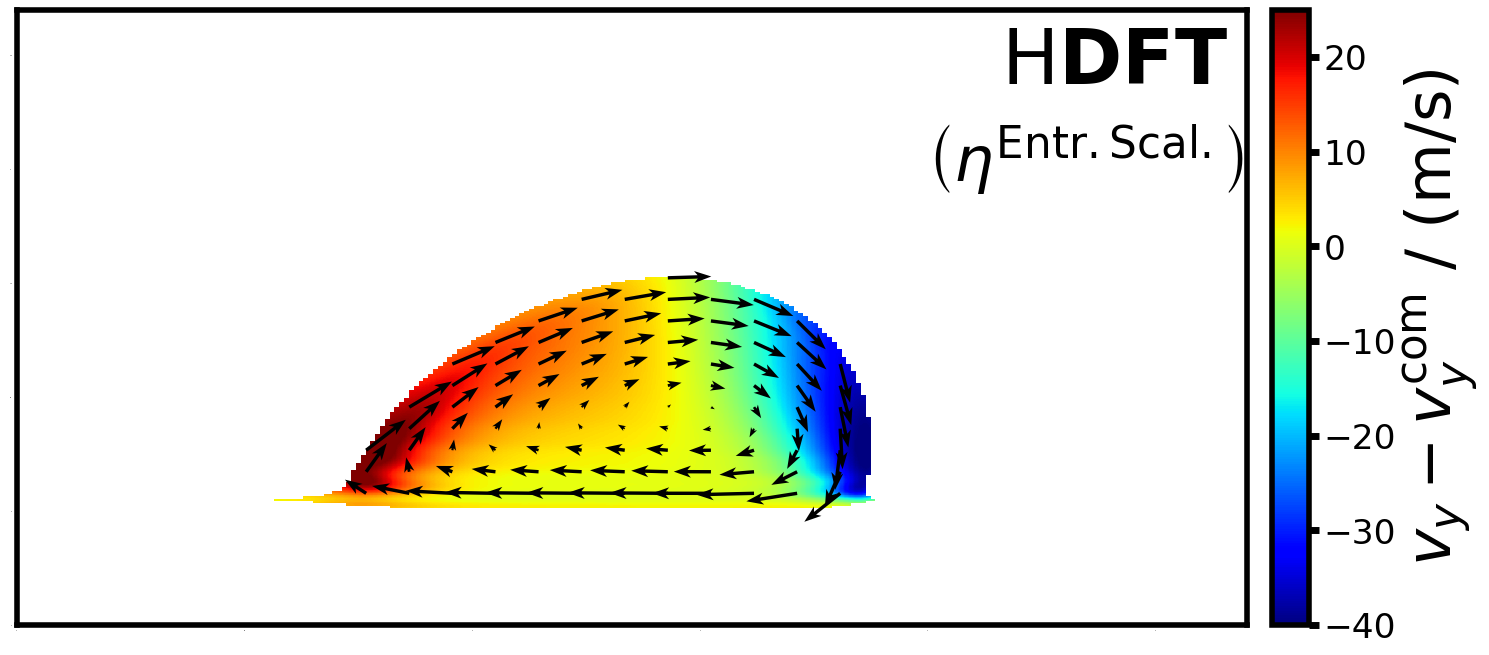}\label{fig:vrel_02_hdft}
 }
 \hfill
  {
      \centering
      \includegraphics[width=0.475\textwidth]{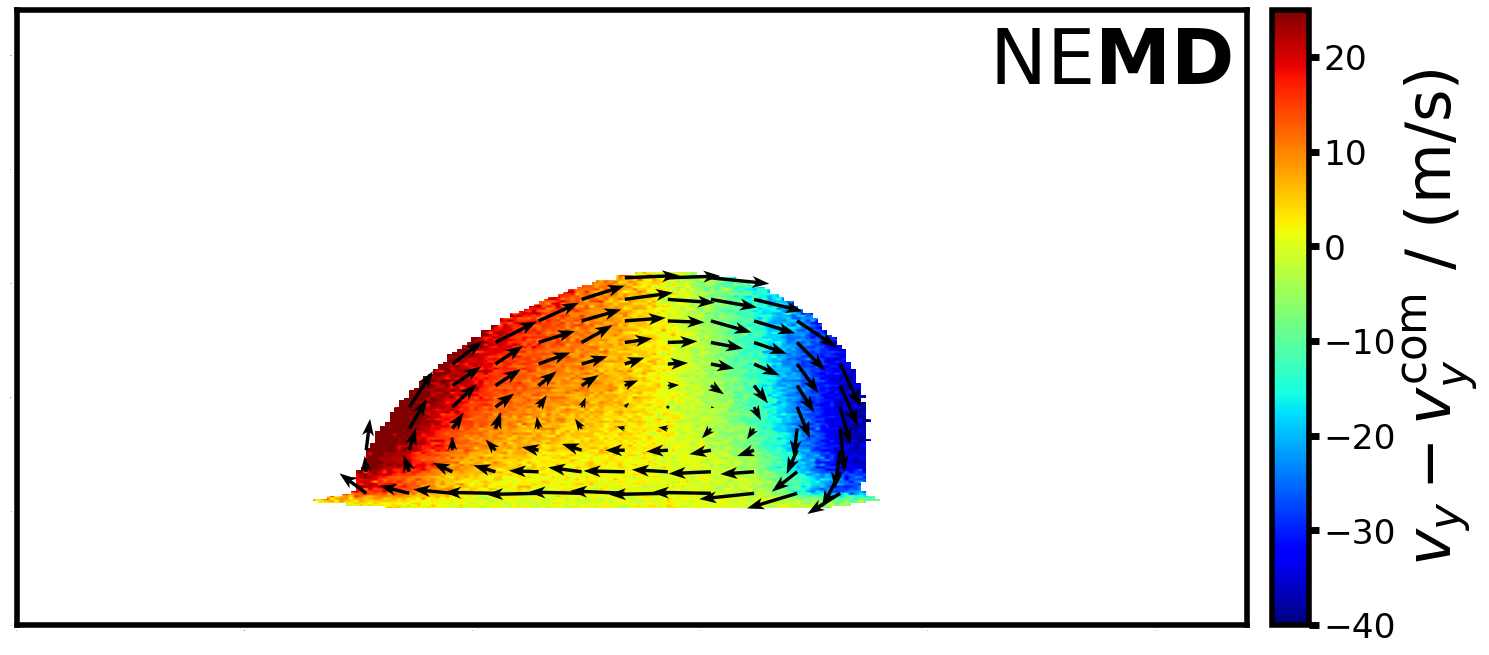}\label{fig:vrel_02_md}
 }
  \caption{Velocity relative to centre of mass velocity of the droplet from hydrodynamic DFT (left) and NEMD (right) for $f_x=\SI{0.112}{\pico\newton}/\mathrm{particle}$ and $T=\SI{120.02}{\kelvin}$ averaged over at least $\SI{700}{\pico\second}$ after a steady state is reached. Arrows denote the direction and magnitude of the flow, whereas the colours correspond to the $y$-component of the velocity.}
  \label{fig:vrel_02}
\end{figure} 
The profiles of velocity with the solid as the reference determined from hydrodynamic DFT and NEMD (top and middle of \cref{fig:v_eta}) reveal that the velocity is close to zero at the solid-fluid interface in the centre of the droplet. 
 However, it is non-zero in the solid-liquid-vapour contact regions.
  This is an important observation as it demonstrates that slip is found in hydrodynamic DFT and in NEMD. 
In addition, 
\cref{fig:vrel_02} provides the velocity vectors relative to the velocity of the droplets centre of mass from hydrodynamic DFT and NEMD. The $y$-component of these vectors is visualised by colours, where positive and negative values correspond to flow  away from and towards the solid, respectively. 
The translational motion of the droplet parallel to the solid-fluid interface is superimposed by 
a clearly visible circular motion as also reported in the literature \citep{fernandez2019molecular,li2018dynamic}. 
According to \cref{fig:vrel_02}, the maximum and minimum values of the relative velocity in $y$-direction are about \SI{25}{\meter\per\second} and \SI{-40}{\meter\per\second}, respectively.  Comparing these values to the maximum values of the absolute velocity in \cref{fig:v_eta}, which are in the range of \SI{100}{\meter\per\second}, demonstrates that the circular motion is quantitatively significant.  
Furthermore, analysing the vorticity of the flow field supports the finding that the droplet exhibits a significant rolling motion (see \cref{sec:appendix_vorticity}).
The presented results demonstrate that hydrodynamic DFT predicts the complex mechanisms of droplet motion at the molecular scale, particularly slip in the contact region and the circular motion of droplets, in good agreement with our NEMD simulations. We emphasise that noise-free velocity profiles are obtained from hydrodynamic DFT, whereas in NEMD substantial averaging is required to generate meaningful velocity profiles. In addition, the good agreement regarding droplet motion further supports the validity of the generalised entropy scaling viscosity model. 

\subsection{ Advancing and Receding Contact Angles } \label{sec:results_hysteresis}

The appearance of different dynamic contact angles is commonly attributed to inhomogeneities of the solid \citep{bonn2009wetting,deGennes2004capillarity,jamet2001second,butt2022contact,voinov1976hydrodynamics,cox1986dynamics,decker1997contact}  and they were found to depend on the velocity of the contact region \citep{cox1986dynamics,voinov1976hydrodynamics}. In literature \citep{butt2022contact}, the term dynamic contact angle hysteresis is sometimes used for the difference between advancing and receding contact angle. Since in a strict thermodynamic sense the term hysteresis is disputable in this context, we will speak of the \emph{difference between advancing and receding contact angles} instead.   
The microscopic mechanisms behind this effect
are studied in several works \citep{bertrand2009influence,blake2015forced,ren2007boundary,lukyanov2016dynamic,lukyanov2017hydrodynamics}, where a particularly convincing explanation is based on a solid-fluid friction force $F_\mathrm{friction}$ \citep{lukyanov2016dynamic}. This friction force occurs mainly in the contact region and depends on its velocity, since it is caused by dynamic interactions of fluid molecules with the solid interaction sites.

From a molecular dynamics point of view, the solid-fluid friction in the contact region can be explained by a net force counteracting the flow of fluid particles. Due to the net flow in positive $x$-direction (parallel to the solid surface), fluid particles are more likely to enter regions, where they experience a repulsive force from the solid atoms acting in the direction opposite to the flow.         
In hydrodynamic DFT the equivalent mechanism is captured by the external potential (cf. \cref{sec:theoryDFT}). The gradient of the external potential multiplied by the density enters the momentum balance in \cref{eq:MomentumBalanceDDFT}. Close to the location of solid atoms  the gradient of the external potential in $x$-direction is negative and thus, effectively reduces the momentum in $x$-direction, which in turn results in the solid-fluid friction force. 
Its velocity dependence results  from the fact, that  higher   velocities in $x$-direction lead to larger densities in the regions where the gradient of the external potential reduces the momentum in $x$-direction. Thus, the product of density and external potential gradient increases with velocity, which leads to a larger reduction of the  momentum in $x$-direction. 
A visual representation of the gradient of the external potential is provided in \cref{sec:appendix_grad_vext}.
The molecular-kinetic theory (MKT), as opposed to our work, assumes a certain mechanism causing this solid-fluid friction force \citep{blake1969kinetics,blake1993dynamic}: the fluid molecules perform random displacement (jumps) of average distance, which occur with a certain temperature-dependent frequency. The solid surface is thought of as a series of potential energy wells, at which the molecules remain for a short period between these displacements.

We compare results from hydrodynamic DFT and NEMD for the transition between the equilibrium and a dynamic steady state to analyse if 
differences between the advancing and receding contact angles are
 observed. 
\begin{figure}
  \centering
  \subfloat[Hydrodynamic DFT after $t=\SI{20}{\pico\second}$]{
      \centering
      \includegraphics[width=0.475\textwidth]{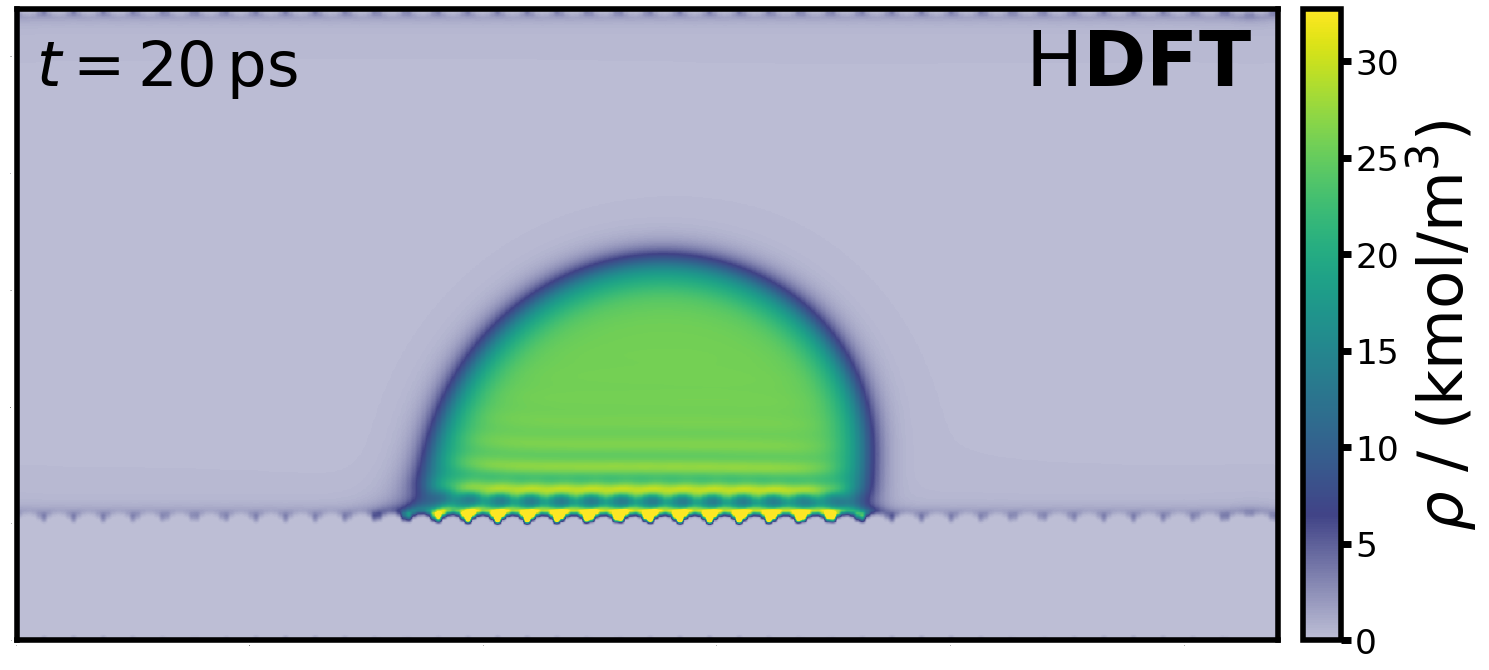}\label{fig:rho_t_dft_20}
 }
 \hfill
  \subfloat[NEMD after $t=\SI{20}{\pico\second}$]{
      \centering
      \includegraphics[width=0.475\textwidth]{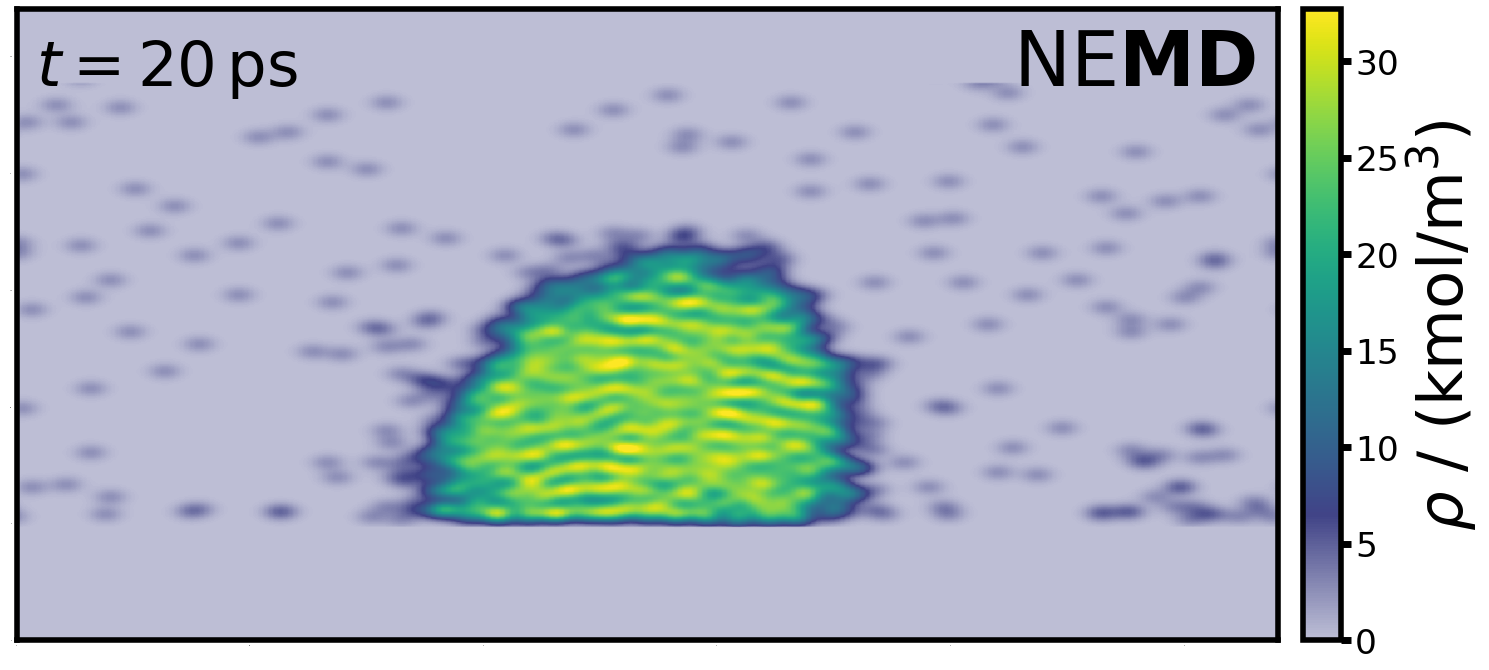}\label{fig:rho_t_nemd_20}
 }
 \vfill
 \subfloat[Hydrodynamic DFT after $t=\SI{40}{\pico\second}$]{
      \centering
      \includegraphics[width=0.475\textwidth]{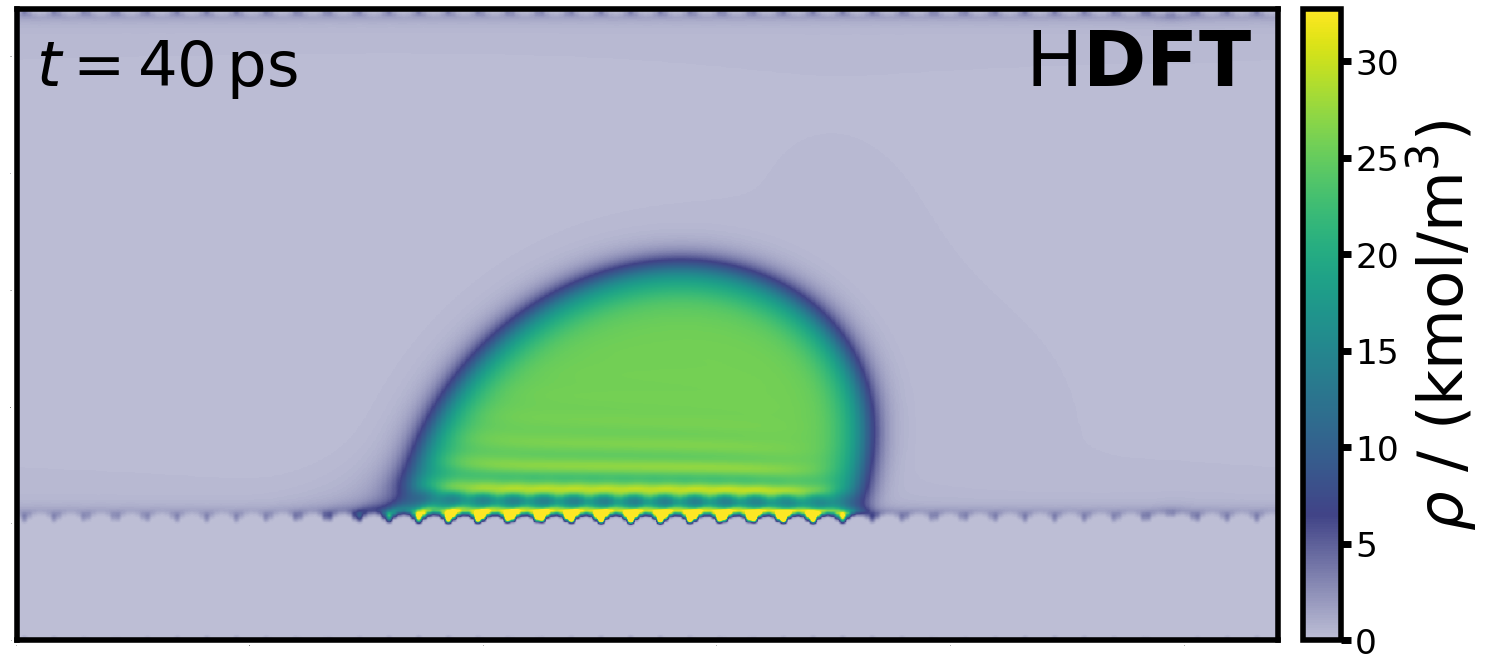}\label{fig:rho_t_dft_40}
 }
 \hfill
  \subfloat[NEMD after $t=\SI{40}{\pico\second}$]{
      \centering
      \includegraphics[width=0.475\textwidth]{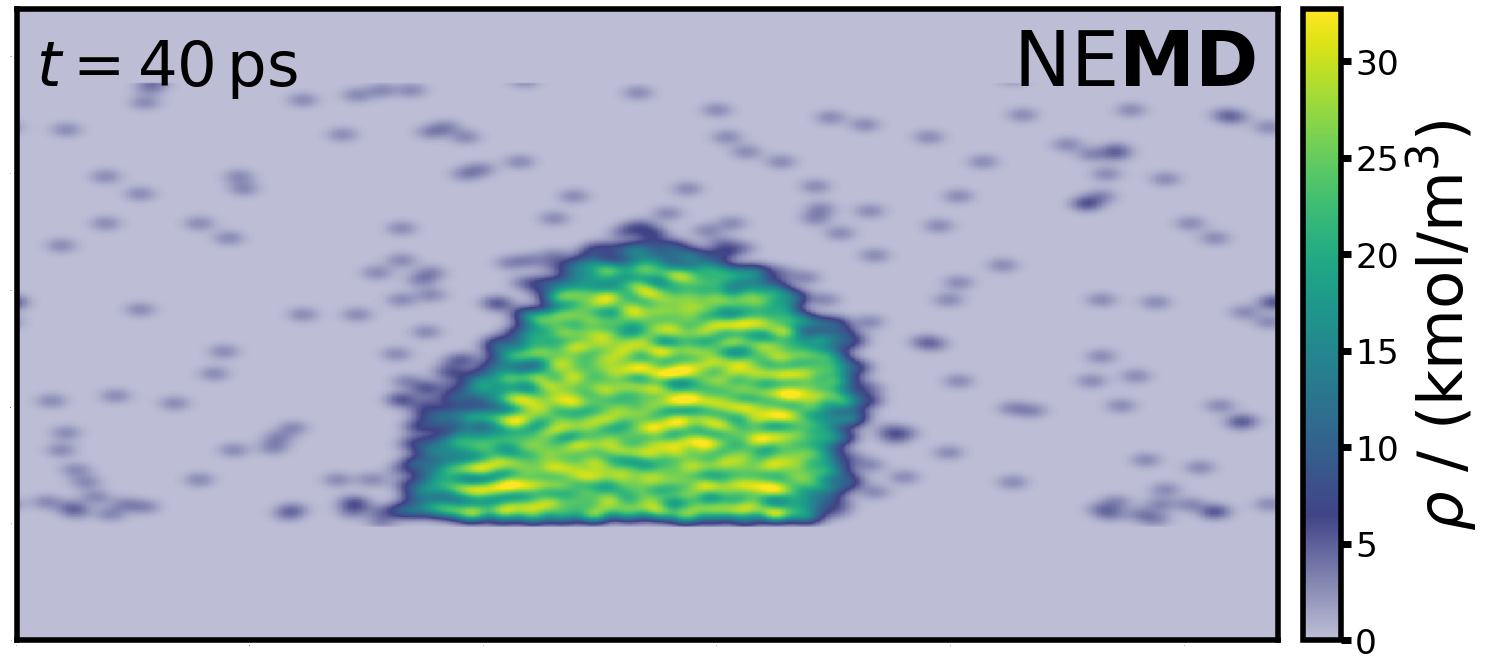}\label{fig:rho_t_nemd_40}
 }
 \vfill
 \subfloat[Hydrodynamic DFT after $t=\SI{60}{\pico\second}$]{
      \centering
      \includegraphics[width=0.475\textwidth]{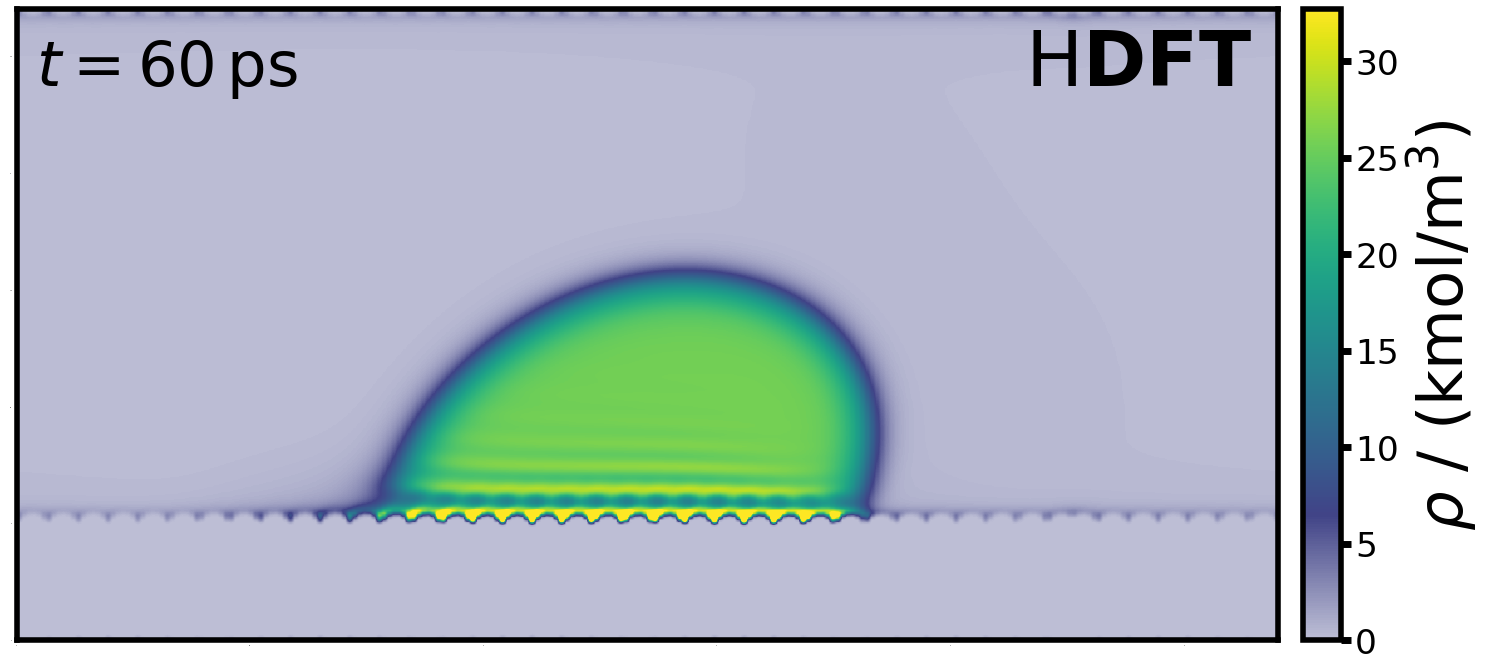}\label{fig:rho_t_dft_60}
 }
 \hfill
  \subfloat[NEMD after $t=\SI{60}{\pico\second}$]{
      \centering
      \includegraphics[width=0.475\textwidth]{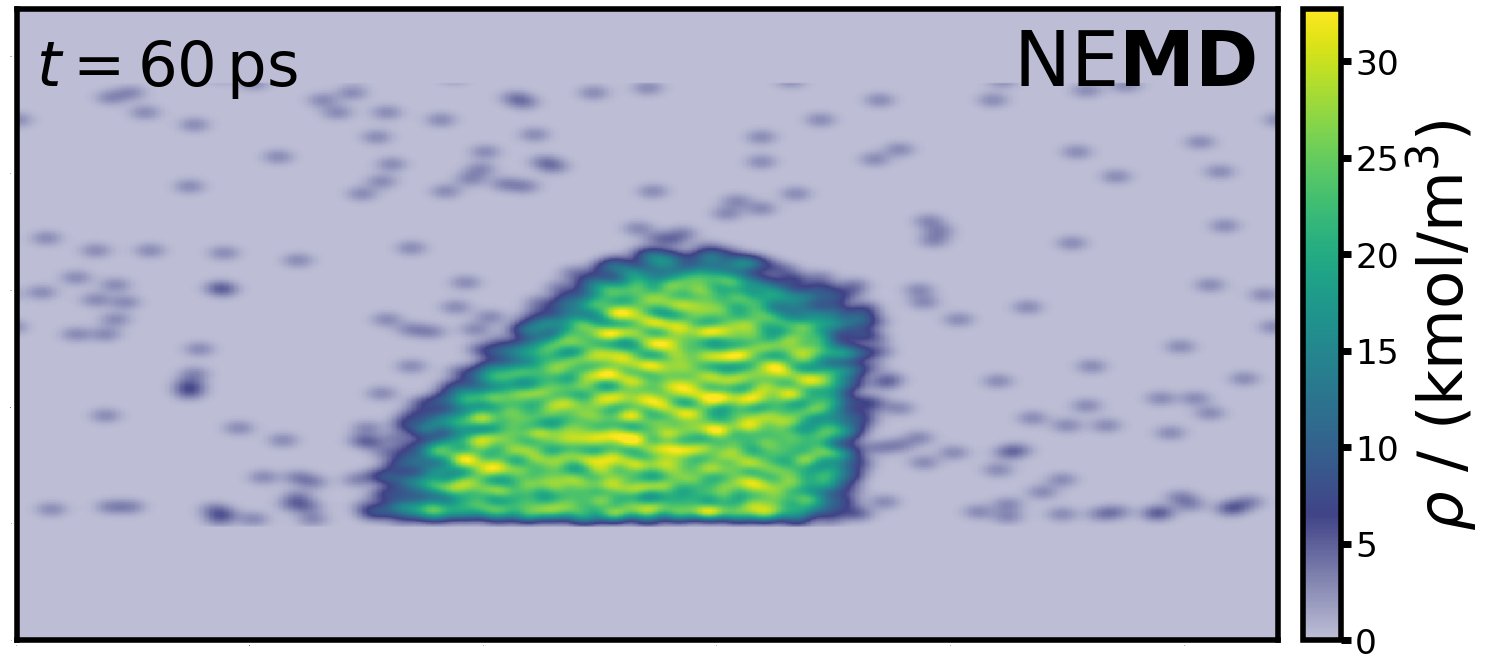}\label{fig:rho_t_nemd_60}
 }
 \vfill
 \subfloat[Hydrodynamic DFT after $t=\SI{100}{\pico\second}$]{
      \centering
      \includegraphics[width=0.475\textwidth]{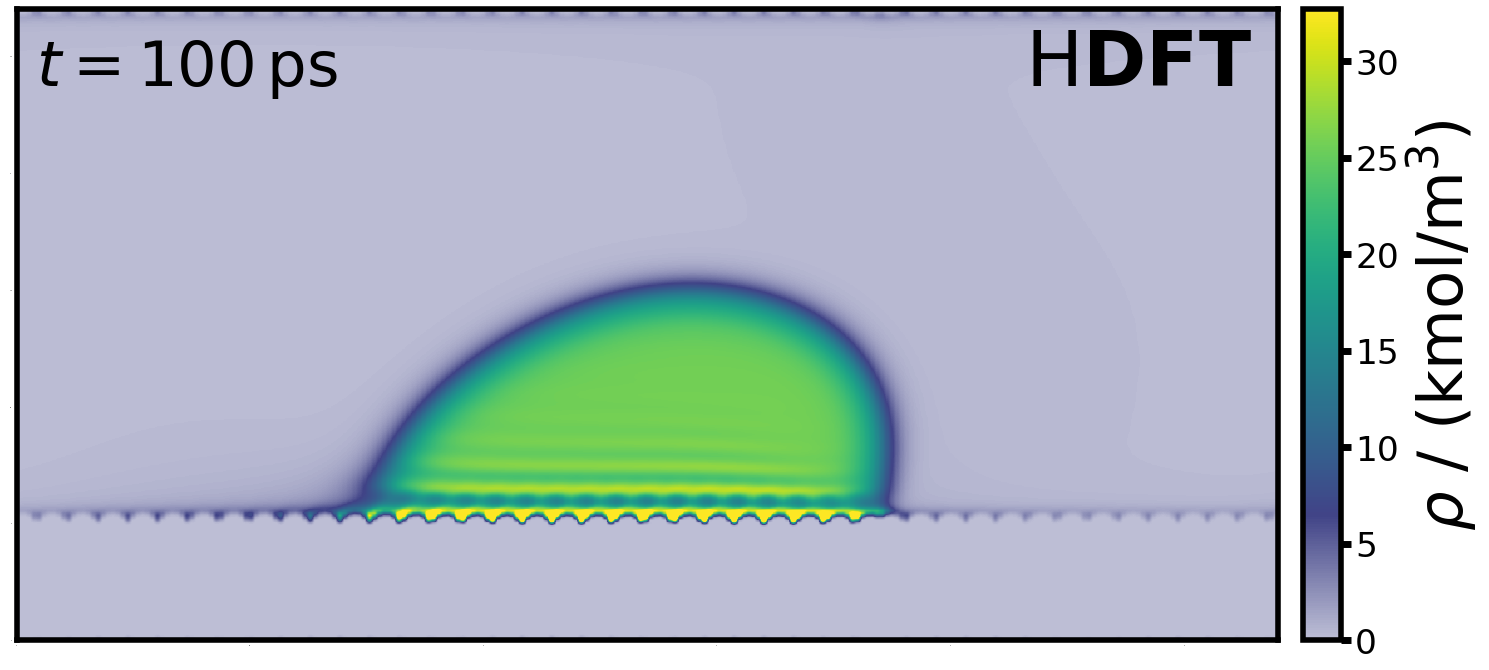}\label{fig:rho_t_dft_100}
 }
 \hfill
  \subfloat[NEMD after $t=\SI{100}{\pico\second}$]{
      \centering
      \includegraphics[width=0.475\textwidth]{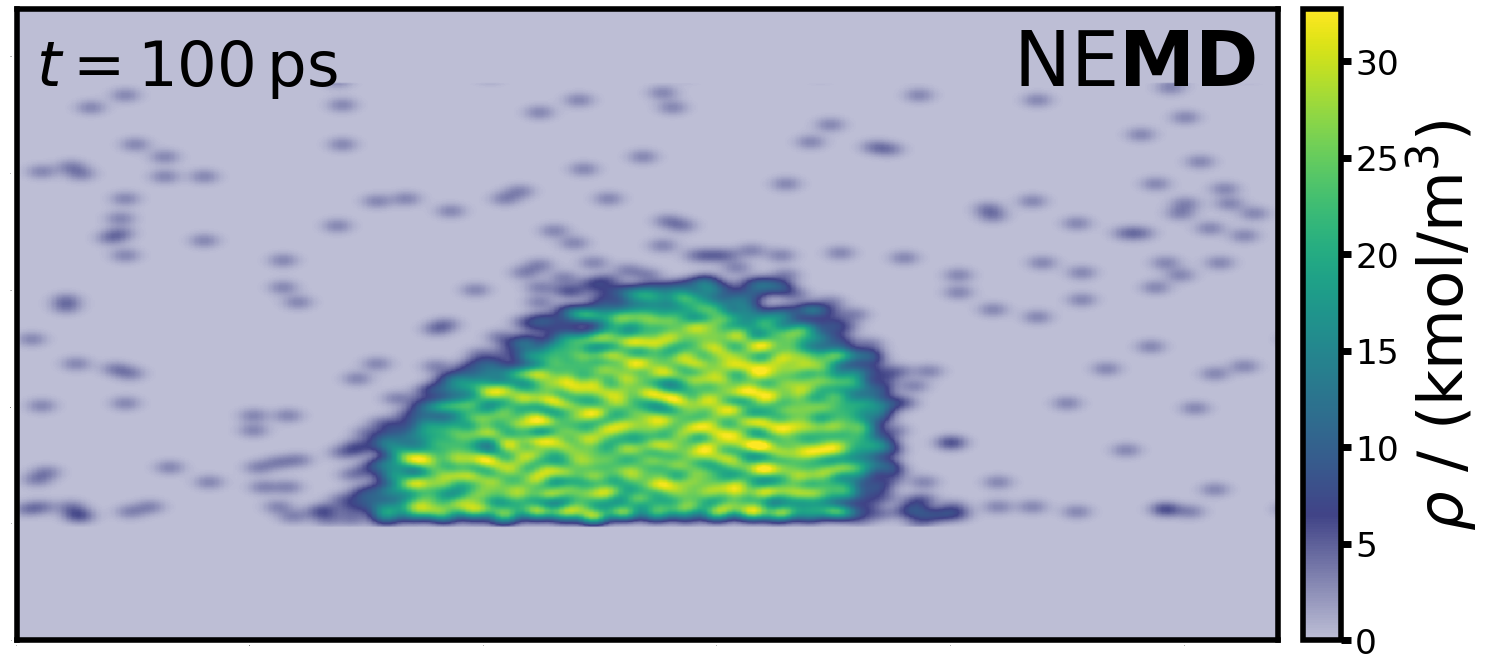}\label{fig:rho_t_nemd_100}
 }
  \caption{Density profiles of droplets moving along the solid-fluid interface with external force $f_x=\SI{0.112}{\pico\newton}/\mathrm{particle}$ and $\varepsilon_\mathrm{sf}^*=0.5$ from hydrodynamic DFT (HDFT) and NEMD   at different simulation times $t$. NEMD results show statistical noise, whereas hydrodynamic DFT provides deterministic density profiles. }
  \label{fig:rho_t}
\end{figure}
\Cref{fig:rho_t} compares density profiles from hydrodynamic DFT and NEMD at different times after the start of the simulation for an external force of $f_x=\SI{0.112}{\pico\newton}$. The results from NEMD show significant statistical noise, since at each time a limited number of particles located at discrete positions enter the calculation of the density profiles (compare the top graph in \cref{fig:system}). The density profiles from NEMD are smoothed following \citet{hong2009static} and all profiles are relocated such that the centre of mass is positioned at $L_x/2$.
Starting from the equilibrium density profile (see \cref{fig:rho_static}), the droplet from hydrodynamic DFT begins to deform after \SI{20}{\pico\second} as shown in \cref{fig:rho_t_dft_20}. With increasing simulation time  the deformations become more pronounced (see \cref{fig:rho_t_dft_40,fig:rho_t_dft_60,fig:rho_t_dft_100}). The molecular layering and adsorption from the gas are observed similarly to the equilibrium case. Advancing contact angles increase  while receding contact angles decrease compared to the static contact angle.
The results from NEMD (\cref{fig:rho_t_nemd_20,fig:rho_t_nemd_40,fig:rho_t_nemd_60,fig:rho_t_nemd_100}) exhibit very similar behaviour, even though the analysis is obscured by fluctuations of the density profile.
The above explanation for 
the difference between advancing and receding contact angles 
and its velocity dependence based on a solid-fluid friction force also applies to the transition behaviour in \cref{fig:rho_t}.
After \SI{100}{\pico\second}, the change in droplet shape is small (not shown), which suggests that a steady state is approached, where the external force is in balance with the solid-fluid friction and viscous forces.

The occurrence of a solid-fluid friction force and thus, of 
differences between the advancing and receding contact angle
is commonly attributed to inhomogeneities of the solid surface. Our results demonstrate that 
these differences appear
 even for inhomogeneities at the molecular scale which is consistent with findings of other studies \citep{blake2015forced,bertrand2009influence,lukyanov2016dynamic,fernandez2019molecular}. The 
 molecular roughness of the solid
  introduced by the solid atoms, that are arranged in a perfect lattice structure, is sufficient to cause 
 different advancing and receding contact angles.
The presented results emphasise that hydrodynamic DFT is capable of predicting 
differences between advancing and receding contact angles 
at the molecular scale. Furthermore, the dynamic behaviour along the path from an initial equilibrium profile to the dynamic steady state is accurately described by hydrodynamic DFT. This includes the interplay between external forces, velocity-dependent solid-fluid friction and viscous forces in the contact region. These findings indicate that generalised entropy scaling provides meaningful values for the viscosity. 

\begin{figure}
  \centering
  \subfloat[Hydrodynamic DFT with $f_x=\SI{0.056}{\pico\newton}/\mathrm{particle}$]{
      \centering
      \includegraphics[width=0.475\textwidth]{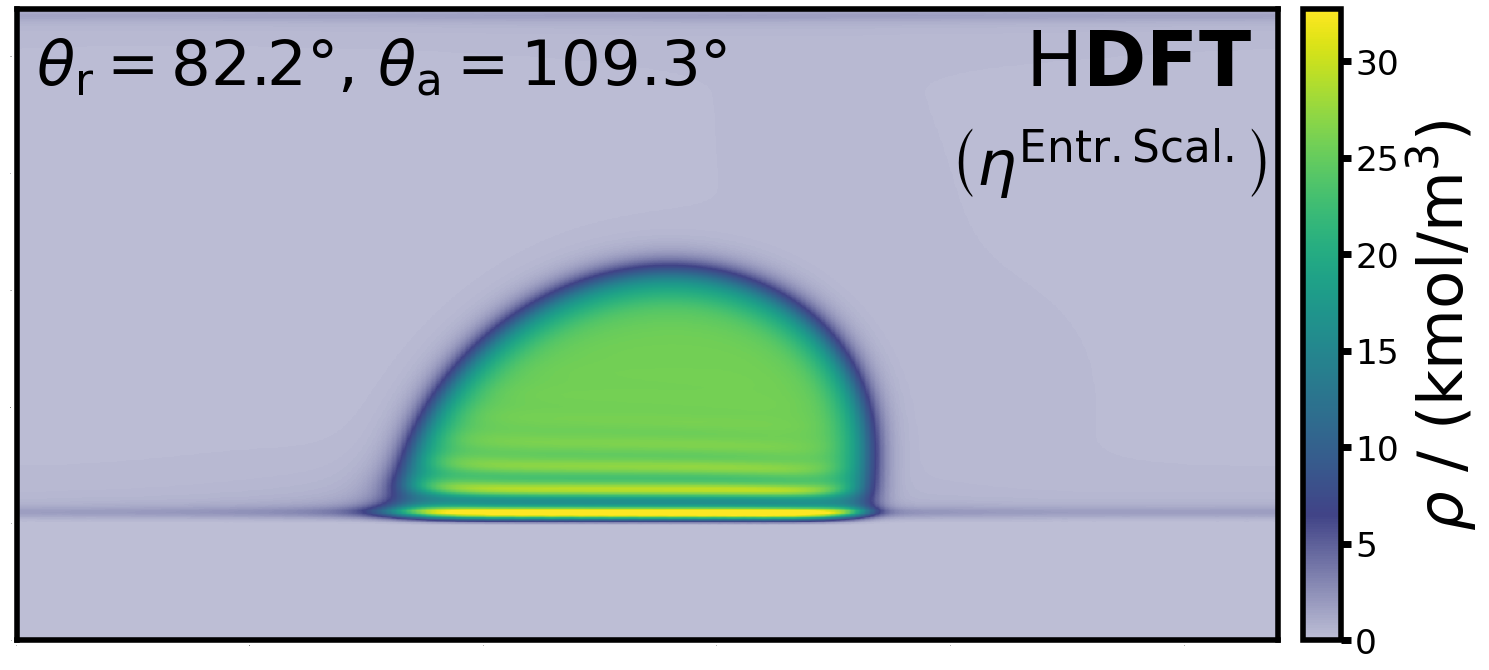}\label{fig:rho_dyn_hdft_01}
 }
 \hfill
  \subfloat[NEMD with $f_x=\SI{0.056}{\pico\newton}/\mathrm{particle}$]{
      \centering
      \includegraphics[width=0.475\textwidth]{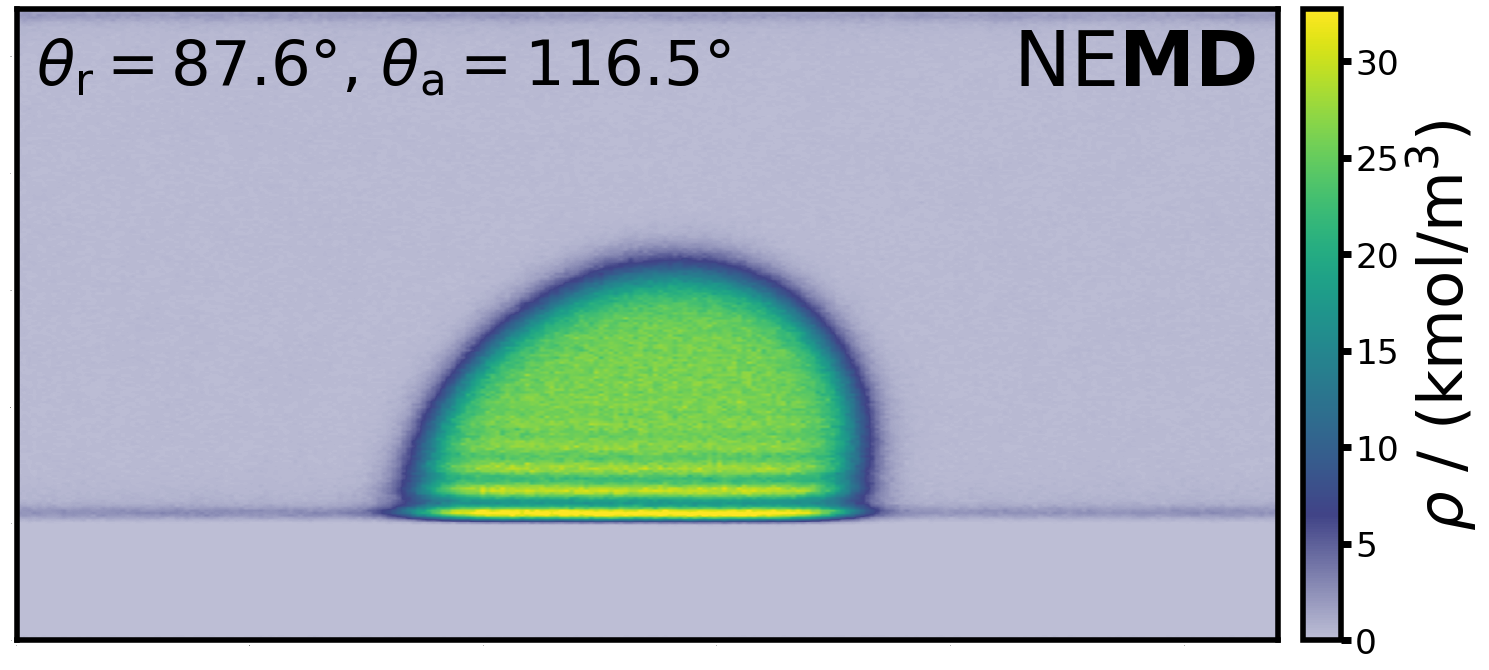}\label{fig:rho_dyn_nemd_01}
 }
 \vfill
 \subfloat[Hydrodynamic DFT with $f_x=\SI{0.112}{\pico\newton}/\mathrm{particle}$]{
      \centering
      \includegraphics[width=0.475\textwidth]{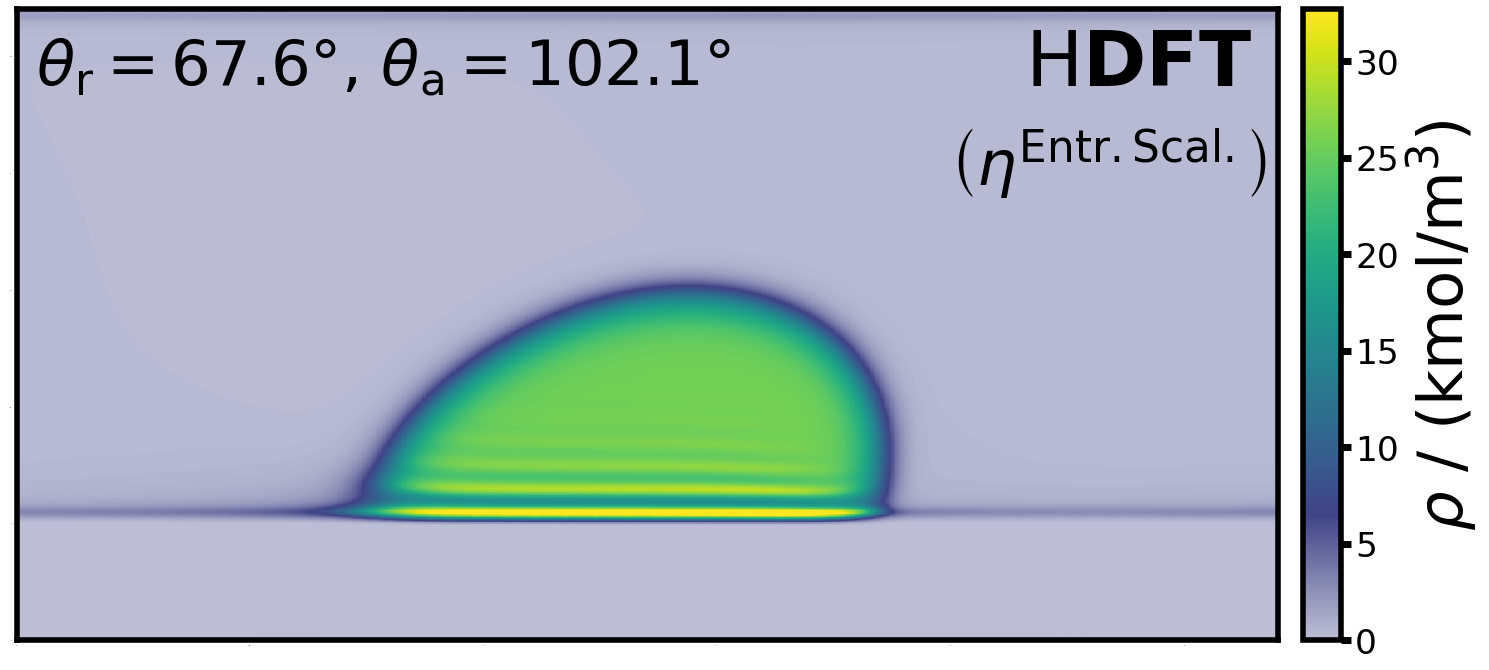}\label{fig:rho_dyn_hdft_02}
 }
 \hfill
  \subfloat[NEMD with $f_x=\SI{0.112}{\pico\newton}/\mathrm{particle}$]{
      \centering
      \includegraphics[width=0.475\textwidth]{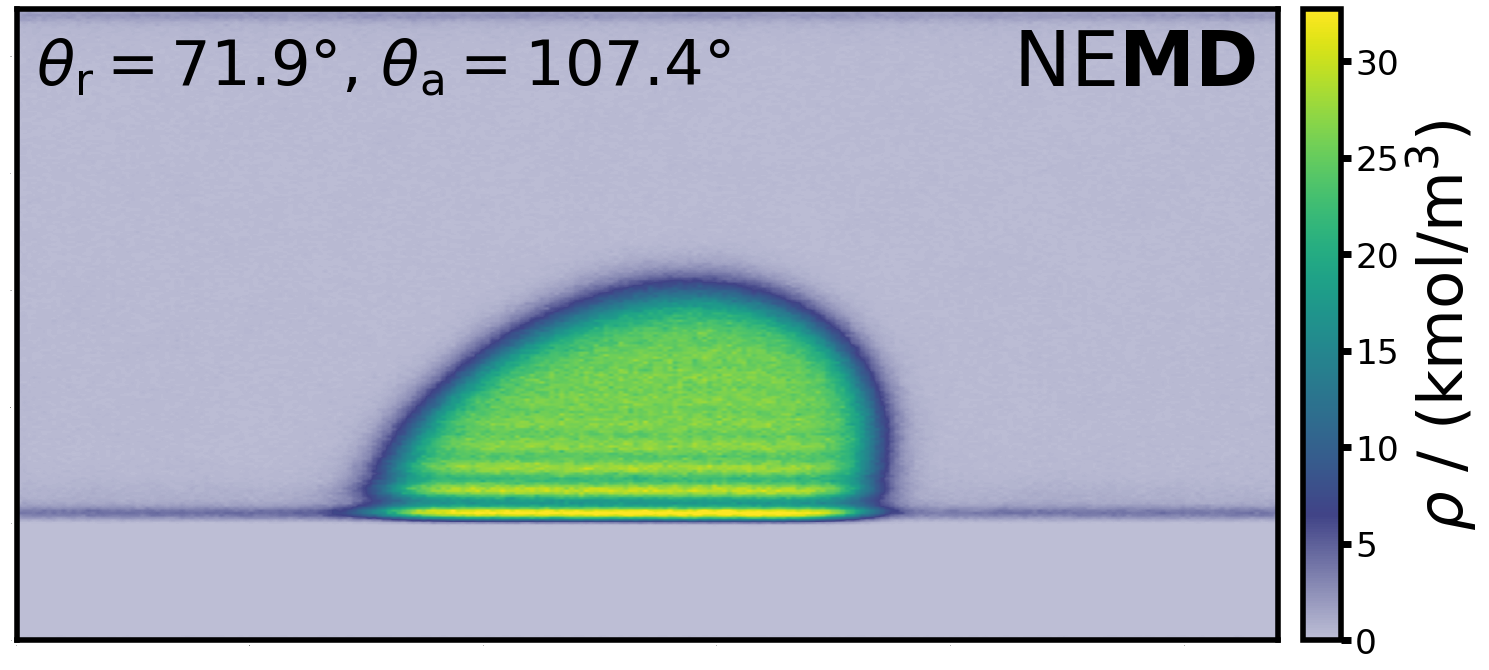}\label{fig:rho_dyn_nemd_02}
 }
 \vfill
 \subfloat[Hydrodynamic DFT with $f_x=\SI{0.224}{\pico\newton}/\mathrm{particle}$]{
      \centering
      \includegraphics[width=0.475\textwidth]{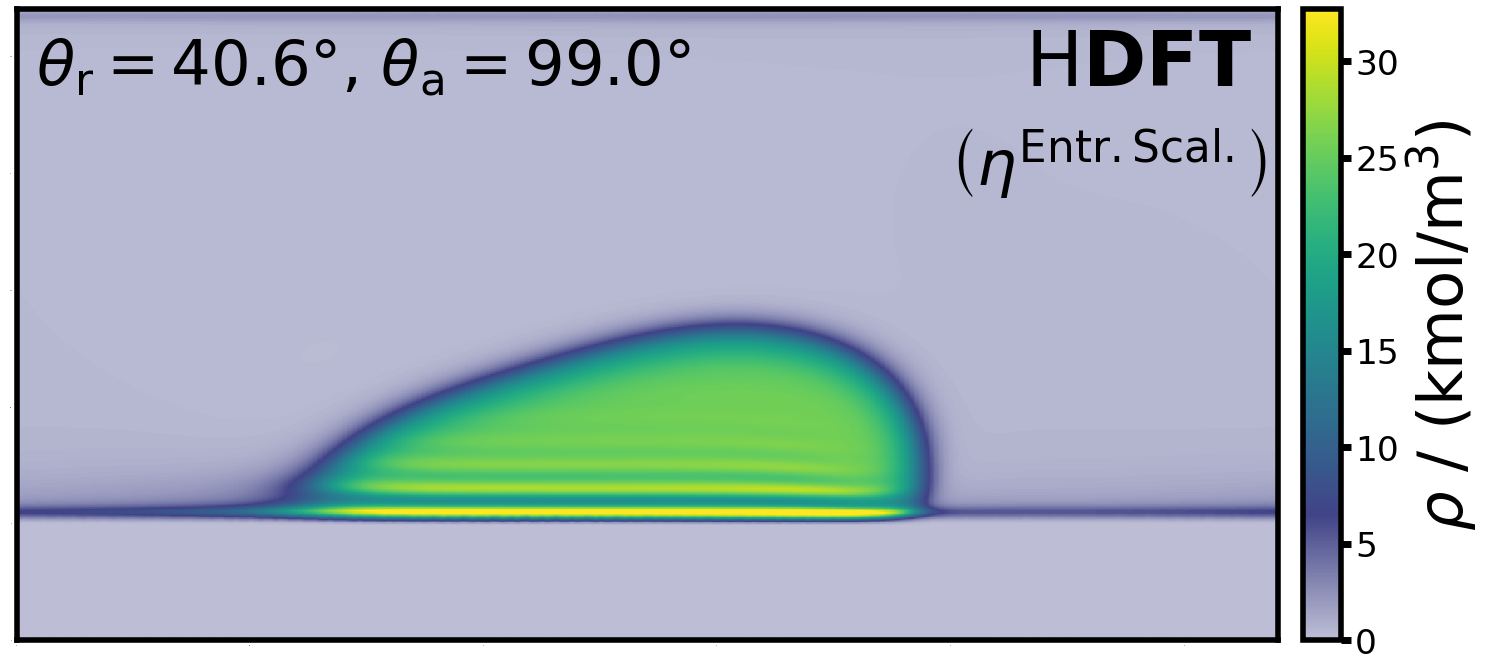}\label{fig:rho_dyn_hdft_04}
 }
 \hfill
  \subfloat[NEMD with $f_x=\SI{0.224}{\pico\newton}/\mathrm{particle}$]{
      \centering
      \includegraphics[width=0.475\textwidth]{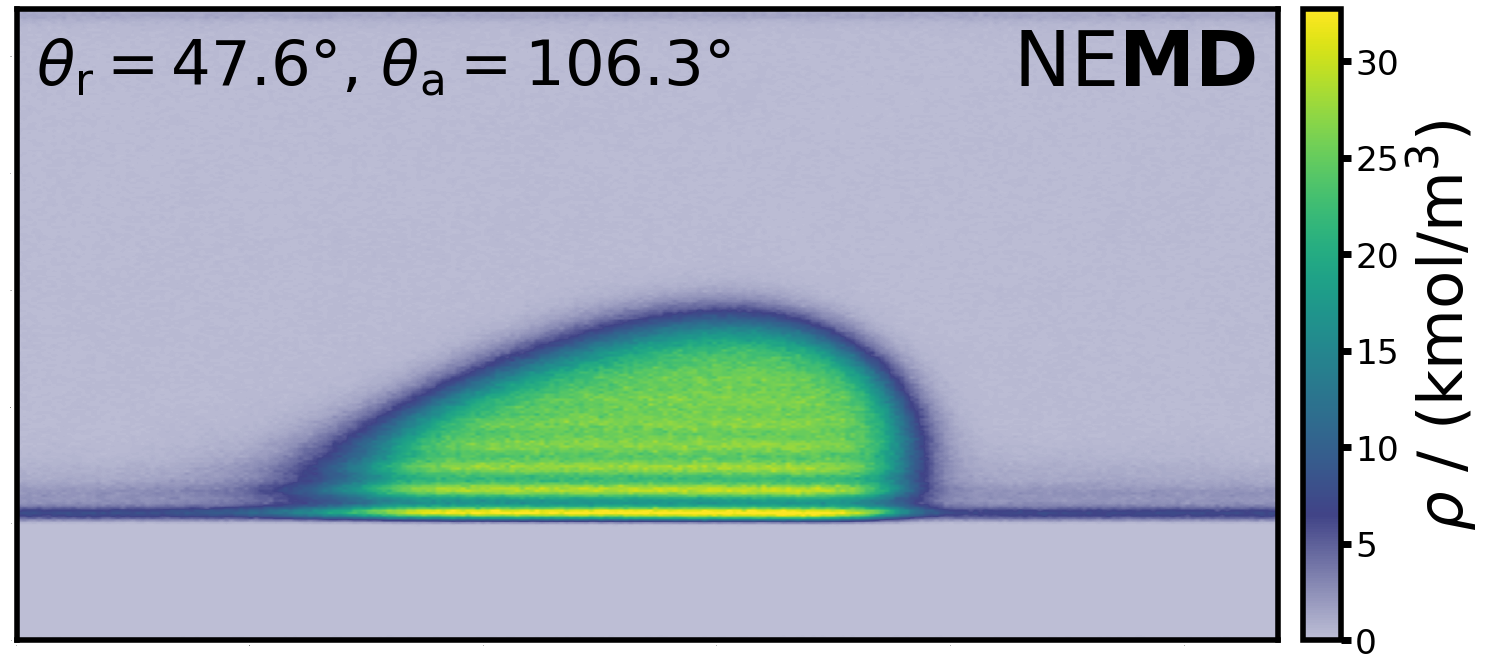}\label{fig:rho_dyn_nemd_04}
 }
  \caption{Density profiles of droplets moving along the solid-fluid interface with different external forces $f_x$ from hydrodynamic DFT (HDFT) and NEMD at $T=\SI{120.02}{\kelvin}$  with $\varepsilon_\mathrm{sf}^*=0.5$  averaged over $\SI{700}{\pico\second}$ after a steady state is reached. 
   }
  \label{fig:rho_dyn}
\end{figure} 

\subsection{Influence of External Force} \label{sec:results_force}

For steady states, larger external forces lead to an increase of contact region velocity, which in turn affects microscopic dynamic contact angles as discussed above and in the literature \citep{ren2007boundary,bertrand2009influence,blake2015forced,lukyanov2016dynamic,lukyanov2017hydrodynamics,fernandez2019molecular}. The general finding is, that advancing contact angles increase and  receding contact angles decrease compared to the equilibrium contact angle.

In \cref{fig:rho_dyn}, results for the density profiles of  moving droplets from hydrodynamic DFT using the generalised entropy scaling viscosity model and NEMD simulations of are depicted for different external forces (results for the bulk viscosity model are provided in the supporting information). Density profiles were averaged over at least \SI{300}{\pico\second} after a steady state was reached, for which the centre of mass of the droplets in each snapshot was shifted to position $L_x/2$. This postprocessing is required for the NEMD results to average out fluctuations, whereas hydrodynamic DFT provides noise free results and averaging is performed solely for better comparability. Due to this averaging, inhomogeneities on the solid-fluid interface are not visible in contrast to  the equilibrium droplets (see \cref{fig:rho_static}) and the snapshots of the accelerating droplets (see \cref{fig:rho_t}).

For the lowest external force given in \cref{fig:rho_dyn_hdft_01,fig:rho_dyn_nemd_01}, 
a difference between advancing and receding contact angle
is observed for both, hydrodynamic DFT and NEMD. 
In particular, advancing contact angles are \SI{109.3}{\degree} and \SI{116.5}{\degree} while receding contact angles are found as \SI{82.2}{\degree} and \SI{87.6}{\degree} from hydrodynamic DFT and NEMD, respectively. The small underestimation of contact angles, which was identified for the equilibrium droplet, transfers to the dynamic case. Furthermore, the viscosity coefficients obtained from generalised entropy scaling and the inherent approximations of hydrodynamic DFT (e.g.\ adiabatic approximation) might introduce additional error.
Nevertheless, these results demonstrate good agreement between hydrodynamic DFT and NEMD for $f_x=\SI{0.056}{\pico\newton}$, qualitatively in terms of the droplet shape given by the density profiles and quantitatively regarding the advancing and receding contact angles. 
 
For an increasing external force (see \cref{fig:rho_dyn_hdft_02,fig:rho_dyn_hdft_04,fig:rho_dyn_nemd_02,fig:rho_dyn_nemd_04}) the deformation of the droplets and consequently, the 
difference between advancing and receding contact angle
, become more pronounced. The density profiles show that the droplets flatten and their width increases.
The adsorption at the solid-vapour interface also becomes stronger with increasing external force. Remarkably, this subtle effect is captured by hydrodynamic DFT. 

\begin{figure}
  \centering
  \includegraphics[width=0.55\textwidth]{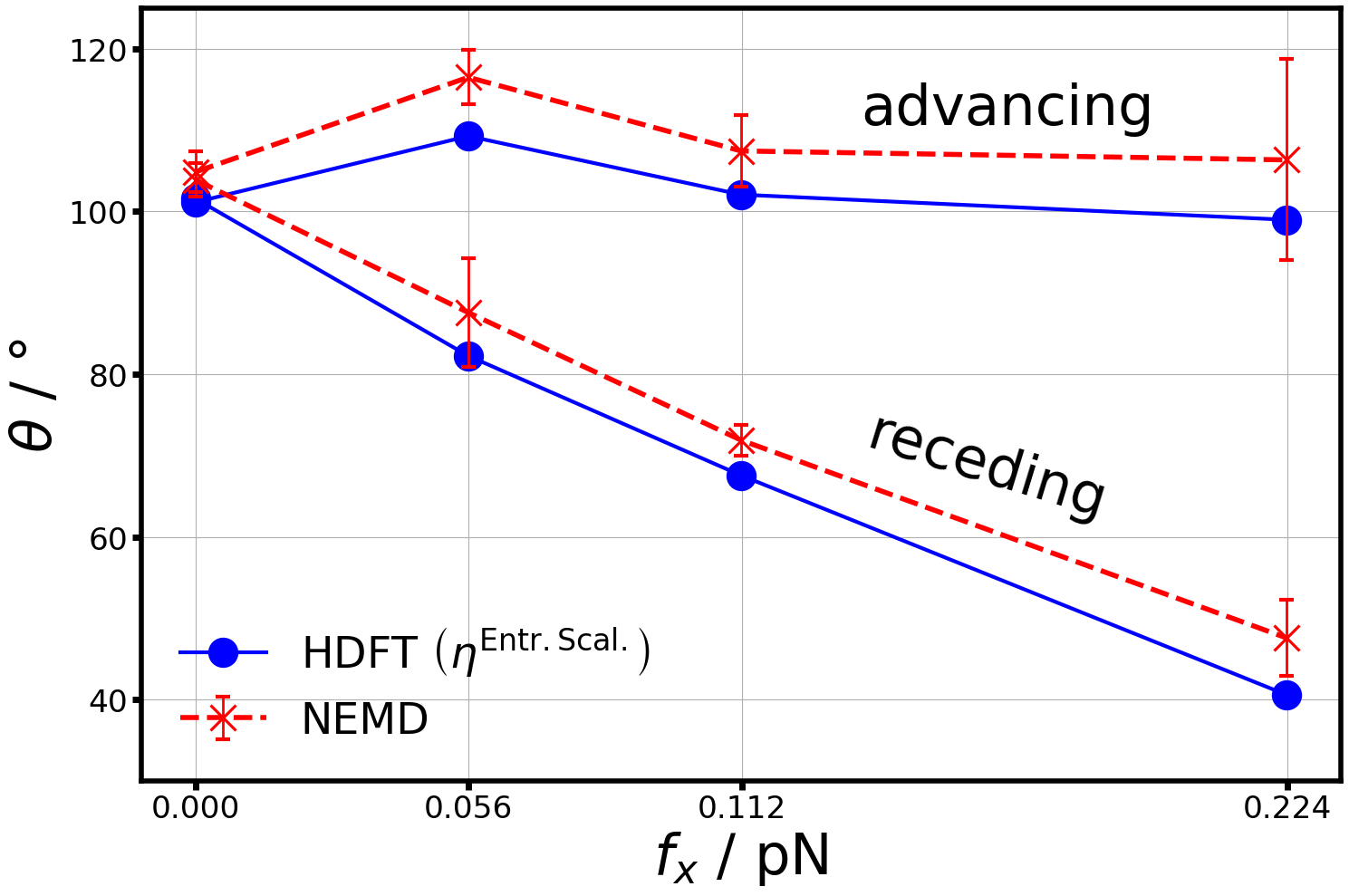}
  \caption{Summary of advancing and receding contact angles from hydrodynamic DFT (blue points) and NEMD (red crosses) for different external forces (per particle) at $T=\SI{120.02}{\kelvin}$  and with $\varepsilon_\mathrm{sf}^*=0.5$. }
  \label{fig:contactAngle_forces}
\end{figure}
\Cref{fig:contactAngle_forces} summarises the contact angles in the system with $\varepsilon_\mathrm{sf}^*=0.5$ for the different external forces. 
In addition, the relation between contact angles and the velocity of the contact region is visualised and discussed in the supplementary data. 
Interestingly, in the system studied here, while the advancing contact angle increases from the equilibrium case to the lowest external force $f_x=\SI{0.056}{\pico\newton}$, it does not increase with further increases in the force, but rather approaches values similar to the equilibrium contact angle. In contrast, the receding dynamic contact angle monotonically decreases with increasing external forces. 
The dependence of the dynamic contact angle and droplet shape on the external force is captured by the hydrodynamic DFT  in good agreement with NEMD simulations with a small systematic underestimation of the contact angles in all results. We expect that this underestimation could be further reduced by improving the agreement in the equilibrium case. 

\subsection{ Influence of Wetting Strength} \label{sec:results_eps}

\begin{figure}
  \centering
  \subfloat[DFT in equilibrium (i.e.\ with $f_x=\SI{0.0}{\pico\newton}/\mathrm{particle}$)]{
      \centering
      \includegraphics[width=0.475\textwidth]{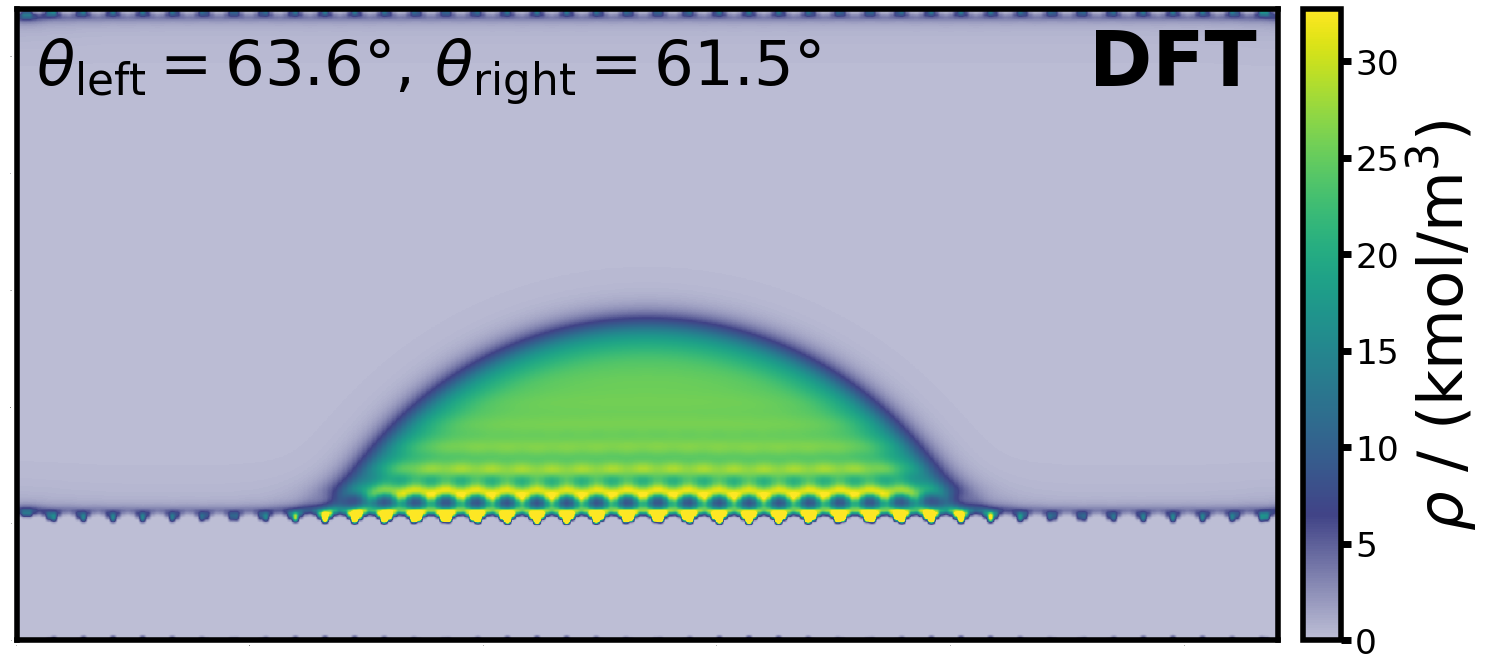}\label{fig:eps07_rho_dft}
 }
 \hfill
  \subfloat[MD in equilibrium (i.e.\ with $f_x=\SI{0.0}{\pico\newton}/\mathrm{particle}$)]{
      \centering
      \includegraphics[width=0.475\textwidth]{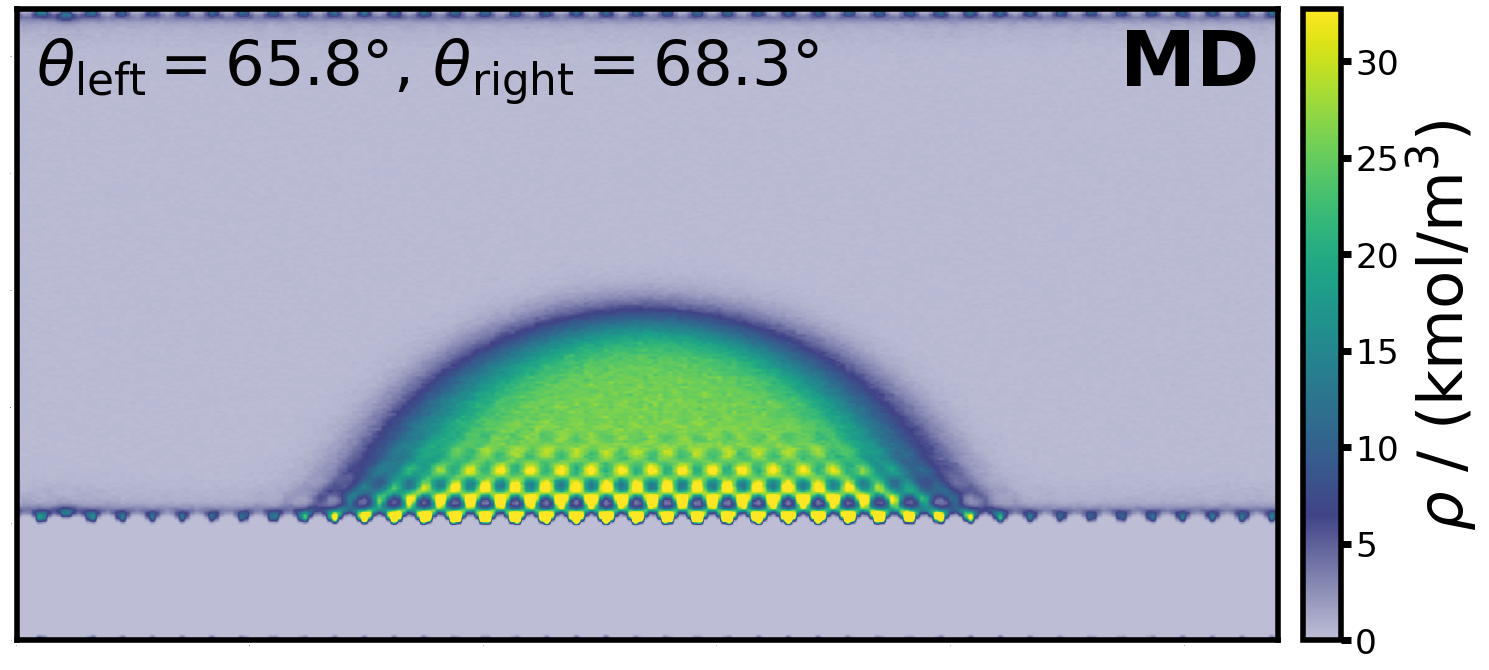}\label{fig:eps07_rho_md}
 }
 \vfill
 \subfloat[Hydrodynamic DFT with $f_x=\SI{0.112}{\pico\newton}/\mathrm{particle}$]{
      \centering
      \includegraphics[width=0.475\textwidth]{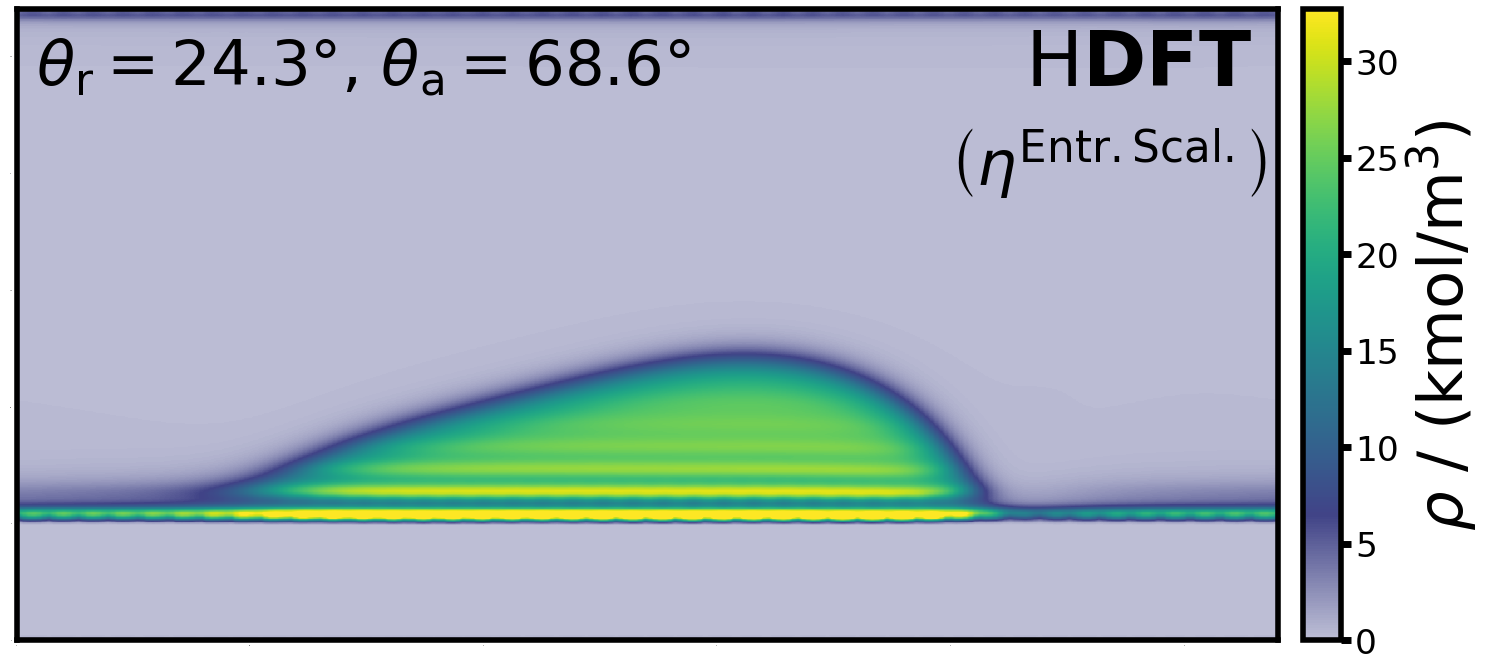}\label{fig:eps07_rho_dyn_hdft_02}
 }
 \hfill
  \subfloat[NEMD with $f_x=\SI{0.112}{\pico\newton}/\mathrm{particle}$]{
      \centering
      \includegraphics[width=0.475\textwidth]{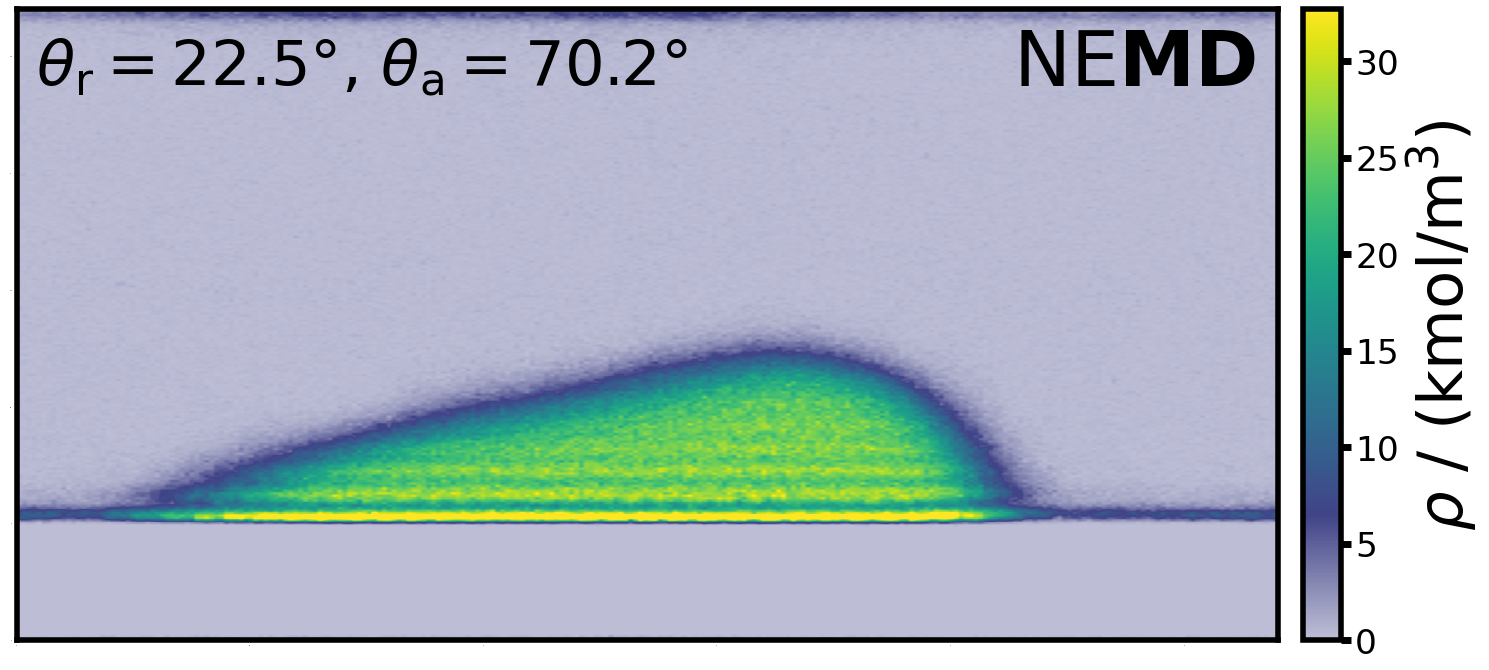}\label{fig:eps07_rho_dyn_nemd_02}
 }
 \vfill
 \subfloat[Hydrodynamic DFT with $f_x=\SI{0.224}{\pico\newton}/\mathrm{particle}$]{
      \centering
      \includegraphics[width=0.475\textwidth]{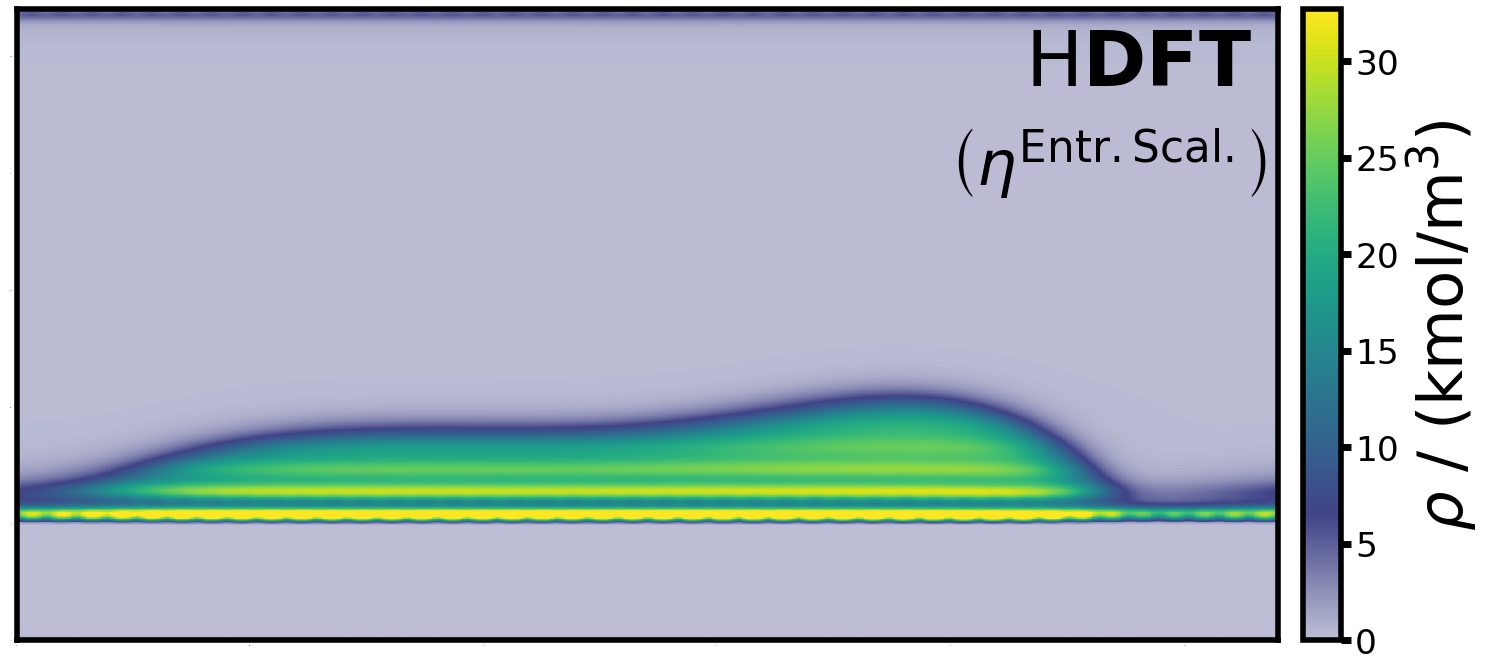}\label{fig:eps07_rho_dyn_hdft_04}
 }
 \hfill
  \subfloat[NEMD with $f_x=\SI{0.224}{\pico\newton}/\mathrm{particle}$]{
      \centering
      \includegraphics[width=0.475\textwidth]{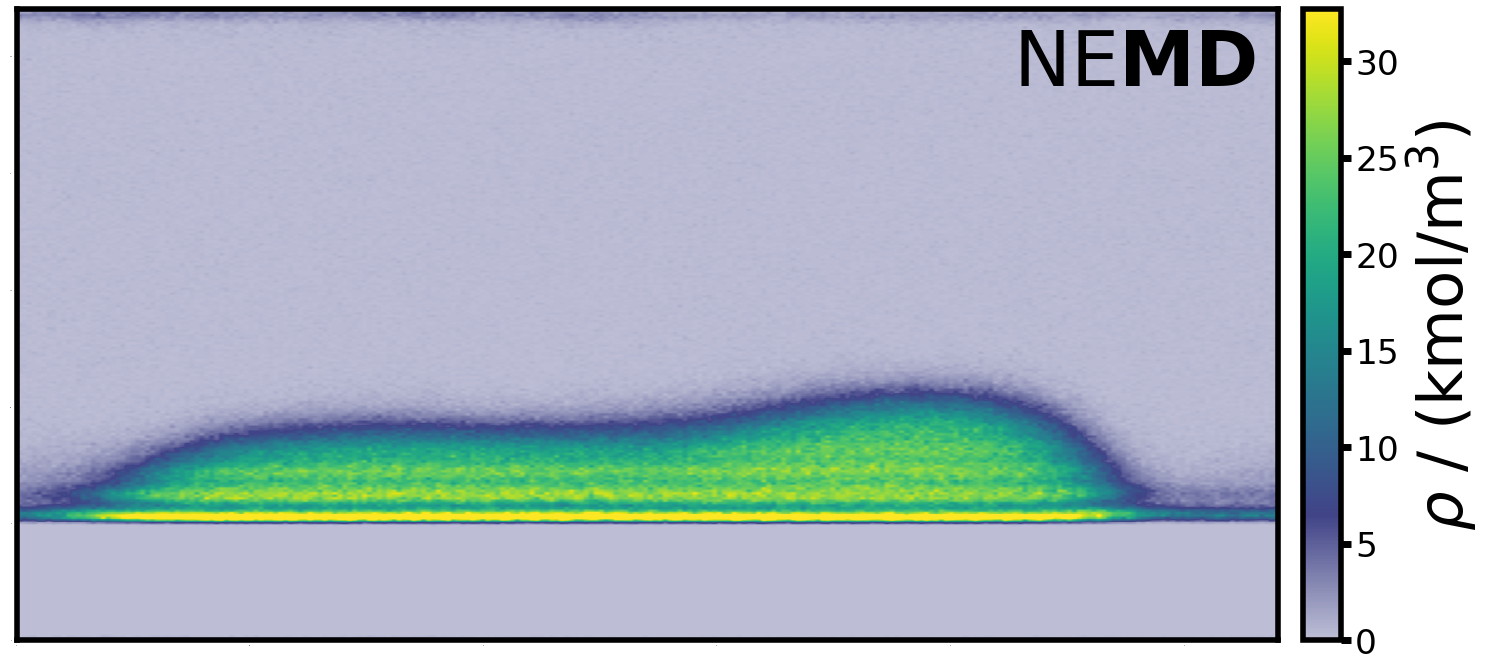}\label{fig:eps07_rho_dyn_nemd_04}
 }
  \caption{Density profiles of droplets moving along the solid-fluid interface with different external forces $f_x$ from equilibrium and hydrodynamic DFT (left) and equilibrium and non-equilibrium MD (right) at $T=\SI{120.02}{\kelvin}$  with $\varepsilon_\mathrm{sf}^*=0.7$  averaged over $\SI{700}{\pico\second}$ after a steady state is reached. 
   }
  \label{fig:eps07_rho_dyn}
\end{figure}

The wetting strength significantly affects the contact angles of droplets. It is modelled in NEMD and hydrodynamic DFT by altering the solid-fluid energy interaction parameter $\epssf$. A large value for $\epssf$ corresponds to strong wetting and small contact angles. In the following, results for stronger wetting, modelled with $\epssf=0.7$, are presented in comparison to the results provided above, where $\epssf=0.5$ was used. The goal is to  analyse if the hydrodynamic DFT model captures the influence of the wetting strength on dynamic contact angles and droplet velocities.   

Density profiles with $\varepsilon_\mathrm{sf}^*=0.7$ for the equilibrium and dynamic case are shown in \cref{fig:eps07_rho_dyn}.
We remind that for equilibrium and hydrodynamic DFT a two-dimensional representation is employed, whereas (NE)MD is simulated in three dimensions. 
The equilibrium density profile from DFT (\cref{fig:eps07_rho_dft}) confirms that increasing the solid-fluid energy interaction parameter to $\epssf=0.7$ leads to a flatter equilibrium droplet with a lower mean contact angle of \SI{62.6}{\degree} compared to the results with  $\epssf=0.5$, where a mean contact angle of \SI{101.4}{\degree} was determined (cf.\ \cref{fig:rho_static_dft}). In addition, the molecular layering at the solid-fluid interface is more pronounced, which is caused by the stronger solid-fluid interactions. The density profile and contact angle from DFT   is in good agreement with results from equilibrium MD (\cref{fig:eps07_rho_md}). Analogously to the results for  $\epssf=0.5$, the mean contact angle from DFT is slightly smaller than from MD (\SI{62.6}{\degree} vs \SI{67.1}{\degree}). 

\Cref{fig:eps07_rho_dyn_hdft_02} visualises the density profile from hydrodynamic DFT  for the medium force ($f_x=\SI{0.112}{\pico\newton}$). The droplet deforms strongly and 
the advancing contact angle differs from the receding one. This difference
 $\Theta_\mathrm{a}-\Theta_\mathrm{r}$ is larger for the case with stronger wetting (\SI{44.3}{\degree} for $\epssf=0.7$ vs.\ \SI{34.5}{\degree} for $\epssf=0.5$). This might be due to a larger solid-fluid friction at higher wetting strengths, as a result of stronger solid-fluid interactions in the contact region. 
At the largest force (\cref{fig:eps07_rho_dyn_hdft_04}), the droplet elongates and the tendency to  separate into two smaller droplets can be observed. This behaviour was not observed in the previous results (cf. \cref{fig:rho_dyn_hdft_04}) and meaningful dynamic contact angles can not be obtained by adjusting a half-circle to the density profiles. This shape deformation as predicted from hydrodynamic DFT agrees well with the results from NEMD simulations as provided in \cref{fig:eps07_rho_dyn_nemd_02,fig:eps07_rho_dyn_nemd_04}.

\begin{figure}
  \centering
  \includegraphics[width=0.55\textwidth]{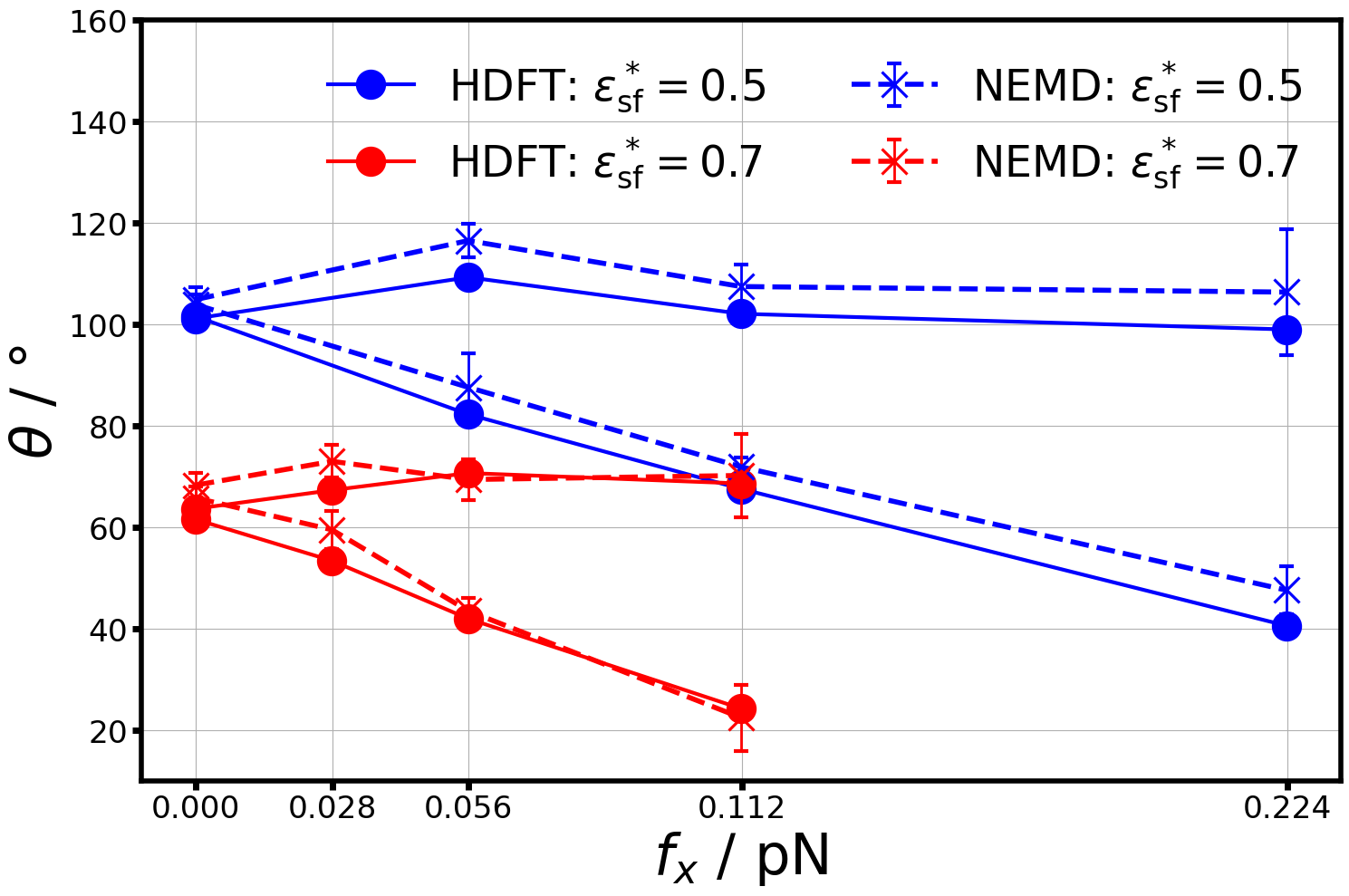}
  \caption{Summary of advancing and receding contact angles for varying solid-fluid interaction parameter $\varepsilon_\mathrm{sf}^*$ determined for different external forces (per particle) from hydrodynamic DFT (HDFT) with entropy scaling viscosity model (circles) and from NEMD (crosses) at $T=\SI{120.02}{\kelvin}$. }
  \label{fig:contactAngle_eps}
\end{figure}
Contact angles determined from these density profiles are summarised and contrasted to previous results for $\epssf=0.5$ in \cref{fig:contactAngle_eps}. For both wetting strengths similar qualitative trends are observed from hydrodynamic DFT and NEMD: the receding contact angles decrease in comparison to the equilibrium contact angles, whereas the advancing contact angles first increase and then approach values similar to the equilibrium contact angles. 
The small underestimation of the contact angles from DFT and hydrodynamic DFT, which appears in the equilibrium case and transfers to the dynamic case, is observed for both wetting strengths. For the larger wetting strength this effect vanishes at larger forces. 
Density profiles and contact angles from hydrodynamic DFT and NEMD are in good quantitative agreement for all forces investigated for both wetting strengths. 

\begin{figure}
  \centering
  \includegraphics[width=0.55\textwidth]{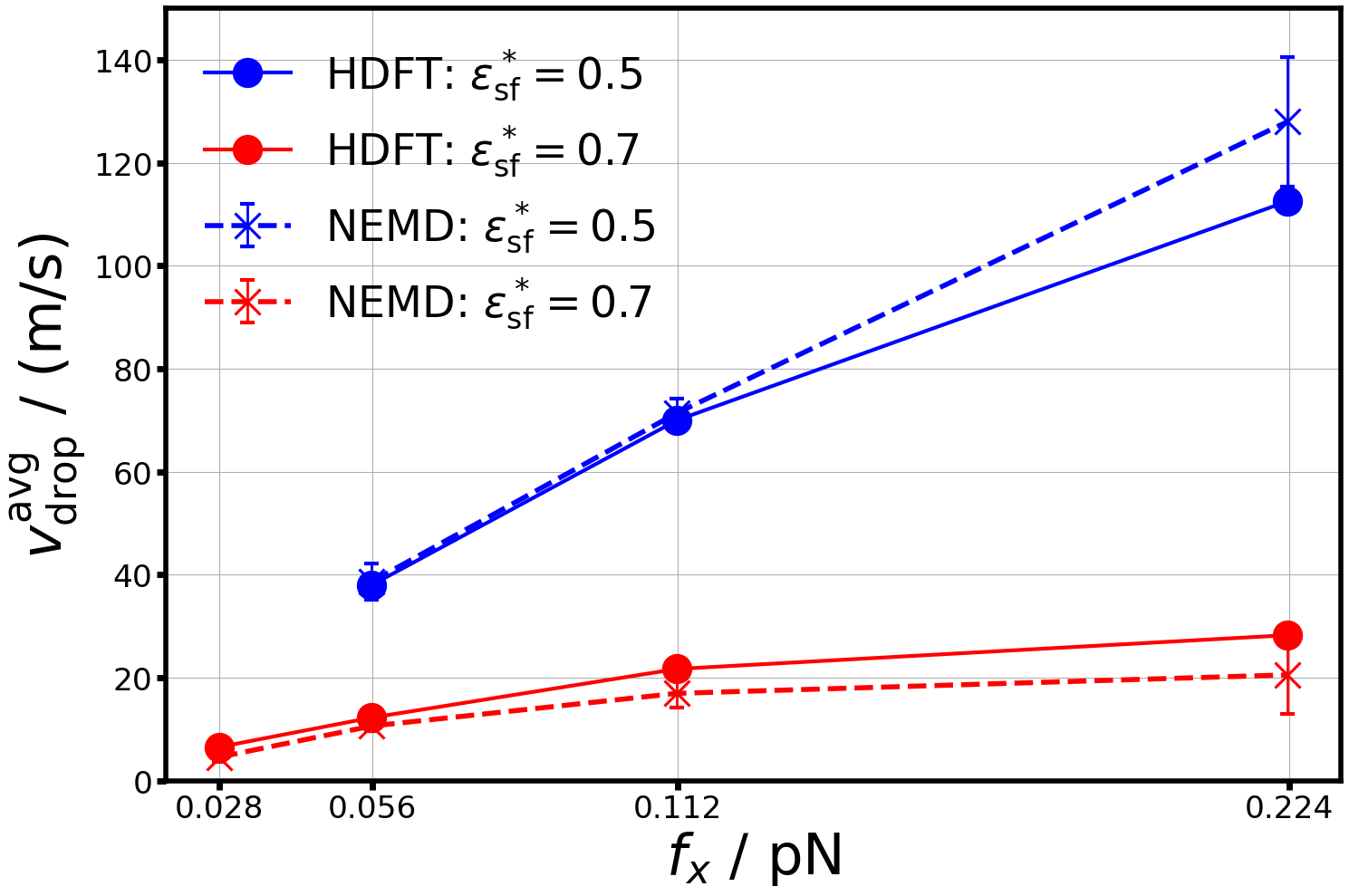}
  \caption{Steady state velocity of the centre of mass of the moving droplet for varying solid-fluid interaction parameter $\varepsilon_\mathrm{sf}^*$ determined for different external forces (per particle) from hydrodynamic DFT (HDFT) with entropy scaling viscosity model (circles) and from NEMD (crosses) at $T=\SI{120.02}{\kelvin}$. }
  \label{fig:vavg_eps}
\end{figure}
According to \cref{fig:vavg_eps}, the steady state velocity  of droplets for stronger wetting  ($\epssf=0.7$)  is much smaller compared to  weaker wetting ($\epssf=0.5$). Furthermore, the steady state velocity increases less strongly between the second largest and largest forces. This is consistent with the more pronounced 
difference between advancing and receding contact angle
for the stronger wetting case, since both can be explained by a larger solid-fluid friction. At the largest force the solid-fluid friction becomes very large leading to a small increase in steady state velocity and the tendency to separate into smaller droplets.
The NEMD results confirm the accuracy of steady state velocities from hydrodynamic DFT.

These results show that hydrodynamic DFT accurately captures the influence of the wetting strength on the wetting behaviour. Importantly, this also illustrates the transferability of the adjustable parameter in the generalised entropy scaling approach.

\subsection{ Influence of Molecular Roughness of the Solid} \label{sec:results_roughness}

\begin{figure}
  \centering
  \includegraphics[width=0.55\textwidth]{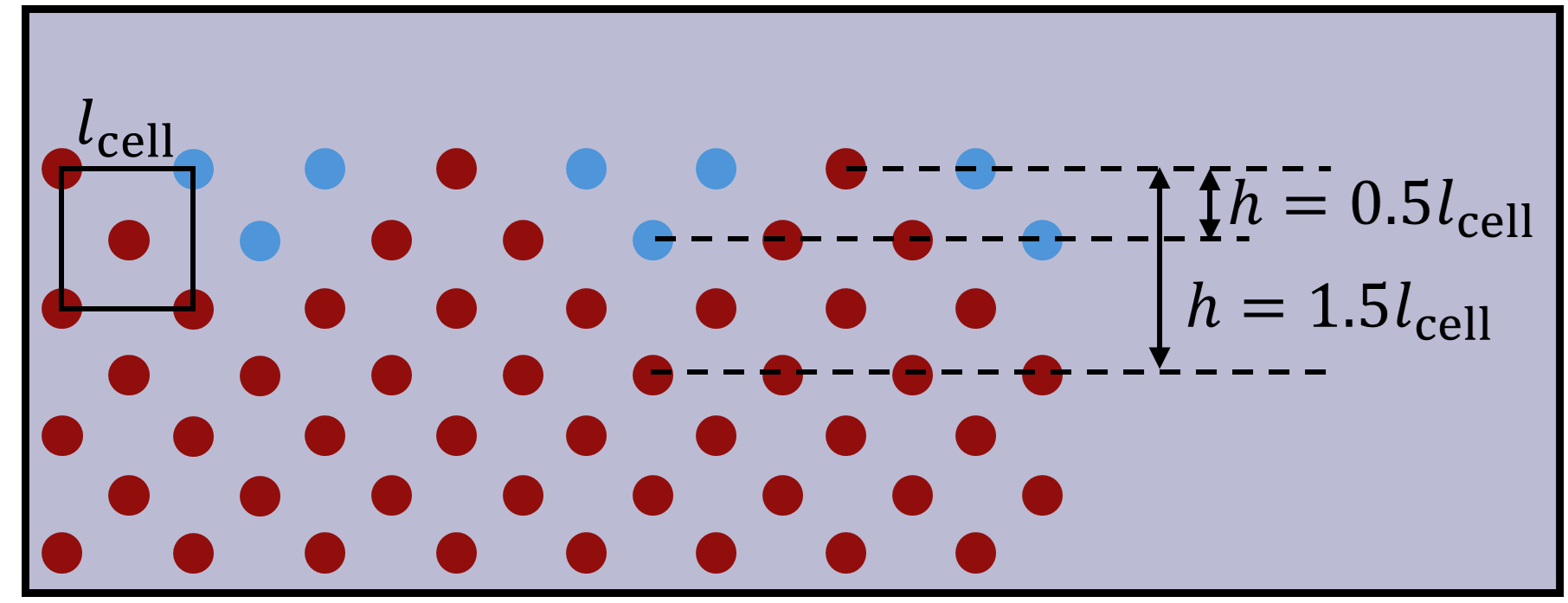}
  \caption{Visualisation of the methodology for modelling different
  molecular 
  roughnesses of the solid. The smoother solid (previous results) contains red and blue atoms; the increased roughness is obtained by removing the blue atoms from the solid. The height difference $h$ between the top and lowest layers which are in contact with the fluid is determined as multiples of the length of a unit cell $l_\mathrm{cell}$.}
  \label{fig:roughness}
\end{figure}

\begin{figure}[h!]
  \centering
  \subfloat[DFT in equilibrium (i.e.\ with $f_x=\SI{0.0}{\pico\newton}/\mathrm{particle}$)]{
      \centering
      \includegraphics[width=0.475\textwidth]{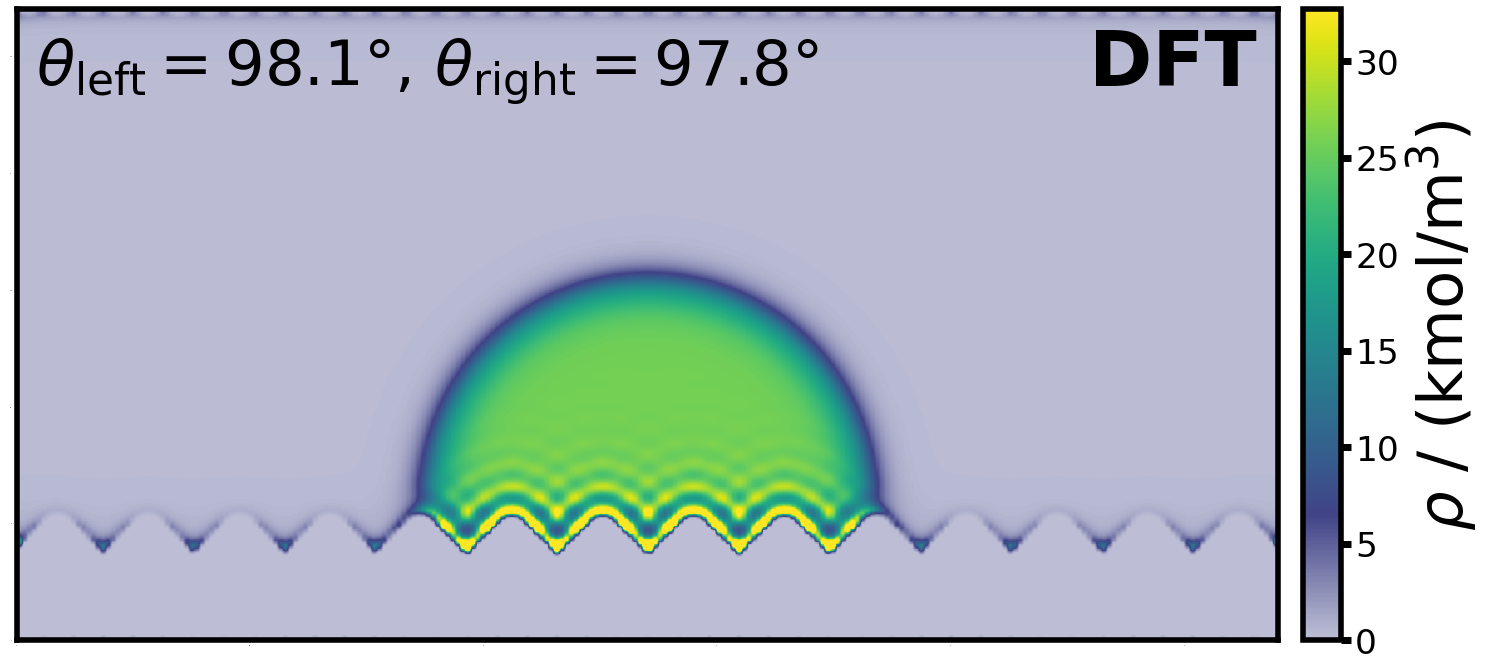}\label{fig:roughness_rho_dft}
 }
 \hfill
  \subfloat[MD in equilibrium (i.e.\ with $f_x=\SI{0.0}{\pico\newton}/\mathrm{particle}$)]{
      \centering
      \includegraphics[width=0.475\textwidth]{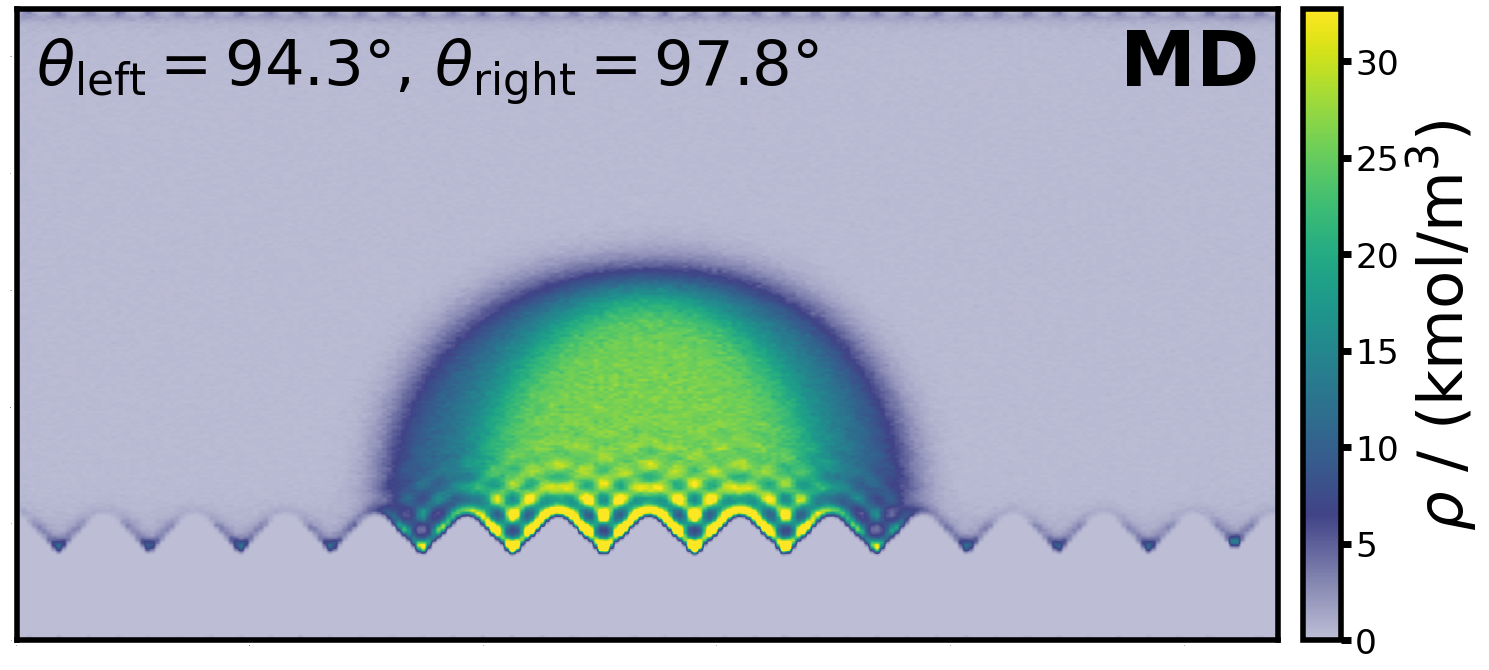}\label{fig:roughness_rho_md}
 }
 \vfill
 \subfloat[Hydrodynamic DFT with $f_x=\SI{0.112}{\pico\newton}/\mathrm{particle}$]{
      \centering
      \includegraphics[width=0.475\textwidth]{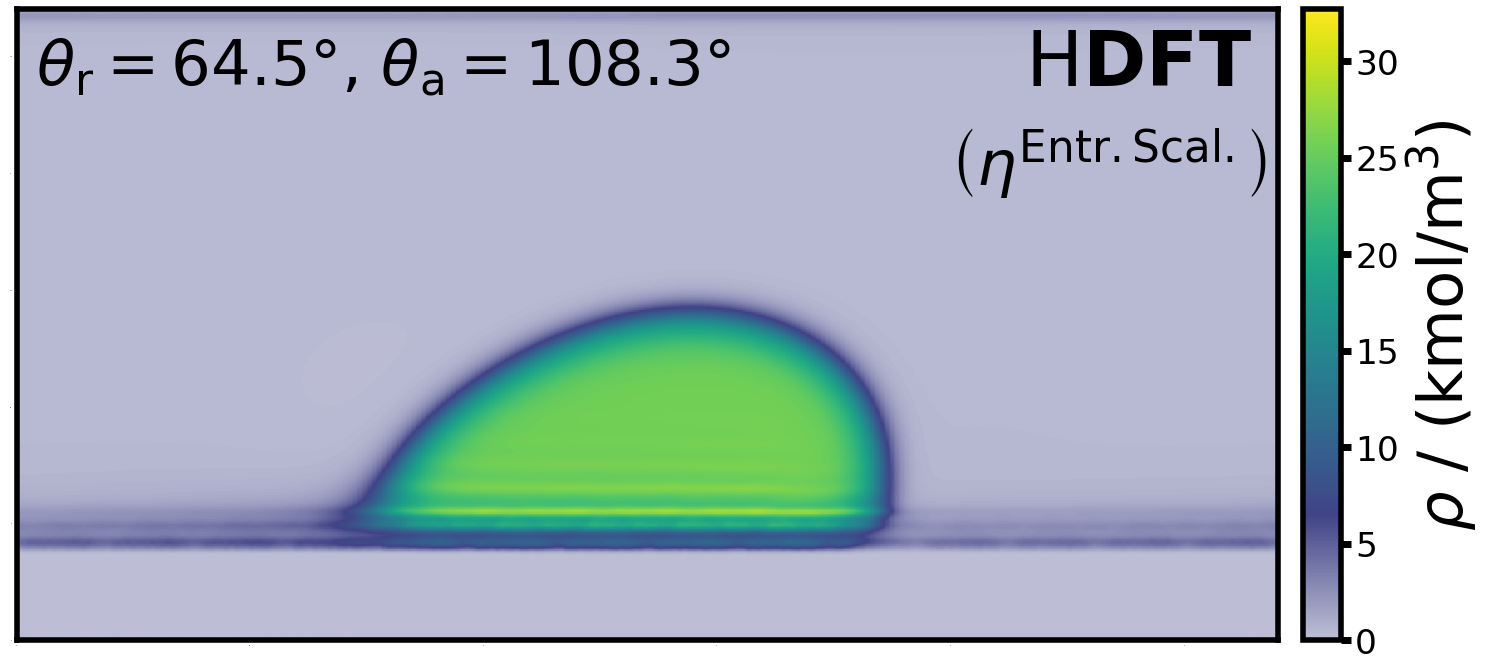}\label{fig:roughness_rho_dyn_hdft_02}
 }
 \hfill
  \subfloat[NEMD with $f_x=\SI{0.112}{\pico\newton}/\mathrm{particle}$]{
      \centering
      \includegraphics[width=0.475\textwidth]{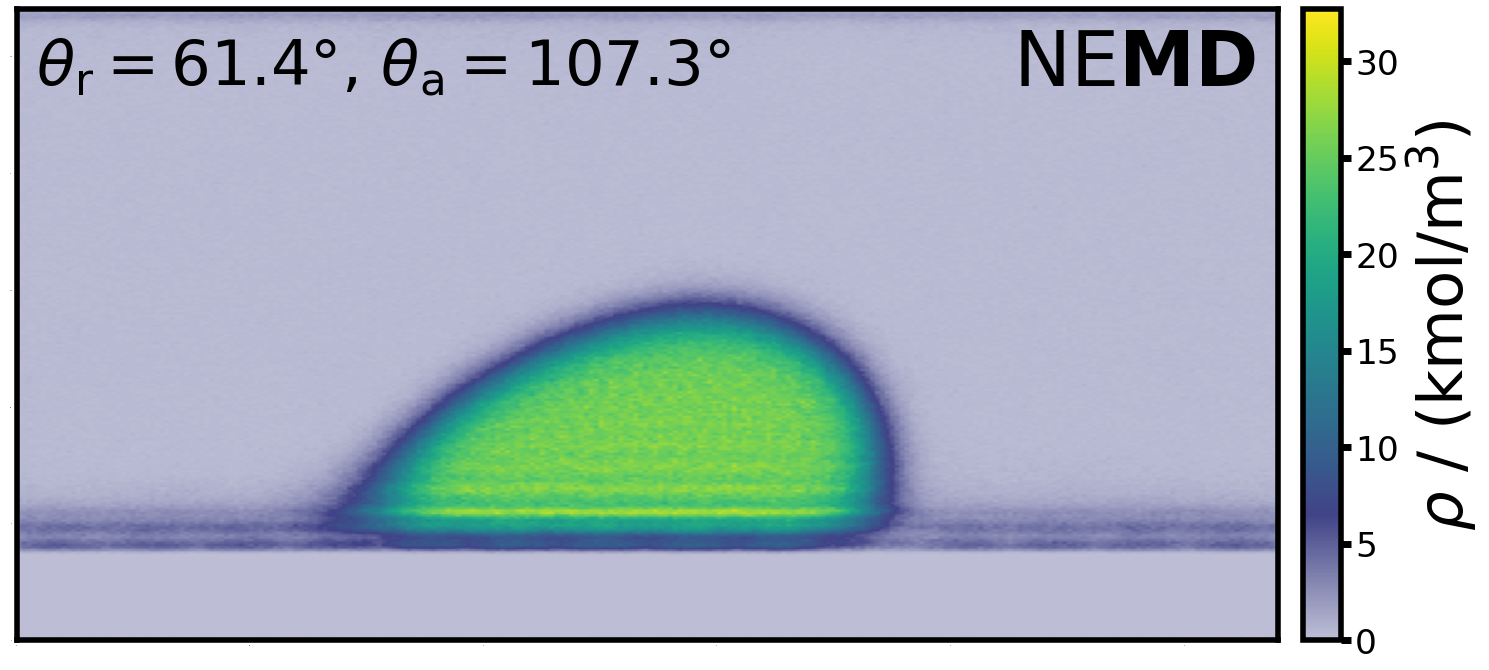}\label{fig:roughness_rho_dyn_nemd_02}
 }
 \vfill
 \subfloat[Hydrodynamic DFT with $f_x=\SI{0.224}{\pico\newton}/\mathrm{particle}$]{
      \centering
      \includegraphics[width=0.475\textwidth]{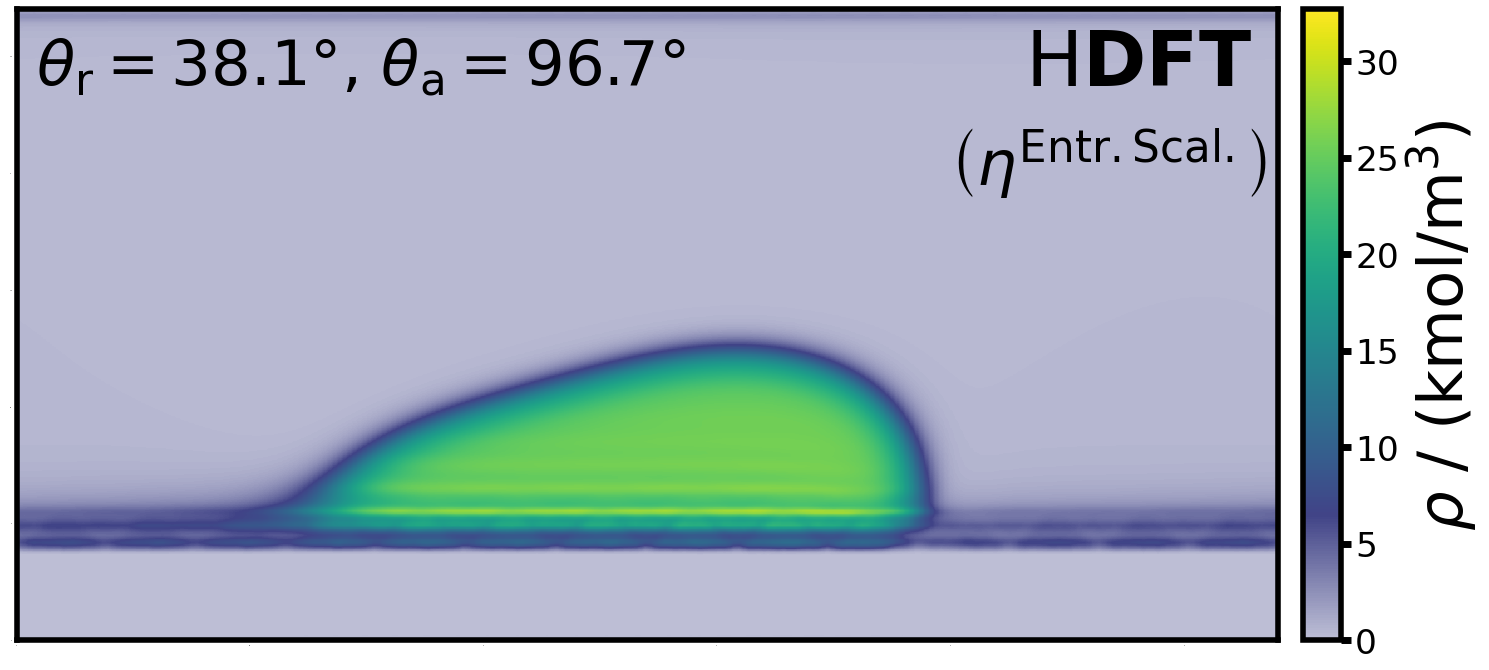}\label{fig:roughness_rho_dyn_hdft_04}
 }
 \hfill
  \subfloat[NEMD with $f_x=\SI{0.224}{\pico\newton}/\mathrm{particle}$]{
      \centering
      \includegraphics[width=0.475\textwidth]{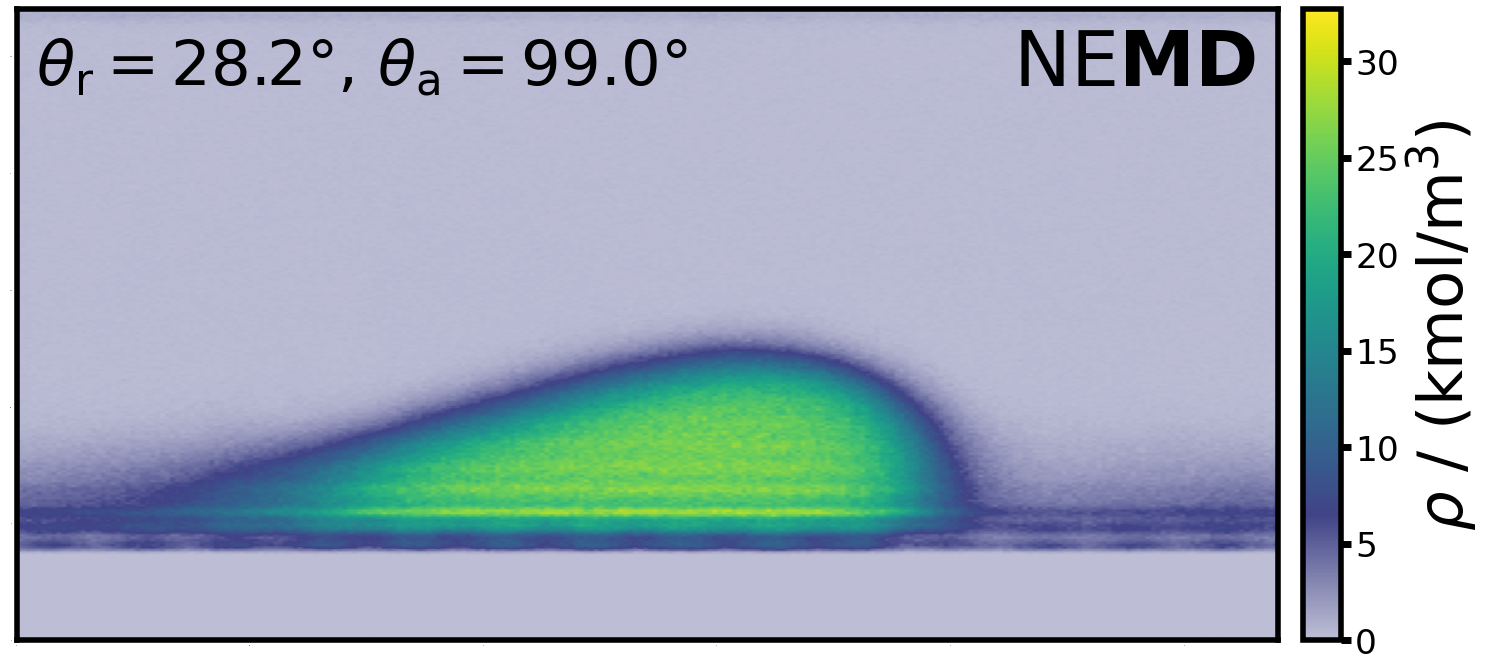}\label{fig:roughness_rho_dyn_nemd_04}
 }
  \caption{Density profiles of droplets moving along the solid-fluid interface for an increased 
  molecular roughness of the solid 
  ($h=1.5l_\mathrm{cell}$) with different external forces $f_x$ from equilibrium and hydrodynamic DFT (left) as well as equilibrium and non-equilibrium MD (right) at $T=\SI{120.02}{\kelvin}$  with $\varepsilon_\mathrm{sf}^*=0.5$  averaged over $\SI{700}{\pico\second}$ after a steady state is reached. 
   }
  \label{fig:roughness_rho_dyn}
\end{figure}

Besides the wetting strength, the 
molecular roughness of the solid
is a major influence factor for the wetting behaviour
as it determines the energetic landscape of the solid-fluid interface, which causes  solid-fluid friction \citep{lukyanov2016dynamic}. 
We employ the term \emph{molecular roughness} to emphasise that this roughness is on the molecular scale and, thus, not directly comparable to the macroscopic roughness of a solid.  
 We use the height difference $h$ between the top and bottom layer of solid atoms exposed to the fluid, as a measure for the 
 molecular roughness of the solid 
 as depicted in \cref{fig:roughness}. By removing solid atoms (visualised in blue) from the lattice, this roughness parameter is increased from $h=0.5l_\mathrm{cell}$ (all previous results) to $h=1.5l_\mathrm{cell}$ (results presented in the following), where $l_\mathrm{cell}$ is the length of a unit cell of the bcc lattice. 

Density profiles for the rougher surface ($h=1.5l_\mathrm{cell}$) are presented in \cref{fig:roughness_rho_dyn}. In the equilibrium density profiles (\cref{fig:roughness_rho_dft,fig:roughness_rho_md}), the increased roughness is clearly visible in the density profiles. While the droplet from DFT is slightly less high, its shape and the determined contact angles agree sufficiently well with results from equilibrium MD. The difference in the shape of the droplets can be explained by the averaging of the external potential, which is required due to the different dimensionalities of the models (two-dimensional hydrodynamic DFT vs. three-dimensional MD) and has a stronger influence on the results for an increased solid roughness.  
The contact angles from DFT and equilibrium MD are slightly smaller compared to the previous results with \SI{98.0}{\degree} and \SI{96.1}{\degree} for  $h=1.5l_\mathrm{cell}$ vs. \SI{101.4}{\degree} and \SI{104.4}{\degree} for $h=0.5l_\mathrm{cell}$. 
On the macroscopic scale the influence of solid roughness on the contact angle can be correlated using the Wenzel equation, while on the microscopic scale its validity remains to be determined. 

In the dynamic density profiles (\cref{fig:roughness_rho_dyn_hdft_02,fig:roughness_rho_dyn_hdft_04,fig:roughness_rho_dyn_nemd_02,fig:roughness_rho_dyn_nemd_04}) the solid roughness is not explicitly visible due to the averaging of density profiles over time, but it leads to a larger gas adsorption which can be observed in all cases. For the medium force ($f_x=\SI{0.112}{\pico\newton}$) the density profiles from hydrodynamic DFT (\cref{fig:roughness_rho_dyn_hdft_02}) and NEMD (\cref{fig:roughness_rho_dyn_nemd_02}) are in very good agreement. While for the strongest force (\cref{fig:roughness_rho_dyn_hdft_04,fig:roughness_rho_dyn_nemd_04}) the agreement with NEMD is still sufficient, the receding contact region is smeared out more strongly in NEMD than in hydrodynamic DFT. As in the equilibrium case, this can be explained by the different dimensionalities of the models. 

\begin{figure}
  \centering
  \includegraphics[width=0.55\textwidth]{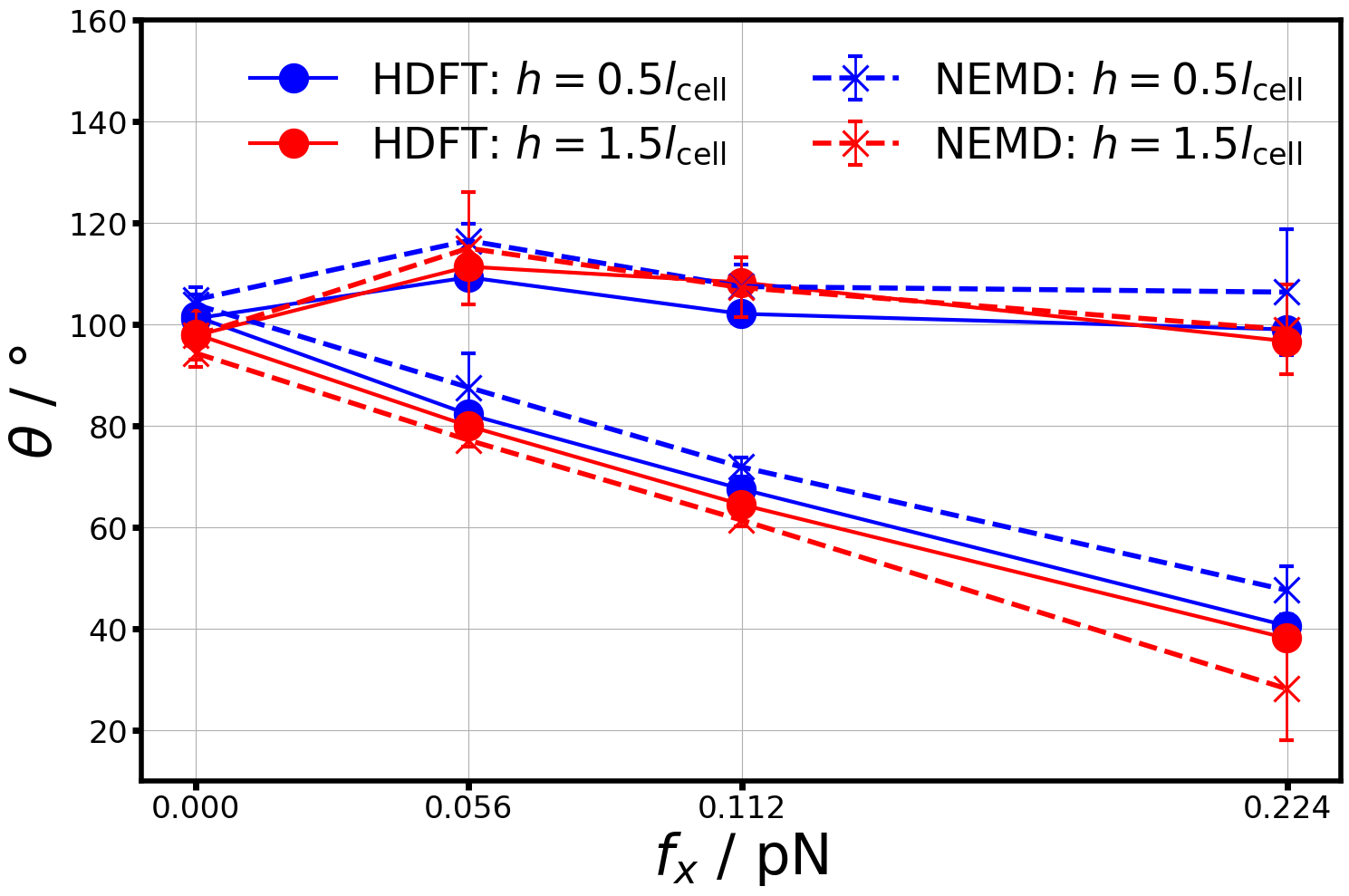}
  \caption{Summary of advancing and receding contact angles for varying solid roughness $h$ determined for different external forces (per particle) from hydrodynamic DFT (HDFT) with entropy scaling viscosity model (circles) and from NEMD (crosses) at $T=\SI{120.02}{\kelvin}$. }
  \label{fig:contactAngle_roughness}
\end{figure}
 Contact angles for the different solid roughness are compared in \cref{fig:contactAngle_roughness}.  Similar values for the contact angles are obtained for the different roughness throughout all forces. For the cases studied here ($h=0.5l_\mathrm{cell}$ and $h=1.5l_\mathrm{cell}$) the solid roughness apparently does not significantly affect the contact angles. Hydrodynamic DFT captures this effect in agreement with NEMD simulations. 

\begin{figure}
  \centering
  \includegraphics[width=0.55\textwidth]{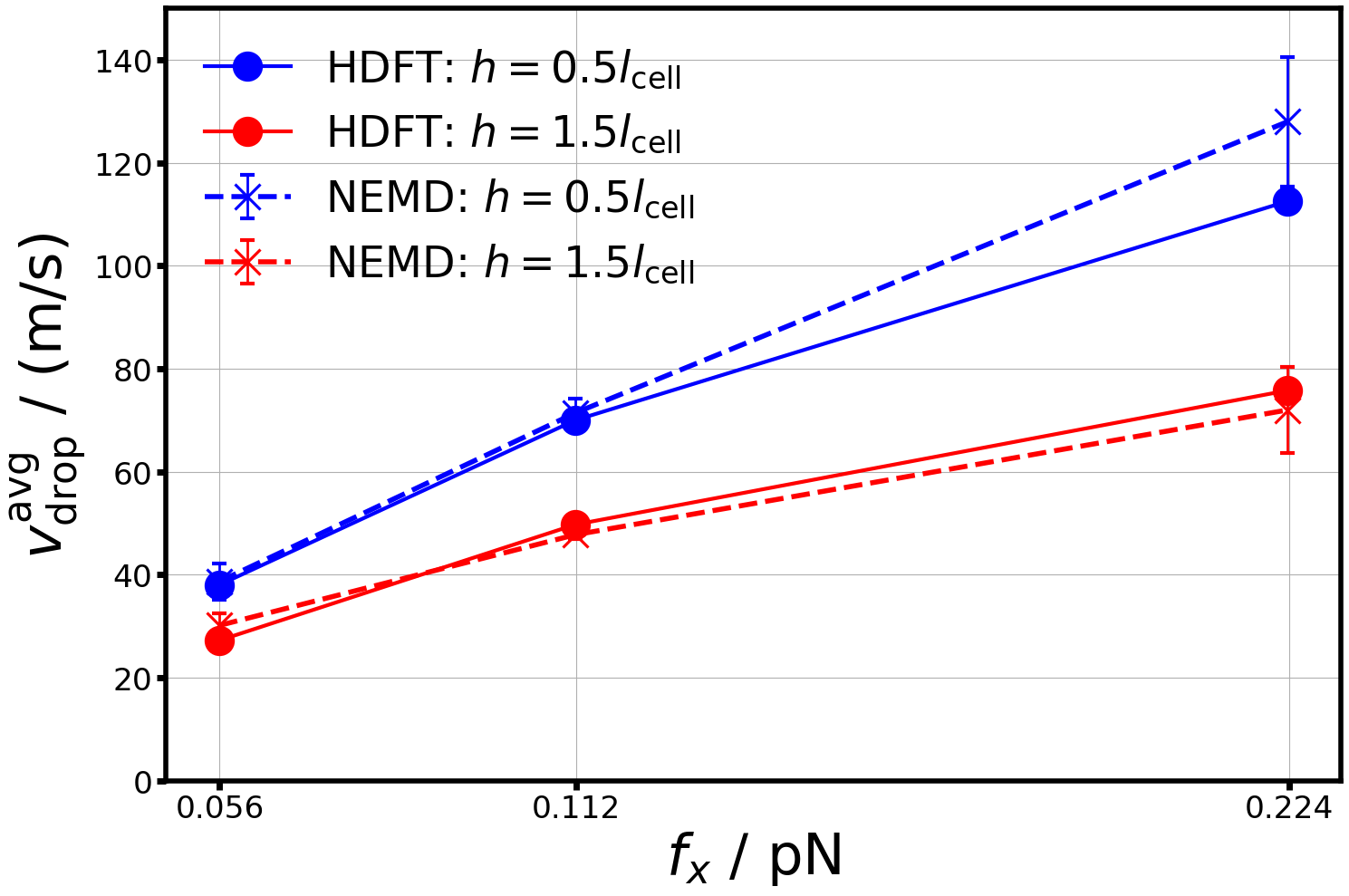}
  \caption{Steady state velocity of the centre of mass of the moving droplet for varying solid roughness $h$ determined for different external forces (per particle) from hydrodynamic DFT (HDFT) with entropy scaling viscosity model (circles) and from NEMD (crosses) at $T=\SI{120.02}{\kelvin}$. }
  \label{fig:vavg_roughness}
\end{figure}
A different behaviour is found for the steady state velocity as depicted in \cref{fig:vavg_roughness}. Smaller steady state velocities are found for the increased solid roughness for all forces, where results from hydrodynamic DFT and NEMD are in very good agreement. The reduced steady state velocity can be explained by increasing attractive solid-fluid interactions acting in the direction opposite of the droplet motion as a result of the changed geometry when increasing the solid roughness.  This increases the solid-fluid friction and leads to smaller steady state velocities. 
The hydrodynamic DFT model correctly describes the influence of the 
molecular roughness of the solid
on contact angles and steady state velocities. Notably, no information on different
molecular roughnesses of the solid 
entered the adjustment of the parameter for the generalised entropy scaling approach. 

\section{\label{sec:conclusion}Conclusion}

In this work we have conducted a two-dimensional and quantitative investigation of the predictive capabilities of hydrodynamic DFT with respect to wetting at the microscopic scale by studying droplets moving along the solid-fluid interface. 
The model captures the  influence of fluid-fluid interfaces on the dynamics by using a DFT term and the influence of the solid by an external potential.  Generalised entropy scaling, an approach which incorporates microscopic detail, is employed for the viscosity, where one transferable parameter  is adjusted to a single NEMD simulation of liquid-phase Poiseuille flow, i.e.\ no wetting information enters the model. 
We have evaluated the hydrodynamic DFT model through  comparison to results from NEMD simulations considering velocity profiles and steady state velocities, as well as density profiles and resulting dynamic contact angles. 

The study demonstrated good quantitative agreement between hydrodynamic DFT and NEMD simulations for a wide range of wetting phenomena and yields three major findings: 
First, the microscopic generalised entropy scaling model for the local viscosity using a single transferable parameter which does not include wetting data, ensures good quantitative agreement with NEMD results and more accurate results than a bulk viscosity model representing a typical continuum approach. 
Second, wetting phenomena at the microscopic scale including differences between advancing and receding contact angles, the transition from equilibrium to steady state and the rolling motion of droplets can be predicted with hydrodynamic DFT. 
Third, hydrodynamic DFT combined with generalised entropy scaling is transferable and captures the influence of different external forces, wetting strengths and 
molecular roughness of the solid
on the wetting behaviour. 

We conclude that hydrodynamic DFT, a unified molecular and continuum mechanics approach, is capable of predicting wetting phenomena at the microscopic scale while reducing to the Navier-Stokes equations far away from interfaces. It is therefore, a suitable candidate for the study of wetting on multiple scales, 
for example with application to mixtures in (multiphase) transport in porous media.

\section*{Funding}
Funded by the Deutsche Forschungsgemeinschaft (DFG, German Research Foundation) -- Project Number 327154368 -- SFB 1313. We thank the German Research Foundation (DFG) for supporting this work by funding EXC 2075/1-390740016 under Germany's Excellence Strategy as well as the Center for Digitalization and Technology Research of the Armed Forces of Germany (dtec.bw) through the project Macro/Micro-Simulation of Phase Decomposition in the Transcritical Regime (MaST); dtec.bw is funded by the European Union--NextGenerationEU\@. We acknowledge support by the Stuttgart Center for Simulation Science (SC SimTech). The authors acknowledge support by the state of Baden-Württemberg through bwHPC. 

\section*{Declaration of interests}
The authors have no conflict to disclose.

\section*{Data availability statement}
\textcolor{red}{The data set is currently unpublished and it will be published with the final version of this manuscript:} 
The data that support the findings of this study are openly available in the data repository of the University of Stuttgart (DaRUS) at \textcolor{red}{[Doi/Url]}, reference number \textcolor{red}{[reference number]}.

\appendix

\section{DFT with constraints}\label{sec:appendix_Nconst}
In this work MD simulations use the canonical ($N,V,T$) ensemble, whereas DFT is formulated in terms of the grand canonical ($\mu, V,T$) ensemble. In DFT, the number of molecules $N$ is not constant and different (but correct) solutions to the Euler-Lagrange equation can be obtained. 
Thus, algorithms keeping the number of molecules constant were proposed. Following \citet{rehner2018surface}, using the Lagrange multiplier $\lambda$, the unconstrained minimisation can be written as 
 \begin{equation}
    \mathcal{L}([\rho(\mathbf{r})], \lambda)=F+\int \rho(\rb) V^\mathrm{ext}(\rb)\dr     + \lambda \cdot\left(N-\int \rho(\mathbf{r}) \mathrm{d} \mathbf{r}\right) \stackrel{!}{=} \mathrm{min}
 \end{equation}
 After rearrangement this provides an additional equation to the Euler-Lagrange equation. The density profiles can be determined from  
 \begin{align}
    \rho & =z e^{-\beta\left(\frac{\delta F^\mathrm{res}}{\delta \rho}+V^{\operatorname{ext}}\right)} \\
    z & =\frac{N}{\int e^{-\beta\left(\frac{\delta F^\mathrm{res}}{\delta\rho}+V^{\mathrm{ext}}\right)} \mathrm{dr}}
 \end{align}
 with the ensemble-averaged number of molecules $N$ and the inverse thermodynamic temperature $\beta=\frac{1}{\kB T}$. Note that this approach can be used for mixtures where the ensemble-averaged number of molecules of each species is constant and with some adaption the total ensemble-averaged number of molecules can be kept constant. We note, that with these equations the system is not canonical; it is a mathematical modification to obtain the solution with the desired ensemble-averaged number of molecules in a grand-canonical ensemble \citep{rehner2018surface}.

 \section{Estimating the External Force} \label{sec:appendix_externalForce}
 An external (body) force is employed in this work to induce the movement of droplets. In NEMD the external force is added to the $x$-component of the force vector for each individual particle. In hydrodynamic DFT the external force~$f_x$ is added to the gradient of the external potential in the momentum balance as an acceleration~$\rho f_x$ according to 
 \begin{equation}
  \frac{\partial (\mw\rho \vb)}{\partial t} + \divd \left(\mw\rho \vb \vb^\intercal \right) = - \rho \nabla  \left( \frac{\delta  F}{\delta \rho} +  V^\mathrm{ext,sf} \right)  - \rho f_x - \divd \boldsymbol{\tau} \label{eq:MomentumBalanceDDFT_force}
 \end{equation}
 where $f_x$ has units of a force (per molecule). If divided by the molecular mass, $f_x/\mw$ is equivalent to the earth's gravitational acceleration $g$. 

 The movement of droplets is governed by the relation of the external driving force to capillary and viscous forces, cf. the momentum balance in \cref{eq:MomentumBalanceDDFT_force}. Compared to macroscopic droplets capillary forces have a much stronger influence on the droplet movement on the  microscopic scale. The reason is that the external force is a body force, which acts on the whole volume of the droplet, while the capillary force acts mostly in the contact region. Since on the microscopic scale, the relation of the contact region to the droplet volume is much larger than at the macroscopic scale, much larger external (body) forces are required to obtain a similar motion of the droplet. The magnitude of the force required to overcome the capillary forces and to cause the droplet to move can be estimated by assuming that the capillary forces act on the contact region only and by neglecting viscous forces. Then the influence of the capillary forces scales with $c/V$, where $c$ is the length of the contact region (or contact line). For both, spherical and cylindrical droplets $c/V \sim R^{-2}$, where $R$ is the radius of the respective droplet. Comparing a macroscopic droplet with $R=\SI{1}{\milli\meter}$ and gravitational acceleration of $g\approx\SI{10}{\meter\per\second\squared}$, then a droplet with  $R=\SI{1}{\nano\meter}$ needs to experience an acceleration of $f_x/\mw\approx\SI{10e12}{\meter\per\second\squared}$ for an equivalent motion. This is roughly the same order of magnitude as the values used for methane used in this study, which are between  \SI{2.1e12}{\meter\per\second\squared} and \SI{8.4e12}{\meter\per\second\squared}.

\section{Adjustment of Entropy Scaling Parameter} \label{sec:appendix_psi}

The parameter $\psi$ scales the convolution radius for the weighted density $\bar{\rho}^\mathrm{ES}$ in \cref{eq:rho_bar_ES_tildepsi} in order to quantitatively capture the influence of solid-fluid interactions on the viscosity close to the solid-fluid interface. As shown in previous work \citep{bursik2024viscosities}, one liquid-phase reference NEMD simulation is sufficient to obtain an accurate and transferable parameter estimate. In this work, we employ a liquid-phase NEMD simulation in the same geometry used for contact angle simulations (cf.\ \cref{fig:system}) at $T=\SI{120.02}{\kelvin}$, $f_x=\SI{0.112}{\pico\newton}$ and $\varepsilon_\mathrm{sf}=0.5$. Since the velocity profiles are not strongly sensitive to the exact numerical value of~$\psi$, we  content ourselves with a rough estimate instead of a rigorous optimisation.

\begin{figure}
  \centering
  \includegraphics[width=0.55\textwidth]{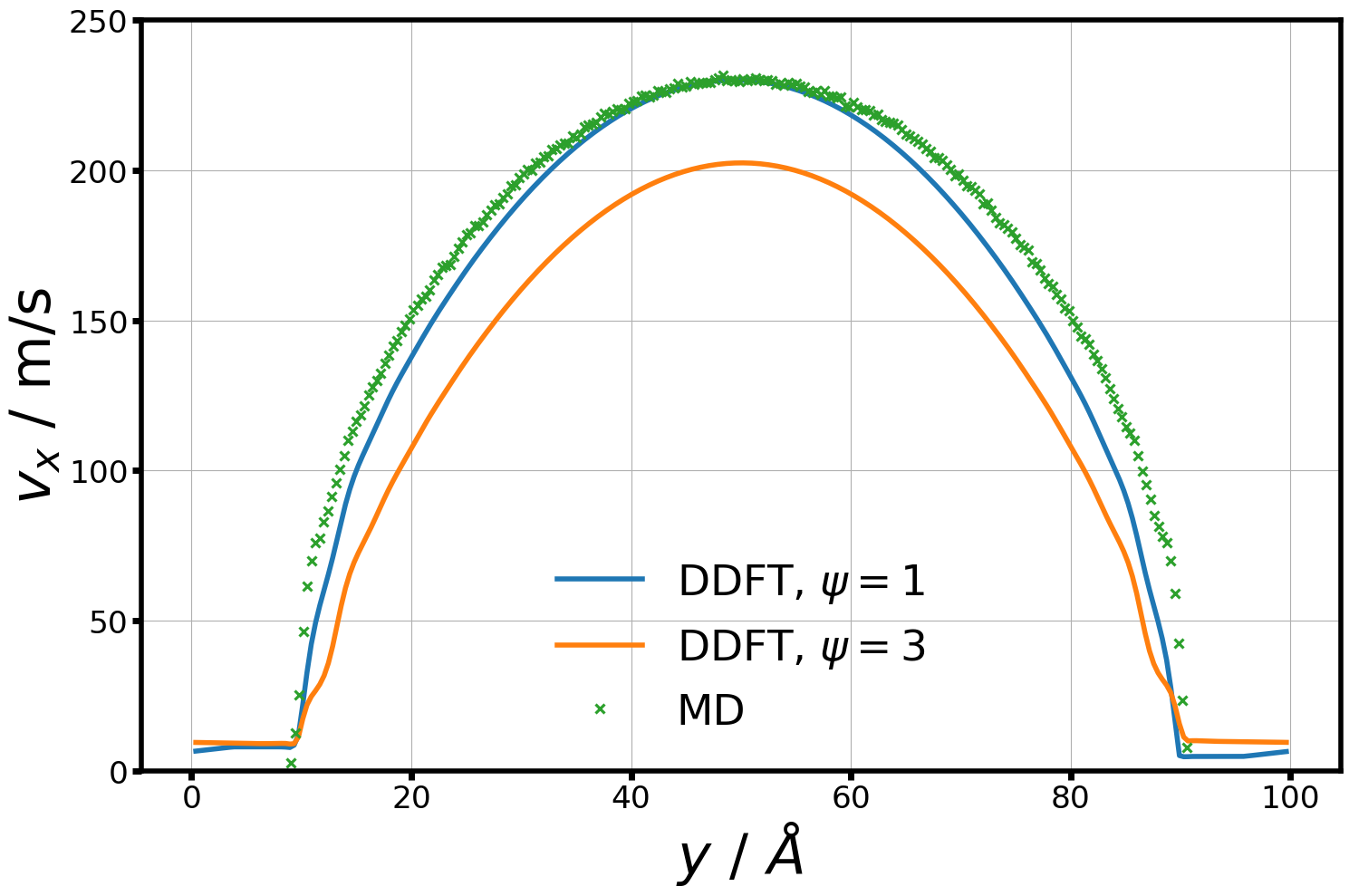}
  \caption{Velocity $v_x$ profiles from hydrodynamic DFT using different values for the parameter $\psi$ and from NEMD for $f_x=\SI{0.112}{\pico\newton}$, $T=\SI{120.02}{\kelvin}$ and $\varepsilon_\mathrm{sf}=0.5$.}
  \label{fig:psi_adjust}
\end{figure}
\Cref{fig:psi_adjust} provides the velocity parallel to the solid-fluid interface $v_x$ over the height of the system. Results are shown for hydrodynamic DFT with different values for $\psi$  and for NEMD. We note that not all values tested are shown here for clarity. If $\psi=3$ hydrodynamic DFT underestimates the velocity profile   compared to NEMD results, which suggests that the influence of solid-fluid interactions on the viscosity are overestimated. On the contrary, using $\psi=1$ the  agreement of the  velocity profiles is considered to be sufficient and $\psi=1$ is used throughout this work.

\section{Mechanism for Appearance of Dynamic Contact Angles} \label{sec:appendix_grad_vext}

\begin{figure}
  \centering
  \includegraphics[width=0.55\textwidth]{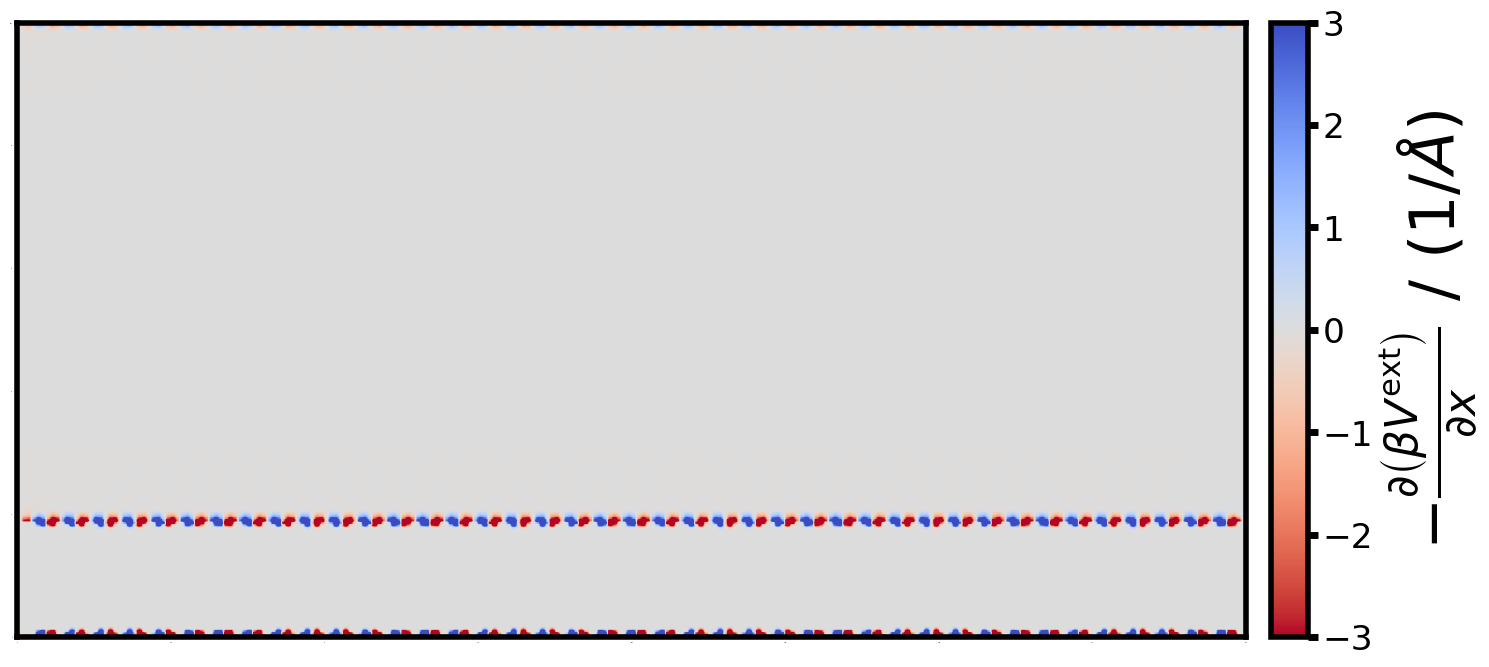}
  \caption{Gradient of the dimensionless external potential $\frac{\partial \left( \beta V^\mathrm{ext}\right)}{\partial x}$ in $x$-direction as used in hydrodynamic DFT. }
  \label{fig:grad_vext}
\end{figure}

In \cref{sec:results_hysteresis}, we argue that in hydrodynamic DFT the mechanism behind the appearance of advancing and receding contact angles  and their speed dependence is captured by the gradient of the external potential. 
\Cref{fig:grad_vext} shows the gradient of the (dimensionless) external potential in $x$-direction. Due to the molecular roughness of the solid, the gradient can become negative which corresponds to a reduction of the momentum in $x$-direction at this position. This is analogous to the potential wells and barriers described by molecular-kinetic theory \citep{blake1969kinetics,blake1993dynamic}.

\section{Quantifying the Degree of Rolling Motion } \label{sec:appendix_vorticity}

\begin{figure}
  \centering
  \includegraphics[width=0.55\textwidth]{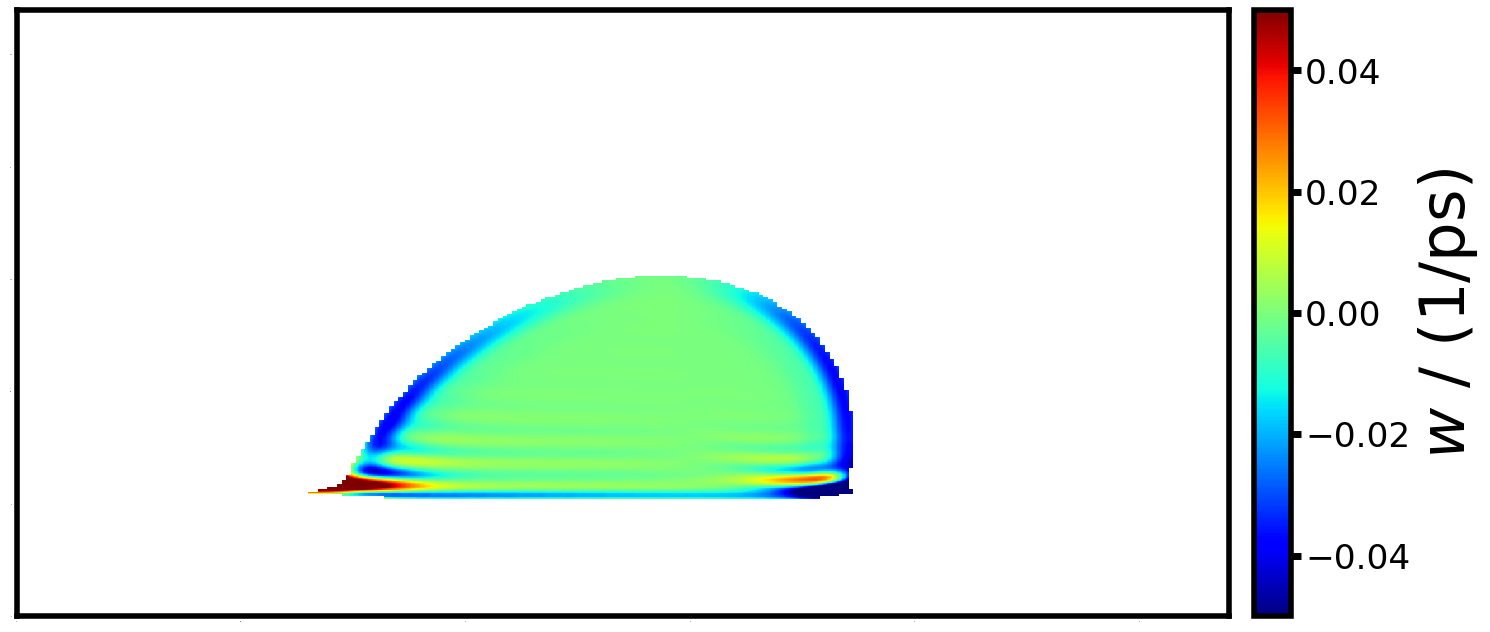}
  \caption{Length of vorticity vector $w = \left( \frac{\partial v_y}{\partial x} -  \frac{\partial v_x}{\partial y} \right)$ of the droplet determined from hydrodynamic DFT. }
  \label{fig:vorticity}
\end{figure}

In \cref{sec:results_v} it is determined that the droplets adhere to a rolling motion. 
This was supported by the relative velocity in $y$ direction, which is larger than $1/3$ of the maximum absolute velocity. 
To further quantify the degree of rolling motion, \cref{fig:vorticity} shows the length of the  vorticity vector $w$ of the two-dimensional flow field according to 
\begin{equation}
  w = \left( \frac{\partial v_y}{\partial x} -  \frac{\partial v_x}{\partial y} \right)
\end{equation}
where negative values correspond to a clockwise rotation. 
The largest negative values are obtained in the advancing and receding part of the droplet close to the vapour-liquid interface and are about \SI{-0.04}{\per\pico\second}. This would correspond to a full rotation in about \SI{25}{\pico\second}; since the droplet for the presented case requires about \SI{300}{\pico\second} to travel through the entire system (\SI{200}{\angstrom}), the rolling motion is significant. 

\section{Temperature in NEMD Simulations} \label{sec:appendix_temperature}

\begin{figure}
  \centering
  \includegraphics[width=0.55\textwidth]{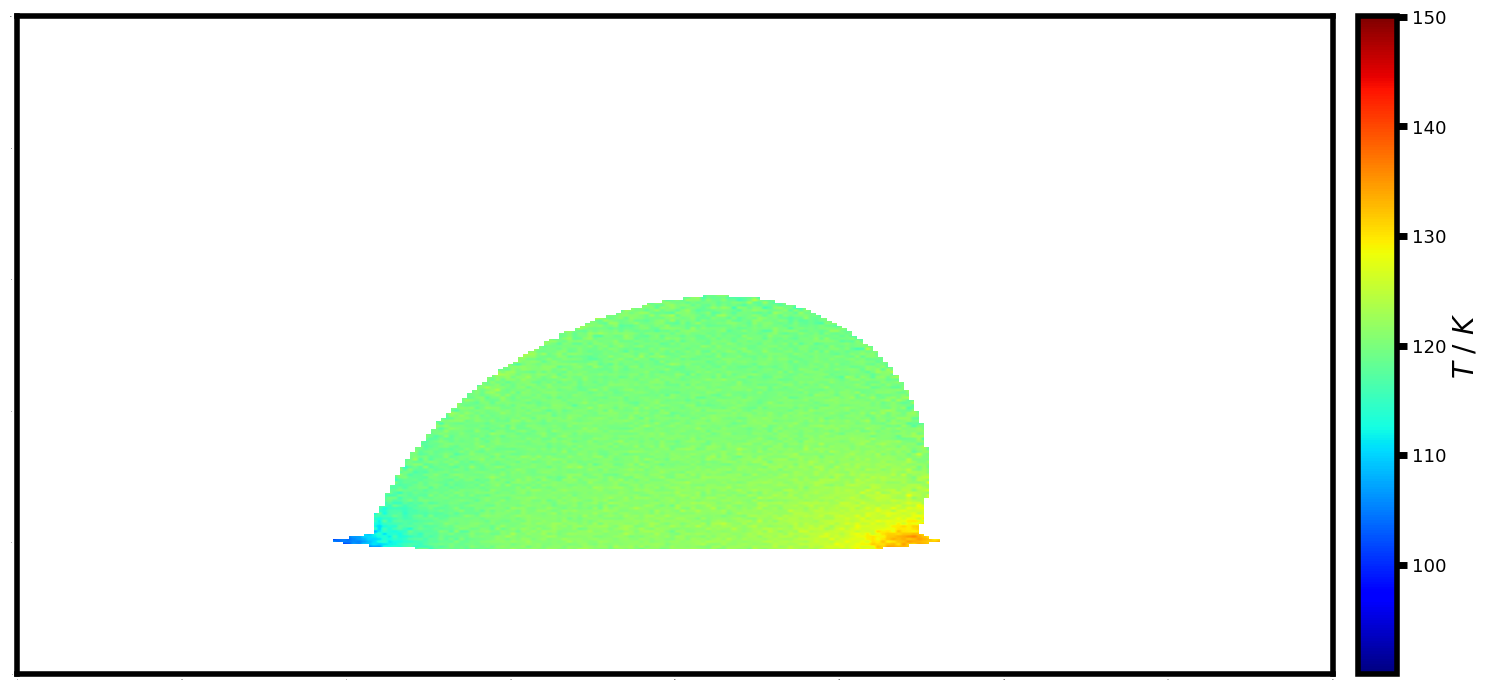}
  \caption{Temperature $T$ of the droplet for $f_x=\SI{0.112}{\pico\newton}$ determined from NEMD simulations and averaged over more than \SI{10000}{\pico\second}. }
  \label{fig:temperature}
\end{figure}

In the NEMD simulations a global Nose-Hoover thermostat is applied to keep the temperature constant at $T=\SI{120.02}{\kelvin}$. In a non-equilibrium simulation, shear effects can lead to local temperatures which deviate from the desired average temperature. 
\Cref{fig:temperature} shows the temperature from NEMD for the medium external force ($f_x=\SI{0.112}{\pico\newton}$). The temperature within the droplet is close to the desired temperature ($T=\SI{120.02}{\kelvin}$). The temperature is determined using relative velocities of molecules, which are obtained by subtracting the mean velocity in each bin from the absolute velocity of the molecules.  In the advancing contact region the temperature is slightly above \SI{130}{\kelvin}, whereas in the receding contact region the temperature is slightly below \SI{110}{\kelvin}. 
Thus, local temperature variations are observed using the global Nose-Hoover thermostat, within about 10\% from the desired value.

\newpage
\section*{References}

\bibliographystyle{jfm}
\bibliography{ddft_dynContactAngles}

\newpage
\section*{Supplementary Data to "Modelling Interfacial Dynamics Using Hydrodynamic Density Functional Theory: Dynamic Contact Angles and the Role of Local Viscosity"}

\subsection{\label{sec:appendix_eta_bulk} Contact Angles Using the Bulk Viscosity Model} 
Regarding velocity profiles and the velocity of the entire droplet, results from hydrodynamic DFT with the generalised entropy scaling model show much better agreement with NEMD results as compared to the bulk viscosity model (see the main text). In the following, we present results for dynamic contact angles using the bulk viscosity model.  

\begin{figure}[h!]
	\centering
	\subfloat[Hydrodynamic DFT with $f_x=\SI{0.056}{\pico\newton}$]{
		\centering
		\includegraphics[width=0.475\textwidth]{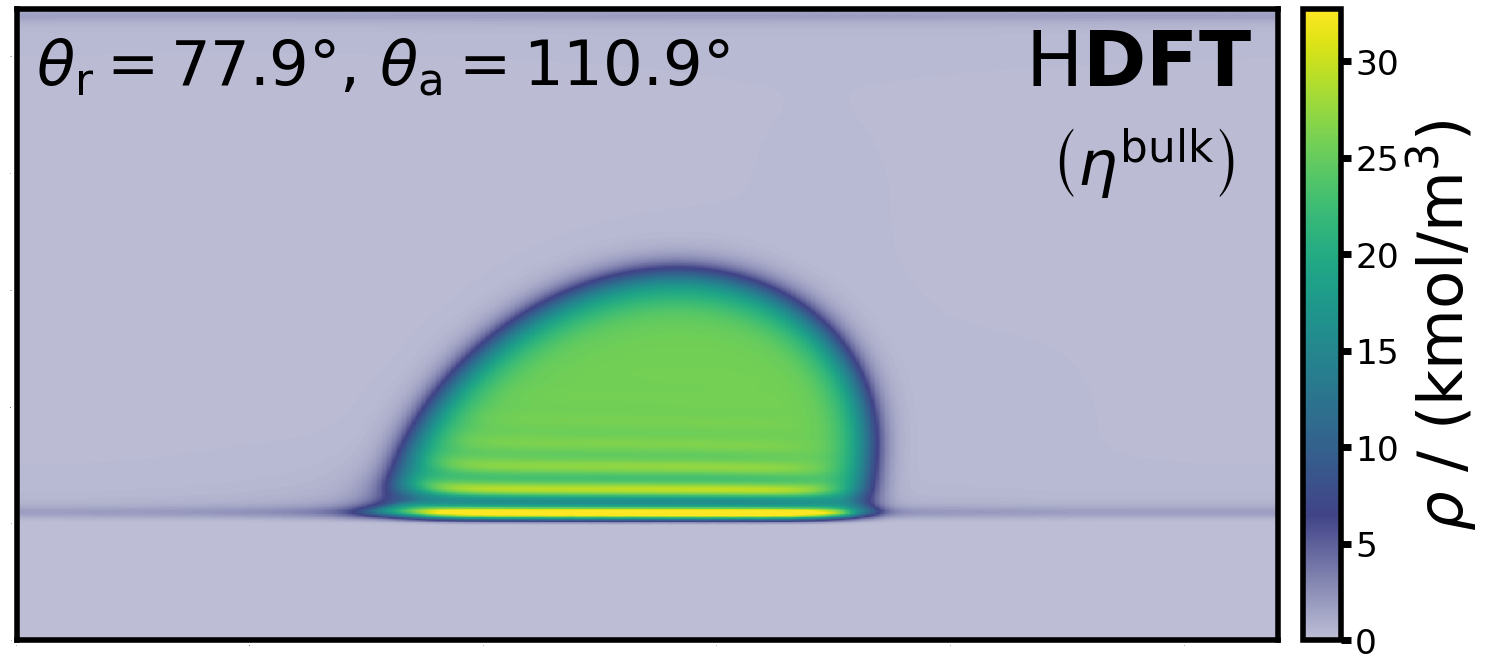}\label{fig:rho_dyn_hdft_01}
   }
   \hfill
	\subfloat[NEMD with $f_x=\SI{0.056}{\pico\newton}$]{
		\centering
		\includegraphics[width=0.475\textwidth]{contactAngle_dynamic_MD.png}\label{fig:rho_dyn_nemd_01}
   }
   \vfill
   \subfloat[Hydrodynamic DFT with $f_x=\SI{0.112}{\pico\newton}$]{
		\centering
		\includegraphics[width=0.475\textwidth]{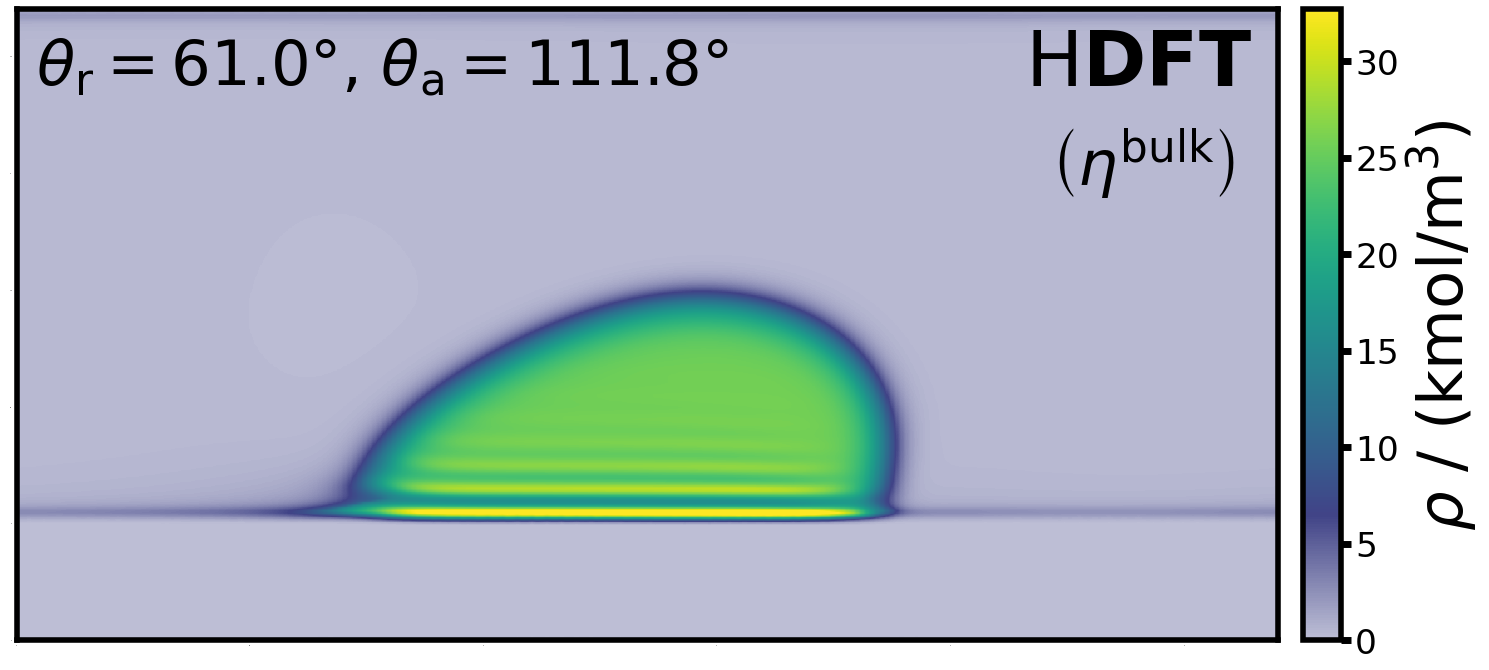}\label{fig:rho_dyn_hdft_02}
   }
   \hfill
	\subfloat[NEMD with $f_x=\SI{0.112}{\pico\newton}$]{
		\centering
		\includegraphics[width=0.475\textwidth]{contactAngle_dynamic_MD_fx002.png}\label{fig:rho_dyn_nemd_02}
   }
   \vfill
   \subfloat[Hydrodynamic DFT with $f_x=\SI{0.224}{\pico\newton}$]{
		\centering
		\includegraphics[width=0.475\textwidth]{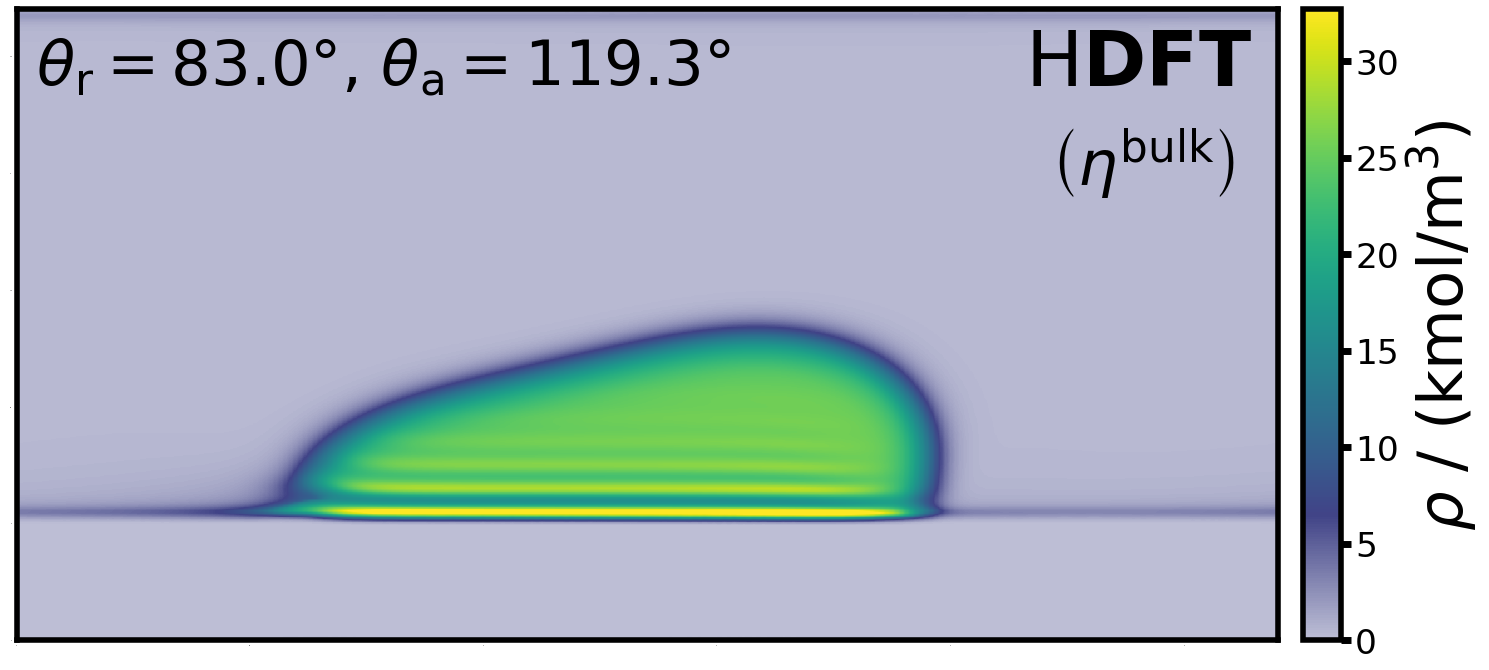}\label{fig:rho_dyn_hdft_04}
   }
   \hfill
	\subfloat[NEMD with $f_x=\SI{0.224}{\pico\newton}$]{
		\centering
		\includegraphics[width=0.475\textwidth]{contactAngle_dynamic_MD_fx004.png}\label{fig:rho_dyn_nemd_04}
   }
	\caption{Density profiles of droplets moving along the solid-fluid interface with different external forces $f_x$ from hydrodynamic DFT (HDFT) using the bulk viscosity model and from NEMD at $T=\SI{120.02}{\kelvin}$  with $\varepsilon_\mathrm{sf}^*=0.5$  averaged over $\SI{700}{\pico\second}$ after a steady state is reached. 
	 }
	\label{fig:SI_rho_dyn}
  \end{figure} 
The effect of the bulk viscosity model on density profiles is presented in \cref{fig:SI_rho_dyn}. At the medium force ($f_x=\SI{0.112}{\pico\newton}$, see \cref{fig:rho_dyn_hdft_02}) the shape of the droplet shows slightly stronger deformations compared to NEMD (see \cref{fig:rho_dyn_nemd_02}) and hydrodynamic DFT with entropy scaling (cf.  the main text).  This can be observed from the decreased height of the droplet and its increased width, i.e.\ the distance between  the advancing and receding contact regions. For the lowest force ($f_x=\SI{0.056}{\pico\newton}$, \cref{fig:rho_dyn_hdft_01} and \cref{fig:rho_dyn_nemd_01}) this effect can be observed to a lesser extent. At the largest external force studied here ($f_x=\SI{0.224}{\pico\newton}$, \cref{fig:rho_dyn_hdft_04}) the droplet, besides being more elongated, also shows a significantly non-spherical shape in strong contrast to results from NEMD (\cref{fig:rho_dyn_nemd_04}). This  manifests itself in an almost straight vapour-liquid  interface in the top-centre of the droplet and an increased curvature at the contact regions, which also renders it difficult to determine the contact angle with the methodology described in the main text.     

  \begin{figure}
	\centering
	\includegraphics[width=0.6\textwidth]{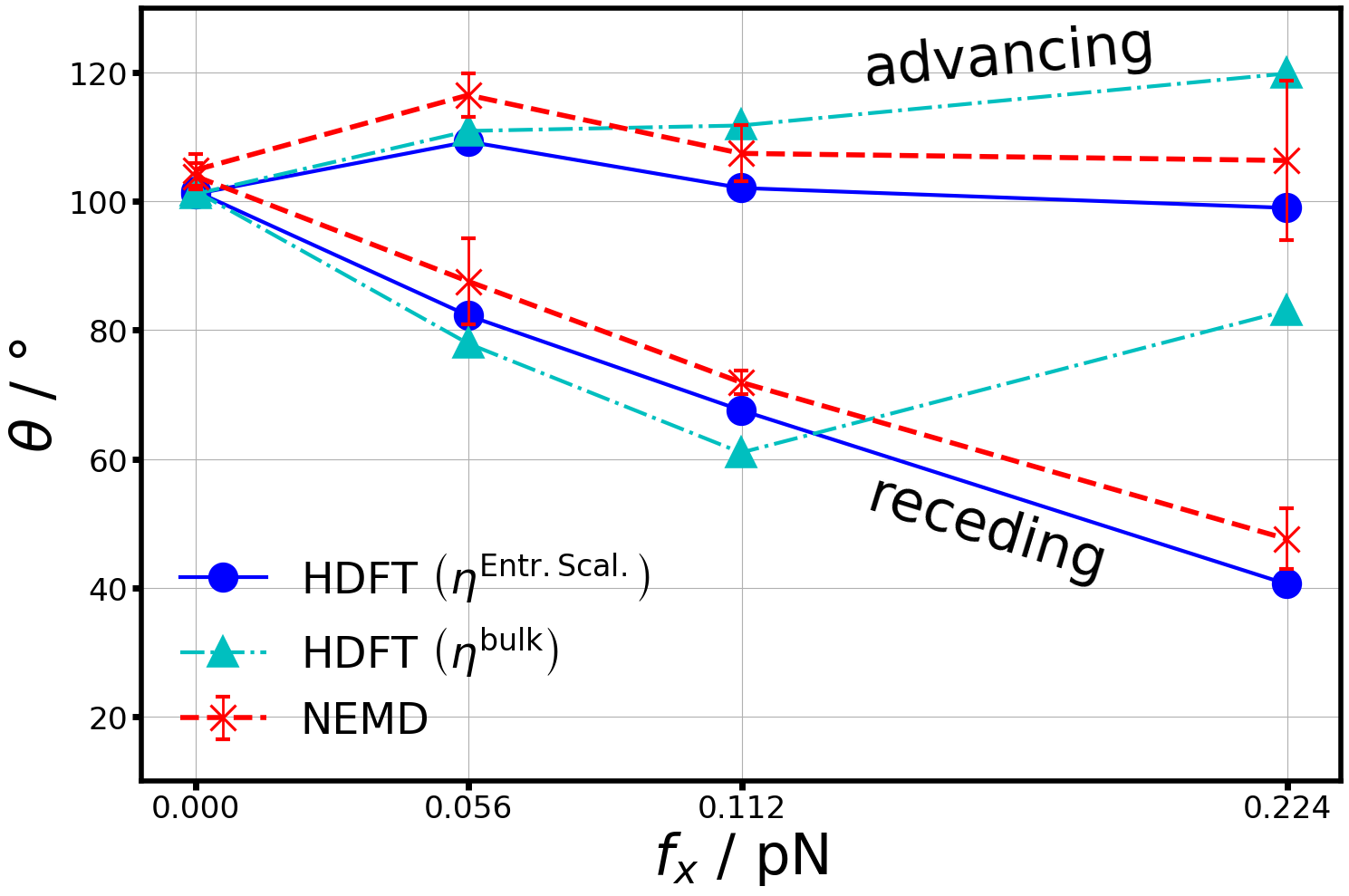}
	\caption{Summary of advancing and receding contact angles from hydrodynamic DFT with generalised entropy scaling viscosity model (blue points) and with the bulk viscosity model (light blue triangles) as well as from NEMD (red crosses) for different external forces at $T=\SI{120.02}{\kelvin}$  and with $\varepsilon_\mathrm{sf}^*=0.5$.}
	\label{fig:contactAngle_forces}
  \end{figure}
From these density profiles, contact angles for the bulk viscosity model are determined. They are contrasted to results from NEMD and hydrodynamic DFT with entropy scaling in \cref{fig:contactAngle_forces}. At the lowest force ($f_x=\SI{0.056}{\pico\newton}$), the contact angles differ only slightly between hydrodynamic DFT with the two viscosity models. At the medium force ($f_x=\SI{0.112}{\pico\newton}$) the receding contact angle is underestimated compared to NEMD. However, the advancing contact angle is overestimated. Consequently, the contact angle hysteresis $\Theta_\mathrm{a}-\Theta_\mathrm{r} = \SI{51,6}{\degree}$ is significantly larger for the simplified model than for both the NEMD results ($\Theta_\mathrm{a}-\Theta_\mathrm{r} = \SI{35,5}{\degree}$) and hydrodynamic DFT with the entropy scaling model ($\Theta_\mathrm{a}-\Theta_\mathrm{r} = \SI{34,5}{\degree}$). This is in agreement with the elongated shape of the droplet provided in \cref{fig:rho_dyn_hdft_02}. At the largest force ($f_x=\SI{0.224}{\pico\newton}$), the advancing contact angle  again overestimates results from NEMD, whereas the receding contact angle deviates strongly from the other models.
According to these results, density profiles and contact angles determined with the bulk viscosity model deviate from the other models. The deviations increase with increasing external force.
As discussed in the main text, a plausible explanation for these results is that the simplified viscosity model does not provide an accurate description of the viscosity in the droplet. 

\subsection{Numerical Values for Dynamic Contact Angles}

The numerical values for dynamic contact angles presented in this work are provided in \cref{table:angles}.

\begin{table}[h!]
  \centering
  \renewcommand{\arraystretch}{1.3}
  \setlength{\tabcolsep}{1.1pt}
    \begin{tabular}{c @{\hspace{3\tabcolsep}} c @{\hspace{3\tabcolsep}} c @{\hspace{3\tabcolsep}} c @{\hspace{3\tabcolsep}} c  c @{\hspace{3\tabcolsep}} c c @{\hspace{3\tabcolsep}} c c @{\hspace{3\tabcolsep}} c c }
       Section & Method & $h/l_\mathrm{cell}$ & $\epssf$ &  \multicolumn{2}{c@{\hspace{3\tabcolsep}}}{$f_x=\SI{0.0}{\pico\newton}$} & \multicolumn{2}{c@{\hspace{3\tabcolsep}}}{$f_x=\SI{0.056}{\pico\newton}$} & \multicolumn{2}{c@{\hspace{3\tabcolsep}}}{$f_x=\SI{0.112}{\pico\newton}$} & \multicolumn{2}{c@{\hspace{3\tabcolsep}}}{$f_x=\SI{0.224}{\pico\newton}$} \\
       &  & & & $\Theta_\mathrm{right}$ & $\Theta_\mathrm{left}$ & $\Theta_\mathrm{a}$ & $\Theta_\mathrm{r}$ & $\Theta_\mathrm{a}$ & $\Theta_\mathrm{r}$ & $\Theta_\mathrm{a}$ & $\Theta_\mathrm{r}$\\
      \midrule
      \multirow{2}{*}{4.4}& HDFT  + Entr. Scal. & 0.5 & 0.5 & \SI{101.1}{\degree} &  \SI{101.6}{\degree} & \SI{109.3}{\degree} &  \SI{82.2}{\degree} & \SI{102.1}{\degree} &  \SI{67.6}{\degree} & \SI{99.0}{\degree} &  \SI{40.6}{\degree} \\
      &NEMD& 0.5 & 0.5  & \SI{104.9}{\degree} &  \SI{103.9}{\degree} & \SI{116.5}{\degree} &  \SI{87.6}{\degree} & \SI{107.4}{\degree} &  \SI{71.9}{\degree} & \SI{106.3}{\degree} &  \SI{47.6}{\degree} \\
      \cref{sec:appendix_eta_bulk} &HDFT + bulk& 0.5 & 0.5 & \SI{101.1}{\degree} &  \SI{101.6}{\degree} & \SI{110.9}{\degree} & \SI{77.9}{\degree}  & \SI{111.8}{\degree} &  \SI{61.0}{\degree} & \SI{119.3}{\degree} & \SI{83.0}{\degree}  \\
      \midrule
      \multirow{2}{*}{4.6}&HDFT  + Entr. Scal.& 0.5  &0.7 & \SI{61.5}{\degree} &  \SI{63.6}{\degree} & \SI{70.7}{\degree} &  \SI{42.0}{\degree} & \SI{68.6}{\degree} &  \SI{24.3}{\degree} & - &  - \\
      &NEMD& 0.5 & 0.7  & \SI{68.3}{\degree} &  \SI{65.8}{\degree} & \SI{69.4}{\degree} &  \SI{43.5}{\degree} & \SI{70.2}{\degree} &  \SI{22.5}{\degree} & - &  - \\
      \midrule
      \multirow{2}{*}{4.7}&HDFT  + Entr. Scal.& 1.5  &0.7 & \SI{97.8}{\degree} &  \SI{98.1}{\degree} & \SI{111.4}{\degree} &  \SI{80.0}{\degree} & \SI{108.3}{\degree} &  \SI{64.5}{\degree} & \SI{96.7}{\degree} &  \SI{38.1}{\degree} \\
      &NEMD& 1.5 & 0.7  & \SI{97.8}{\degree} &  \SI{94.3}{\degree} & \SI{115.0}{\degree} &  \SI{77.2}{\degree} & \SI{107.3}{\degree} &  \SI{61.4}{\degree} & \SI{99.0}{\degree} & \SI{28.2}{\degree} \\
  \end{tabular}
  \caption{Contact angles for different driving forces from hydrodynamic DFT (HDFT) using either the entropy scaling or the bulk viscosity model and from NEMD. }
  \label{table:angles}
\end{table} 

\subsection{Relation Between Contact Angles and Contact Region Velocity } 

\begin{figure}[h!]
  \centering
  \includegraphics[width=0.55\textwidth]{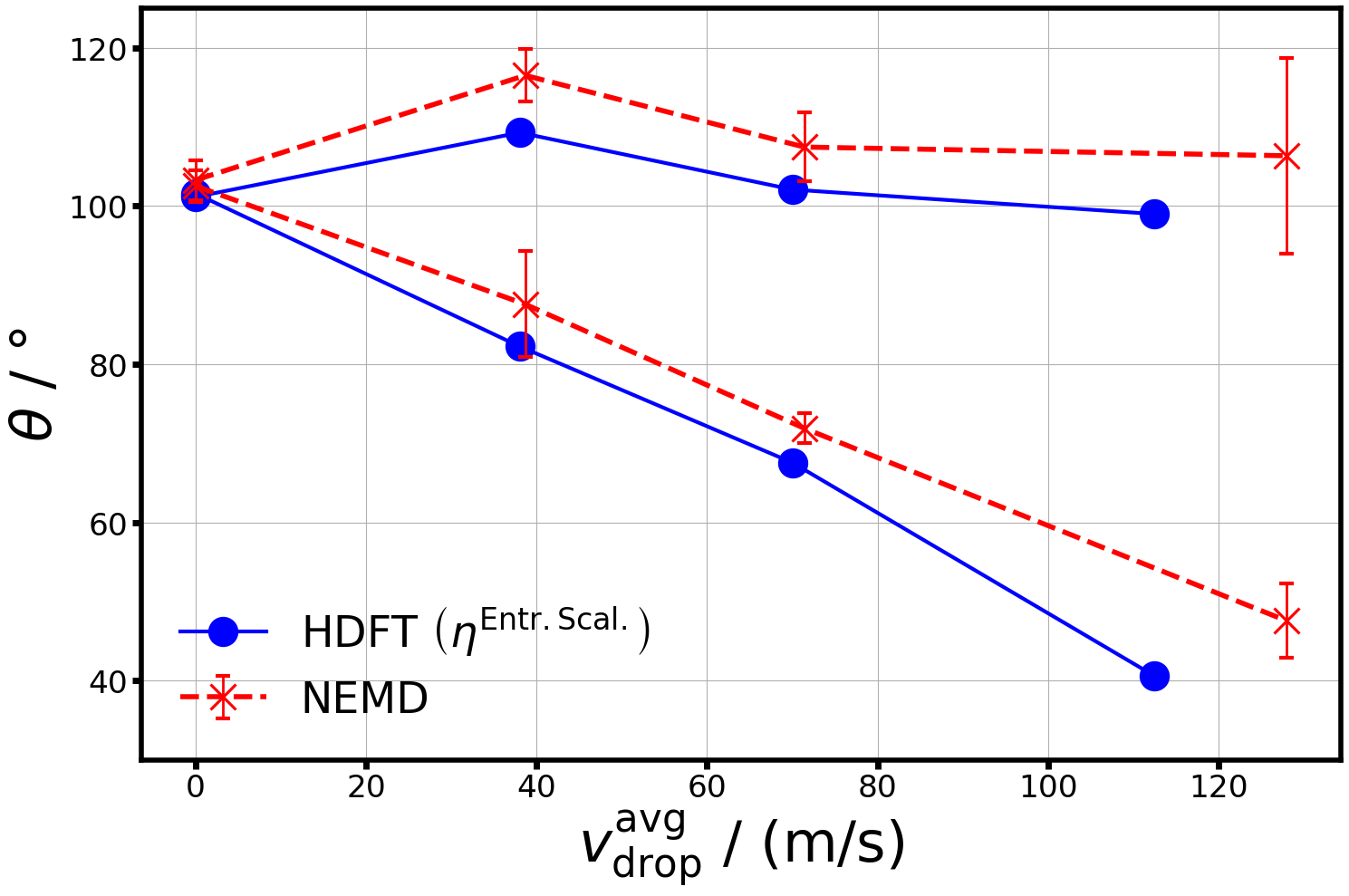}
  \caption{Contact angles as a function of the steady state velocity of the contact region from hydrodynamic DFT (HDFT) with entropy scaling viscosity model (circles) and from NEMD (crosses) at $T=\SI{120.02}{\kelvin}$. }
  \label{fig:theta_vs_vavg}
\end{figure}

\begin{figure}[h!]
  \centering
  \includegraphics[width=0.55\textwidth]{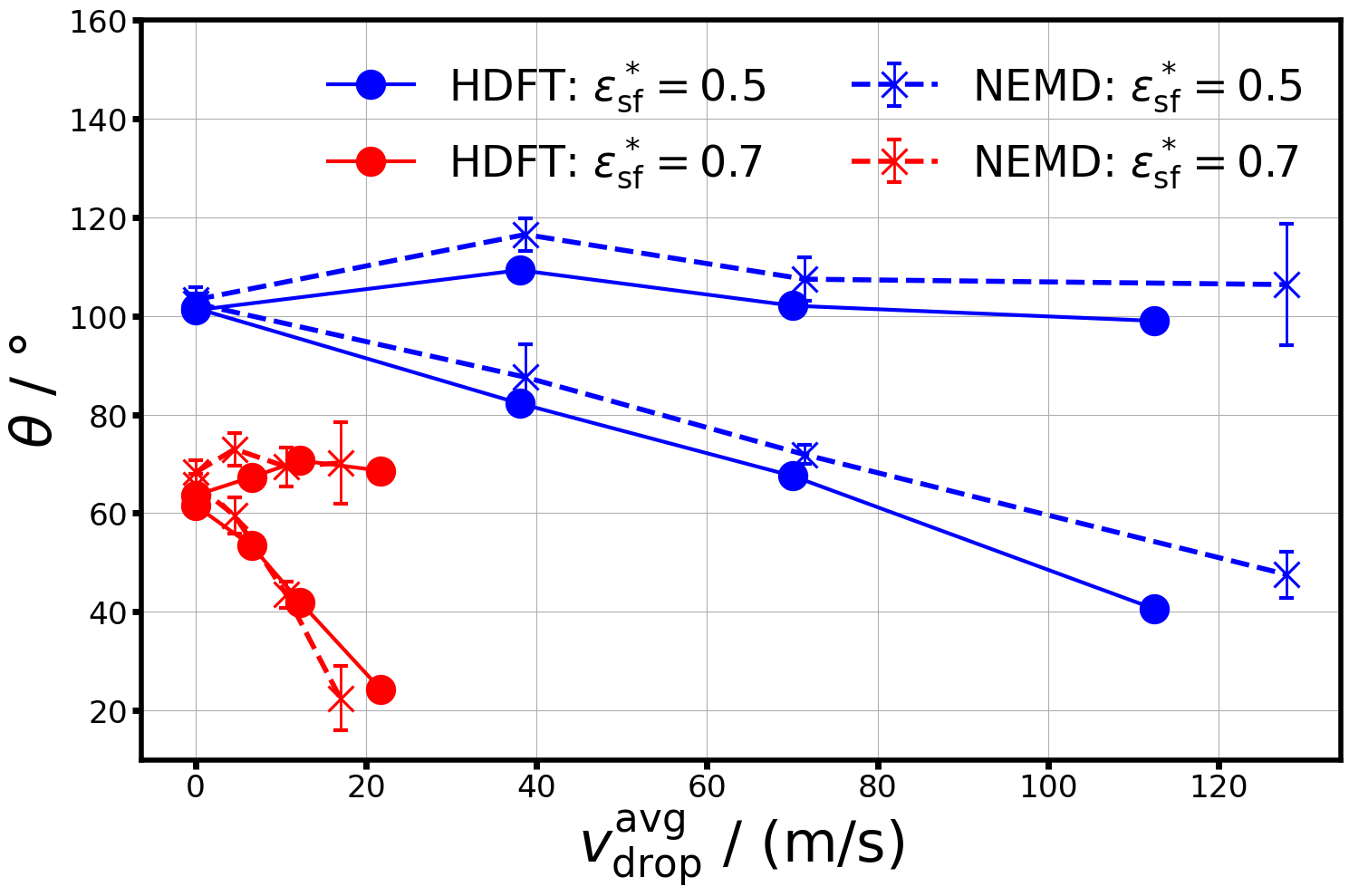}
  \caption{Contact angles as a function of the steady state velocity of the contact region for varying solid-fluid interaction parameter $\varepsilon_\mathrm{sf}^*$ from hydrodynamic DFT (HDFT) with entropy scaling viscosity model (circles) and from NEMD (crosses) at $T=\SI{120.02}{\kelvin}$. }
  \label{fig:theta_vs_vavg_wetting}
\end{figure}

\begin{figure}[h!]
  \centering
  \includegraphics[width=0.55\textwidth]{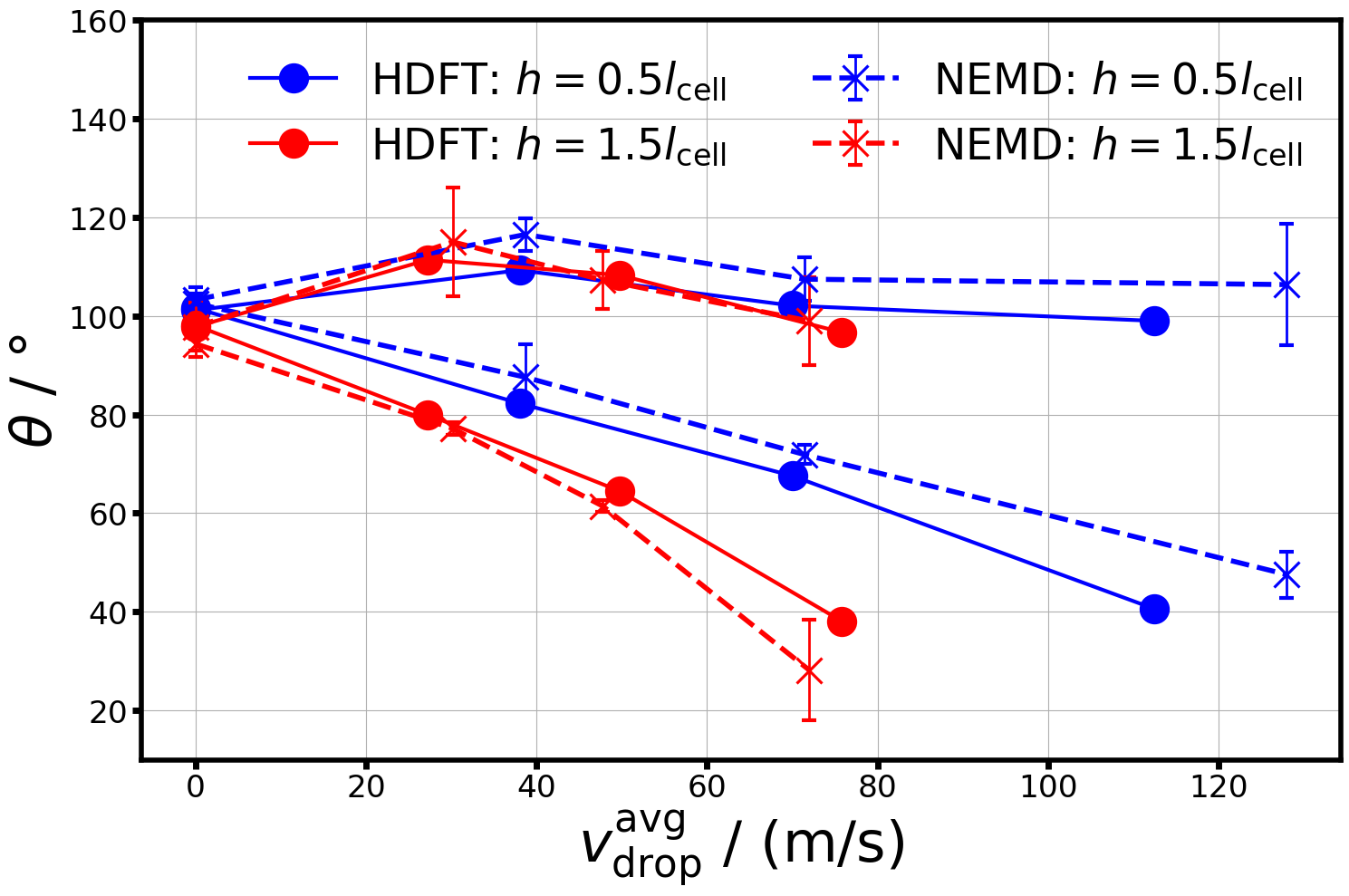}
  \caption{Contact angles as a function of the steady state velocity of the contact region for varying solid roughness $h$ from hydrodynamic DFT (HDFT) with entropy scaling viscosity model (circles) and from NEMD (crosses) at $T=\SI{120.02}{\kelvin}$. }
  \label{fig:theta_vs_vavg_roughness}
\end{figure}

Macroscopic studies are often concerned with the dependence of the (dynamic) contact angles on the velocity of the contact region (or contact line). At steady state the contact region moves with the same velocity as the centre of mass of the droplet. Thus, the dependence of contact angles on the velocity of the contact region can readily be visualised for the droplets studied in this work (see figures 20-22). Similar trends are observed as in the case where the contact angles are shown as a function of external force. This is expected, since the average velocity increases  monotonically  with the external force. In all cases (for varying solid-fluid interaction energy and varying solid roughness), the agreement between results from hydrodynamic DFT and NEMD is satisfactory.

\end{document}